%% file: ms.tex
\documentclass[12pt,preprint]{aastex}

\slugcomment{Accepted to AJ, 2006 September 23}

\shorttitle{Ultracool Luminosity Function.}

\shortauthors{Cruz et al.}

\received{2006 July 28}
\begin{document}
\newcommand{\LF}{$\Phi(M_J)$ and $\Phi(M_{K_S})$}
\newcommand{\LFj}{$\Phi(M_J)$}
\newcommand{\LFk}{$\Phi(M_{K_S})$}
\newcommand{\LFobjects}{99}
\newcommand{\LFsystems}{91}

\newcommand{\multiple}{10}
\newcommand{\add}{9}

\newcommand{\newspectra}{198}

\newcommand{\latenear}{12}
\newcommand{\latefar}{94}
\newcommand{\earlyfar}{34}
\newcommand{\giants}{33}
\newcommand{\carbon}{17}

\title{Meeting the Cool Neighbors. IX. The Luminosity Function of M7--L8
Ultracool Dwarfs in the Field}

\author{Kelle L. Cruz\altaffilmark{1,2,3,4}, I. Neill
Reid\altaffilmark{3,4,5}, J. Davy Kirkpatrick\altaffilmark{6},
Adam J. Burgasser\altaffilmark{7}, James
Liebert\altaffilmark{3,8}, Adam R. Solomon\altaffilmark{1,9,10},
Sarah J. Schmidt\altaffilmark{1,11,12}, Peter R.
Allen\altaffilmark{3,13}, Suzanne L. Hawley\altaffilmark{12}, and
Kevin R. Covey\altaffilmark{12} }

\altaffiltext{1}{Department of Astrophysics, American Museum of
Natural History, Central Park West at 79th Street, New York, NY
10024; \email{kelle@amnh.org}}

\altaffiltext{2}{NSF Astronomy and Astrophysics Postdoctoral
Fellow}

\altaffiltext{3}{Visiting Astronomer, Kitt Peak National
Observatory, National Optical Astronomy Observatory, which is
operated by the Association of Universities for Research in
Astronomy, Inc. (AURA) under cooperative agreement with the
National Science Foundation.}

\altaffiltext{4}{Visiting Astronomer, Cerro Tololo Inter-American
Observatory, National Optical Astronomy Observatories, which are
operated by the Association of Universities for Research in
Astronomy, under contract with the National Science Foundation.}

\altaffiltext{5}{Space Telescope Science Institute, 3700 San
Martin Drive, Baltimore, MD 21218 }

\altaffiltext{6}{Infrared Processing and Analysis Center, MS
100-22, California Institute of Technology, Pasadena, CA 91125}

\altaffiltext{7}{Massachusetts Institute of Technology, Kavli
Institute for Astrophysics and Space Research, Building 37, Room
664B, 77 Massachusetts Avenue, Cambridge, MA 02139}

\altaffiltext{8}{Department of Astronomy and Steward Observatory,
University of Arizona, Tucson, AZ 85721}

\altaffiltext{9}{John F. Kennedy High School, 3000 Bellmore
Avenue, Bellmore, NY 11710}

\altaffiltext{10}{Department of Astronomy, Yale University, P.O.
Box 208101, New Haven, CT 06520}

\altaffiltext{11}{Department of Physics and Astronomy, Barnard
College, Columbia University, 3009 Broadway, New York, NY 10027}

\altaffiltext{12}{Department of Astronomy, University of
Washington, Box 351580, Seattle, WA 98195}

\altaffiltext{13}{Department of Astronomy and Astrophysics,
Pennsylvania State University, 525 Davey Lab, University Park, PA
16802}

\begin{abstract}

We present a 20-pc, volume-limited sample of M7--L8 dwarfs created
through spectroscopic follow-up of sources selected from the Two
Micron All Sky Survey (2MASS) Second Incremental Release Point
Source Catalog. In this paper, we present optical spectroscopy of
\newspectra\ candidate nearby ultracool dwarfs, including
\latenear\ late-M and L dwarfs likely to be within 20~pc of the
Sun and \latefar\ more distant late-type dwarfs. We have also
identified five ultracool dwarfs with spectral signatures of
low-gravity. Combining these data with previous results, we define
a sample of \LFobjects\ ultracool dwarfs in \LFsystems\ systems
within 20~pc. These are used to estimate the \textit{J}- and
\textit{K}-band luminosity functions for dwarfs with optical
spectral types between M7 and L8 ($10.5 < M_J < 15$, $9.5< M_{K_S}
<13$). We find a space density of $4.9\times10^{-3}$~pc$^{-3}$ for
late-M dwarfs (M7--M9.5) and a lower limit of
$3.8\times10^{-3}$~pc$^{-3}$ for L dwarfs.

\end{abstract}


\section{Introduction}

Over the last several years, we have been using the Two Micron All
Sky Survey \citep[2MASS,][]{2MASS} to carry out a census of
late-type dwarfs in the vicinity of the Sun. In particular, we
have undertaken the first comprehensive survey of ultracool M and
L dwarfs in the immediate Solar Neighborhood \citep[hereafter
Paper~V]{Cruz03}. \defcitealias{Cruz03}{Paper~V} This population
of cool, low-luminosity dwarfs includes both very low-mass stars
and substellar-mass brown dwarfs
\citep{Chabrier00_Review,Burrows01}.

One of the primary aims of our program is the derivation of the
luminosity function (the number of objects as a function of
absolute magnitude) across the hydrogen-burning limit. The local
field luminosity function has been well measured for objects as
cool as late-M dwarfs \citep{PMSU4}, but studies probing to
fainter magnitudes have been limited by poor statistics and low
sensitivity \citep{Reid99,K99,K00,NN,Burgasser02_thesis}. Pushing
to lower luminosities and larger volumes is crucial to measuring
the contribution of brown dwarfs to the local space density.

The initial results from our program to create a statistically
well-defined sample of nearby ultracool dwarfs were presented in
\citetalias{Cruz03}. In that paper, we described how we
constructed a sample of 630 candidate nearby late-M and L dwarfs
from the 2MASS Second Incremental Data Release Point Source
Catalog, designated the ``2MU2'' sample. We also presented
follow-up spectroscopic observations of a subset of the ultracool
candidates. In this paper, we present the remainder of the far-red
optical spectroscopic observations and combine these results with
previous observations to construct the \textit{J}- and
\textit{K$_S$}-band luminosity functions. These data include many
newly discovered late-type dwarfs, including twelve additions to
the core ``20-pc 2MU2'' sample of M7--L8 dwarfs estimated to be
within 20~parsecs and used to measure the luminosity function.

This is the ninth paper in the series and follows directly from
\citetalias{Cruz03}. The first three papers and Papers~VII
and~VIII \citep{Paper1, Paper2, Cruz02,Paper7,Paper8} concentrate
on proper-motion selected K and M dwarfs. Paper IV \citep{Paper4}
describes the discovery of \object{2MASS J18353790+3259545}, an
M8.5 dwarf within 6~pc of the Sun (which is included in the
present analysis). Results from a search for ultracool dwarfs
lying close to the Galactic Plane are presented in \citet[Paper
VI]{Paper6}.

Here, we build on \citetalias{Cruz03} and present the remainder of
the far-red optical spectroscopic follow-up of the 2MU2 sample and
derive the \textit{J}- and \textit{K$_S$}-band luminosity
functions based on objects estimated to be within 20~pc. In
\S~\ref{sec:9sample} we overview the characteristics of the 2MU2
sample including its creation and follow-up status. Spectroscopic
observations from seven telescopes over sixteen runs are described
in \S~\ref{sec:9obs}. The results of the spectroscopy and the
methods used to estimate spectral types, absolute magnitudes, and
distances are in \S~\ref{sec:9results}. Interesting individual
objects are presented in \S~\ref{sec:individual}. In
\S~\ref{sec:9LF} we examine the completeness of the 20-pc 2MU2
sample and derive the luminosity function. Finally, we discuss
these results in \S~\ref{sec:9discussion} and briefly summarize
our findings in \S~\ref{sec:9summary}.

\section{The 2MU2 Sample}
\label{sec:9sample}

The creation of the 2MU2 sample is meticulously discussed in
\citetalias{Cruz03}; however, we briefly summarize the selection
criteria and procedures here, making particular note of the minor
changes that have since been applied. Eleven million point sources
were selected from the 2MASS Second Incremental Data Release Point
Source Catalog (PSC) with $|b|>10\degr$, $(J-K_S)>1.0$, and
$J<16.5$. The sample was narrowed to 1672 objects using optical
and near-infrared color-color and color-magnitude
criteria\footnote{\citetalias[Equation~1][]{Cruz03} incorrectly
stated the slope of the $J, (J-K_S)$ selection criterion as 1.5
where the correct value is 3. The line is plotted correctly in
Figure~1 of \citetalias{Cruz03}.}, and excluding sources with
positional coincidence with star formation regions (e.g., Taurus,
Orion), high star density regions (e.g., LMC, M31), and other
reddening regions \citepalias[listed in][Table 3]{Cruz03}. Below
we discuss the further reduction of the sample to 630 objects, and
the follow-up status of the 518 candidates requiring additional
observations.

\subsection{Narrowing and Refining the Sample}

As discussed in \citetalias{Cruz03}, 588 of the 1672 sources in
the 2MU2 sample have $J\leq9$. Those bright sources were
considered separately in the appendix of \citetalias{Cruz03}, and
they include no new, nearby late-type dwarfs. Ninety-nine of the
remaining 1084 candidates were eliminated based either on previous
observations listed in the SIMBAD
database\footnote{\url{http://simbad.u-strasbg.fr/}} (24),
positional coincidence with known interstellar clouds (41), or
near-infrared colors consistent with mid-type M dwarfs (34 tagged
``bland''). We have eliminated a further 355 stars based on the
following criteria:

\begin{description}

\item{Visual inspection.---} We originally identified 211 sources
in \citetalias{Cruz03} as either artifacts or as being associated
with large galaxies or globular clusters. A further eight sources
have been added to this list, giving a total of 219 sources
eliminated.

\item{$(F-J)/(J-K_S)$.---} We cross-referenced the sample against
the HST Guide Star Catalog v2.2 \citep{GSC}, and used the
photographic \textit{F}-band (red) magnitudes to eliminate 137
sources as too blue in ($F-J$). However, one object, \object{2MASS
J07003664+3157266}, proved to be matched incorrectly. Observations
by \citet{TK03} show that this object is a nearby L dwarf, and it
is included in the present sample. We have double-checked the
remaining 136 ($F-J$) rejectees, paying particular attention to
outliers in the ($J-H$)/($H-K_S$) plane. No additional ultracool
candidates were recovered.

\end{description}

Of the remaining 630 ultracool candidates, 112 objects were
already known as nearby, late-type dwarfs at the time of
publication of \citetalias{Cruz03}, thus leaving 518 sources
requiring follow-up observations.

\subsection{Follow-up Status}

We presented far-red optical spectroscopy of 298 sources in
\citetalias{Cruz03}; optical and/or near-infrared spectroscopy has
since been obtained for the remaining 220 targets. If
near-infrared data suggested that a source is an L dwarf within
20~pc, additional optical spectroscopy was obtained since $M_J$ of
M and L dwarfs is better correlated with optical spectral types
than infrared types.

\begin{figure}
\plotone{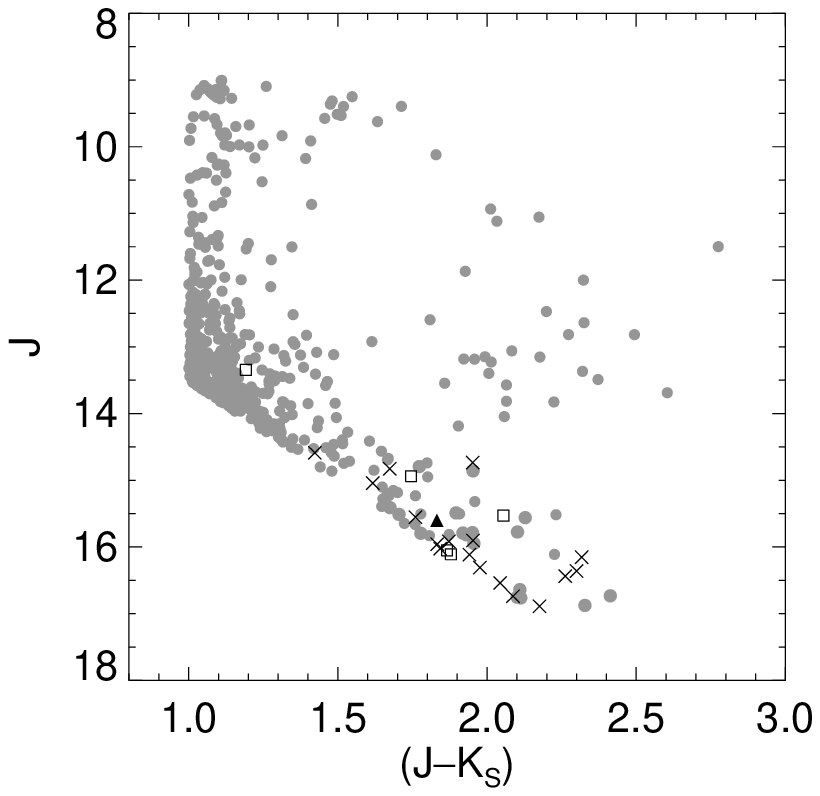}
\caption[Status of 2MU2 follow-up] {Status of the follow-up
observations of the 518 candidate ultracool dwarfs in the 2MU2
sample. Far-red optical spectra (\textit{circles}) have been
obtained for 495 objects in the sample. Near-infrared spectra have
been used to eliminate 17 fainter targets from the 20-pc 2MU2
sample (\textit{crosses}) and to identify one object near the
20-pc distance limit that requires supplemental far-red data to
determine its membership (\textit{triangle}). Five objects have
been classified using data from the literature
(\textit{squares}).\label{fig:obs_stat}}
\end{figure}

Figure~\ref{fig:obs_stat} summarizes the current status of our
observations. Optical spectroscopy has been obtained for 495
objects. Based on near-infrared spectroscopy, we eliminate
seventeen sources as either extragalactic objects or ultracool
dwarfs well beyond our 20-pc distance limit. One source,
\object{2MASSI J0028208+224905}, has near-infrared spectroscopy
indicative of possible membership in the 20-pc 2MU2 sample and
still requires optical observations. For five objects, we list
data from the literature---three from Keck (\citealt{K01};
\citeauthor[in prep.]{DavyLs}) and two from the Sloan Digital Sky
Survey \citep[SDSS,][]{Fan00,Hawley02}.

Our infrared spectroscopic observations of ultracool dwarfs will
be discussed in a future paper (\citeauthor[in prep.]{CruzNIR}).
Here, we present far-red optical spectra for 197 sources from the
2MU2 sample that previously lacked sufficient data.

\section{Far-red Spectroscopic Observations}
\label{sec:9obs}

We have obtained new optical spectra for 198 objects from the 2MU2
sample (including one re-observation of an object originally
reported in \citetalias{Cruz03}) using NOAO facilities and the
3.5-m Apache Point Observatory\footnote{Spectra will be made
available from \url{DwarfArchives.org} and upon request from
K.~L.~C, \texttt{kelle@amnh.org}.}. The instrumental setups and
data reduction techniques used for these data are the same as
those described in \citetalias{Cruz03}. The objects we used as
spectral standards are listed in Table~\ref{tab:standards}.

For all observations,
Tables~\ref{tab:9latenear}--\ref{tab:9carbon} list the coordinates
(as the 2MASS designation) and the near-infrared photometry from
the 2MASS Second Incremental Data Release PSC\footnote{There are
usually small differences between the astrometry and photometry
given in the 2MASS Second Release and the values in the final
All-Sky Data Release.}, date of observation, and telescope used.
Several objects were observed on multiple observing runs to
improve the signal-to-noise of the spectra or to monitor for
H$\alpha$ emission line variability; all observations are listed
and the active objects are noted.

Observations were obtained with the RC Spectrograph on the Kitt
Peak 2.1-m telescope (KP 2.1~m) during four runs: 2001
November~1--6, 2002 July~3--8, 2003 March~13--15, and 2003 Oct
8--12. For all runs, the 400~line~mm$^{-1}$ grating, blazed at
8000~\AA, was used with the OG~550 order blocking filter to give
spectra covering 6000--10000~\AA. Observations were made using a
1$\farcs$2--1$\farcs$5 slitwidths to accommodate various
conditions. We obtained an average resolution of 5.8~\AA\ (3.1
pixels). Internal quartz flats and HeNeAr arcs were taken at each
position to correct for fringing for all targets except for the
one object observed during 2001 November (\objectname[LHS
2215]{2MASSI~J0959560+200234}) where calibration frames were taken
nightly.

The Double Imaging Spectrograph (DIS II) on the 3.5-m telescope at
Apache Point Observatory (APO) was used on 2002 April 10, 2002 May
14, 2002 May 30, and 2002 July 10--11. A 1$\arcsec$-wide slit and
the medium resolution grating with 300~line~mm$^{-1}$ on the red
camera was used to cover 6000--10000~\AA\ at a resolution of
7.3~\AA\ (2.4~pixels). Conditions were mostly clear with
0\farcs8--1\farcs2 seeing.

The MARS instrument on the Mayall 4-m telescope (KP 4~m) on Kitt
Peak was used for three runs: 2002 September~25--28, 2003
July~9--14, and 2004 February~10--12. The VG0850-450 grism was
used for all three runs with 1$\farcs$5--2$\arcsec$ slitwidths, to
cover 6300--10000~\AA\ at a spectral resolution $\sim$10~\AA\
(3.3~pixels). An older CCD detector was used in 2002 September
than in the two subsequent runs. The main difference between the
two detectors is the cosmic-ray event rate, which does not affect
the spectral data presented here. The conditions in 2003 July were
significantly hampered by the combination of a nearly full moon
and smoke from the nearby Aspen fire. The conditions were clear
for the other runs.

Observations were obtained on 2003 April 20--23 and 2006
January~13--14 with the RC spectrograph and Loral 3K CCD on the
Blanco 4-m telescope (CT 4~m). We used a 1$\arcsec$-wide slit, an
OG~515 filter to block higher orders, and a 315~line~mm$^{-1}$
grating blazed at 7500~\AA\ to cover the range 5500--10000~\AA\
with a resolution of 7~\AA\ (3.5 pixels). All six nights were
clear with seeing ranging from 0\farcs5 to 1\farcs5.

Data were obtained with the CTIO 1.5-m (CT 1.5~m) telescope on
2003 November 7--11 with a Loral 1K CCD and the RC Spectrograph. A
total lunar eclipse on 9 November enabled the observations of
several fainter objects. We employed 1\farcs5-wide slit, an OG~530
filter to block higher orders, and a 400~line~mm$^{-1}$ grating
blazed at 8000~\AA\ to cover 6300--9000~\AA\ at a resolution of
6.5~\AA\ (3 pixels). Conditions were clear and the seeing ranged
from 0\farcs7 to 0\farcs9.

The Gemini Multi-Object Spectrometer \citep[GMOS,][]{GMOS} was
used on both Gemini North (GN) and South (GS) during queue
observations taken during 2004 September--2005 March (Program IDs:
GN-2004B-Q-10 and GS-2004B-Q-30). The RG610\_G0307 filter and
R400\_G5305 disperser was used on GN, while on GS the RG610\_G0331
filter and R400\_G5325 disperser was implemented to cover
6000--10000~\AA. On both telescopes, the nod and shuffle mode was
used with a 0\farcs75-wide slit to provide good sky subtraction
and a resolution of 5.5~\AA\ (4 pixels). Additionally, two
consecutive observations were taken with different central
wavelengths to obtain spectral coverage over the chip gaps.

\begin{figure}
\epsscale{0.75}
\plotone{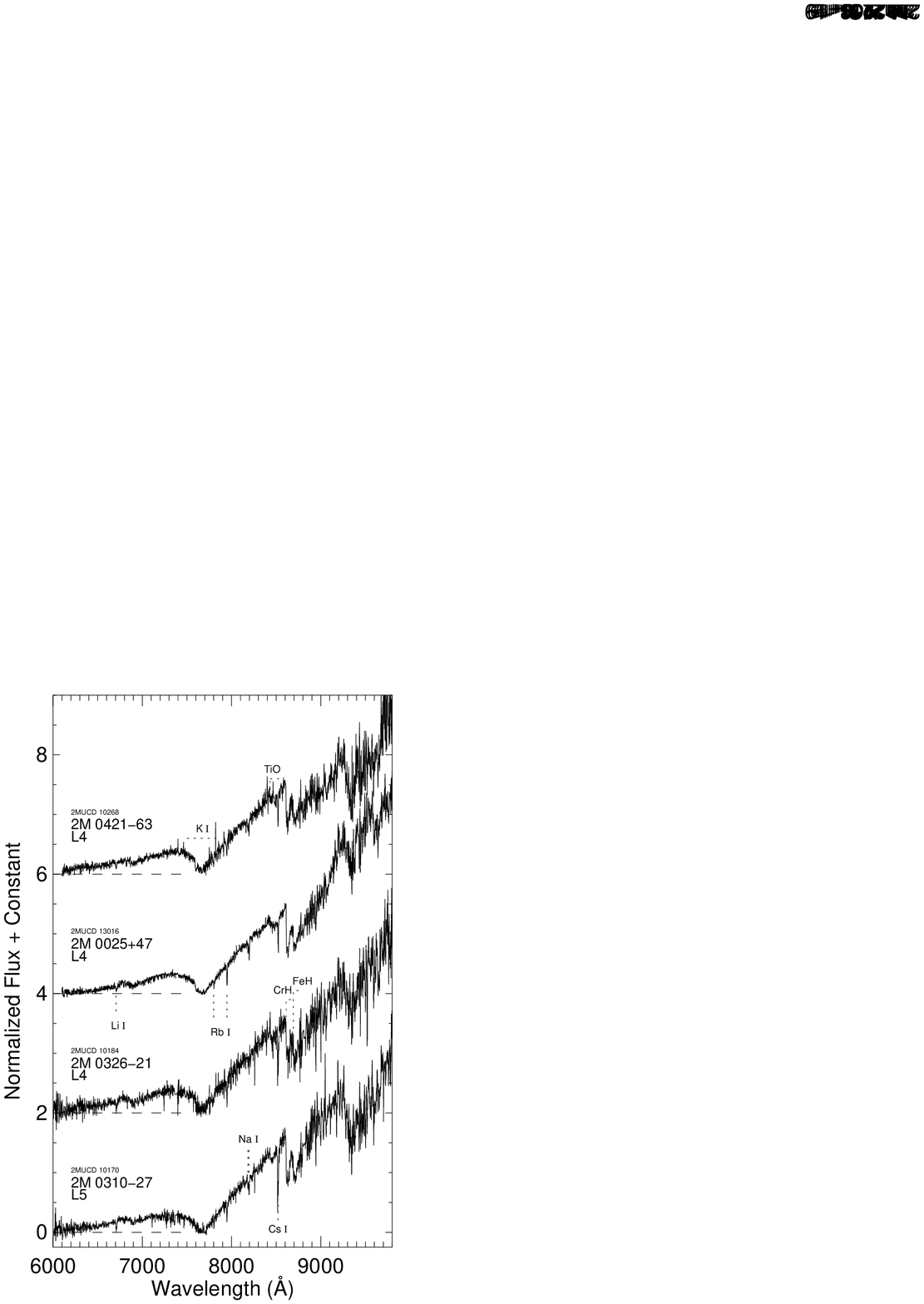}
\caption[Gemini spectra]{Gemini spectra of mid-L dwarfs where
lithium absorption is detected. From top to bottom, the objects
shown are: \objectname[2MASSI J0421072-630602]{2M~0421$-$63},
\objectname[2MASSI J0025036+475919]{2M~0025+47},
\objectname[2MASSI J0326422-210205]{2M~0326$-$21}, and
\objectname[2MASSI J0310140-275645]{2M~0310$-$27}. The steep red
slope of 2M~0025+47 is due to a problem with our Gemini North flux
calibration. The bottom spectrum is not offset and the zero points
of the offset spectra are shown by dashed lines. \label{fig:spec}}
\end{figure}

A comparison between the Gemini North data from semester 2004B and
other observations of ultracool dwarfs reveals an inconsistency in
the flux calibration. For example, 2M~0025+47 (L4) was observed
with Gemini North and, as can be seen in Figure~\ref{fig:spec},
has a significantly steeper spectral slope longward of 8700~\AA\
than the two other L4 dwarfs plotted, both of which were observed
with Gemini South. While we are continuing to investigate the
source of this problem, with the aim of correcting it, our ability
to spectral type is not hindered since we observed spectral
standards with the same observational setup and data reduction
procedure. (This sound methodology also enabled us to recognize
that it was a systematic problem and not a true property of the
new objects observed.)

All of the various data were bias-subtracted, flat-fielded,
wavelength-calibrated, and flux-calibrated using
IRAF\footnote{IRAF is distributed by the National Optical
Astronomy Observatories, which are operated by the Association of
Universities for Research in Astronomy, Inc., under cooperative
agreement with the National Science Foundation.}. The CCDRED
package and the DOSLIT routine were used for the KPNO, CTIO, and
APO data while the Gemini GMOS package was used for the Gemini
data.

To wavelength calibrate, HeNeAr or CuAr lamps were taken nightly
for most of the observations, at each position for the KP 2.1~m
data, and only a few times over the semester for the Gemini queue
observations. Flux-calibration was done using observations of the
flux standards \object{BD +26 2606}, \object{BD +17 4708},
\object{Feige 56}, \object{Feige 110}, \object{HD 19445},
\object{Hiltner 600}, \object{G191-B2B}, and \object{LTT 2415}
\citep{OG83,Massey88,Massey90,Hamuy}. None of the spectra were
corrected for telluric absorption.

\section{Results: Spectral Types, Magnitudes, and Distances}
\label{sec:9results}

We have measured spectral types, estimated absolute magnitudes,
and derived distances for all the observed dwarfs in the same
manner as described in \citetalias{Cruz03}. For the late-type
dwarfs (spectral types M7 and later), we have assigned a unique,
five digit ``2MUCD'' (2MASS Ultracool Dwarf) reference number in
addition to the full 2MASS designation. These data are listed in
Tables~\ref{tab:9latenear}--\ref{tab:9earlyfar}.

Spectral types are determined via side-by-side comparison with
spectra of spectral standards. The spectral standards we used are
listed in Table~\ref{tab:standards}. Data for many spectral
standards were taken during the course of our
observations\footnote{Data for these and other spectral standards
are available from
\url{http://research.amnh.org/$\sim$kelle/M\_standards/}}.
However, we supplemented our data with high signal-to-noise Keck I
+ LRIS data available for L dwarf spectral
standards\footnote{Available from
\url{http://DwarfArchives.org}.}.

The resulting uncertainty on spectral type is $\pm0.5$ subtypes
except where low signal-to-noise data result in uncertainties of 1
or 2~types, noted in the tables by a single or double colon
respectively\footnote{In previous papers in this series, a
question mark was used to indicate uncertainty in the spectral
type due to a low signal-to-noise spectrum. Here we adopt the the
more standard notation of a colon.}. The spectral types of objects
with multiple observations were estimated from the higher
signal-to-noise data---typically this is the second observation
where the spectrum was obtained with a larger aperture telescope
than the first.

For objects with spectral types M6 and later, $M_J$ is estimated
from the spectral type/$M_J$ calibration derived in
\citetalias{Cruz03}. This $M_J$ is combined with photometry from
the 2MASS Second Incremental Data Release PSC to yield $M_{K_S}$
and spectrophotometric distances. The uncertainties in both the
estimated absolute magnitudes and distance are dominated by the
uncertainty in the spectral type. Table~\ref{tab:9latenear} lists
the data for \latenear\ objects with types M6 and later that
appear to be within 20~pc. Data for \latefar\ more distant
late-type objects are listed in Table~\ref{tab:9latefar}.

Absolute \textit{J} magnitudes for objects with spectral types
earlier than M6 are estimated using the TiO5, CaH2, and CaOH
spectral indices as described in \citet[Paper~III]{Cruz02}. For
four objects (\objectname[2MASSI
J0510239-280053]{2MASSI~J0510239$-$280053}, M4; \objectname[2MASSI
J0544167-204909]{2MASSI~J0544167$-$204909}, M5; \object{2MASSI
J1242271+445140}, M5; \objectname[2MASSI
J1436418-153048]{2MASSI~J1436418$-$153048}, M5), at least two of
the three spectral-index relations yield two estimates for $M_J$.
Since these objects fall well outside our distance limit and
spectral type criteria, we have chosen to adopt $M_J=8.5\pm0.7$
and list the distance as a range. We list data for \earlyfar\
distant, early-to-mid M dwarfs in Table~\ref{tab:9earlyfar}.

Not surprisingly, spectroscopy revealed a number of the candidates
to be distant giants or carbon stars. Rough spectral types
($\pm1$) for the \giants\ giants were measured via comparison to
giant standards \citep{KHI97,Garcia89} and are listed in
Table~\ref{tab:9giants}. The \carbon\ carbon stars are listed in
Table~\ref{tab:9carbon}.

In Table~\ref{tab:newdata} we list new spectral data for six
dwarfs that were presented in \citetalias{Cruz03}. For five of
these, we previously listed data from the literature but we have
since reobserved them with our instrumental setup to maintain
consistency in the sample. One object, \objectname[2MASSI
J0326422-210205]{2MASSI J0326422$-$210205}, was reobserved to
obtain a higher signal-to-noise spectrum than that presented in
\citetalias{Cruz03}; these new data revealed a lithium absorption
feature (discussed in \S~\ref{sec:lithium}). There is general
agreement between the our new spectral types and those previously
quoted.

Five objects with spectral features indicative of low gravity are
listed in Table~\ref{tab:9young} and discussed below in
\S~\ref{sec:young}. Included among the low-gravity spectra is
\objectname[DENIS J043627.8-411446]{DENIS J043627.8$-$411446},
which was listed in \citetalias{Cruz03} with data from the
literature and \object{2MASSI J04433761+0002051}, which was
previously classified as a normal dwarf in \citetalias{Cruz03}.

\section{Interesting Individual Objects}
\label{sec:individual}

\subsection{Lithium Detections}
\label{sec:lithium}

We have detected the \ion{Li}{1} absorption line at 6708~\AA\ in
six objects in the entire 2MU2 sample. Two of these detections
were included in \citetalias{Cruz03} and are in the 20-pc sample:
\object{2MASSI J0652307+471034} and \objectname[2MASSI
J2057540-025230]{2MASSI J2057540$-$025230}.

Here we present four new detections, although all lie beyond 20
parsecs. Figure~\ref{fig:spec} displays the spectra of these
objects. All four have strong lithium absorption and are spectral
type L4--L5: \object{2MASSI J0025036+475919} with EW=10$\pm$2~\AA
(also candidate wide companion, see \S~\ref{sec:widebinaries}
below), \objectname[2MASSI J0421072-630602]{2MASSI
J0421072$-$630602} with EW=6$\pm$2~\AA, \objectname[2MASSI
J0310140-275645]{2MASSI J0310140$-$275645} with EW=10$\pm$1~\AA,
and \objectname[2MASSI J0326422-210205]{2MASSI J0326422$-$210205}
with EW=11$\pm$5~\AA.

The DUSTY theoretical models by \citet{Chabrier00} suggest that
the continued presence of undepleted lithium indicates that these
dwarfs are less than one~gigayear old. Given the absolute
magnitudes and effective temperatures inferred from the spectral
types, we find that undepleted lithium suggest these objects are
500~Myr old with masses of 50~M$_{Jup}$.

Our original observations of 2M~0326$-$21\footnote{Source
designations in this article are abbreviated in the manner
2M~hhmm$\pm$dd, where 2MASS has been further abbreviated as 2M;
the suffix is the sexagesimal right ascension (hours and minutes)
and declination (degrees) at J2000.0 equinox.}, with the CTIO
4-meter, lacked sufficient signal-to-noise to permit detection of
the lithium line and, based on those data, we estimated a spectral
type of L5:. New observations with Gemini South not only reveal
lithium absorption, but also provide an improved spectral type of
L4 (Table~\ref{tab:newdata}).

It is noteworthy that all of these new detections were obtained
with Gemini 8-m telescopes. The majority of our observations of L
dwarfs were obtained with 4-m class telescopes which are simply
not sensitive enough to reliably detect the relatively weak
\ion{Li}{1} absorption line.

\subsection{L/T Transition Objects}

\begin{figure}
\epsscale{1}
\plotone{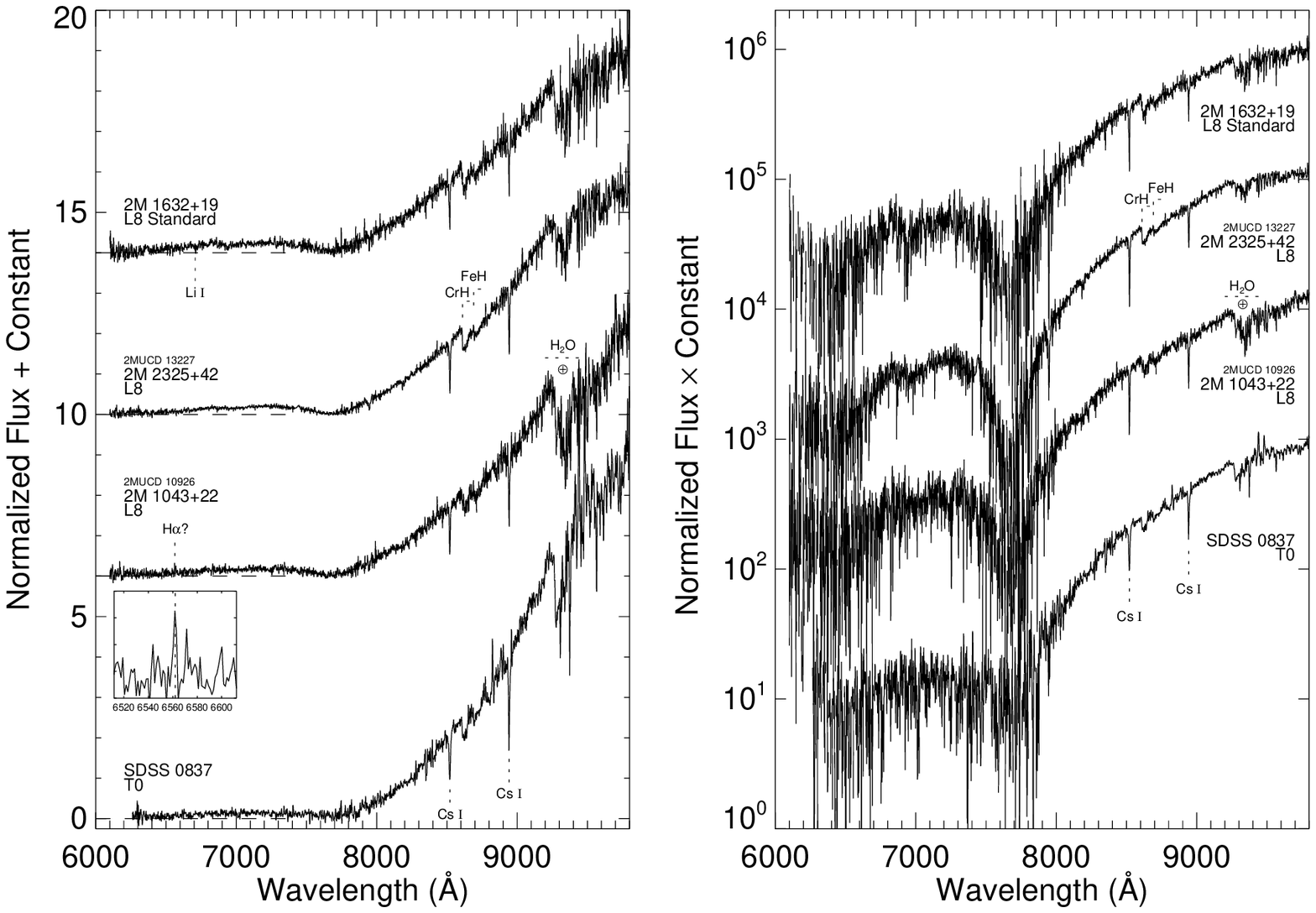}
\caption[late spectra]{Spectra of the two latest-type objects in
the present sample, \objectname[2MASSI
J2325453+425148]{2M~2325+42} (\textit{second-to-top}) and
\objectname[2MASSI J1043075+222523]{2M~1043+22}
(\textit{second-to-bottom}), with the L8 spectral standard
\objectname[2MASS J16322911+1904407]{2M~1632+19}
\citep[\textit{top},][]{K99} and early-T dwarf \objectname[SDSS
J083717.22-000018.3]{SDSS~0837$-$00}
(\citealt{Leggett00,Burgasser03_optical}; \citeauthor[in
prep.]{DavyTs}). The bottom spectrum is not offset and the zero
points of the offset spectra are shown by dashed lines.
\label{fig:latespec}}
\end{figure}

The optical spectra of the two latest objects in the new data
presented here, \object{2MASSI J1043075+222523} and \object{2MASSI
J2325453+425148}, are shown in Figure~\ref{fig:latespec}. The
spectrum of 2M~2325+42 is consistent with a spectral type of L8.
On the other hand, the red spectrum of 2M~1043+22 is significantly
steeper than that of both 2M~2235+42 and the L8 spectral standard.
(All three spectra were obtained with the same instrumental setup
and data reduction procedure.)

Infrared spectra show that 2M~1043+22 lacks significant methane
absorption in the \textit{H} band, thus ruling out a T spectral
type (\citeauthor[in prep.]{CruzNIR}). As a result, we adopt a
spectral type of L8 for 2M~1043+22. We note, however, that it is
the reddest L8 dwarf (where the spectral type is based on optical
data).

One possible explanation for the unusual spectrum of 2M~1043+22 is
that it is an unresolved binary similar to \objectname[SDSS
J042348.57-041403.5]{SDSS J042348.57$-$041403.5}
\citep{Burgasser05_0423} and \objectname[DENIS-P
J225210.7-173013]{DENIS-P J225210.7$-$173013} \citep{Reid06_2252}.
The binary frequency among late-type L dwarfs/early-type T dwarfs
has been found to twice as high as that of all other L and T
dwarfs \citep{Burgasser06_binary, Liu06} and high resolution
imaging observations of 2M~1043+22 are warranted.

Additionally, our optical spectrum of 2M~1043+22 shows a feature
near the wavelength of H$\alpha$ that might be consistent with
emission. This would be similar to the unusual T dwarfs
\object{2MASS J10475385+2124234} and \objectname[SDSS
J125453.90-012247.4]{SDSS J125453.90$-$012247.4} found with weak
H$\alpha$ emission by \citet{Burgasser03_optical}. Further
observations are required to verify the reality of the emission in
2M~1043+22.

\subsection{Low-Gravity Objects}
\label{sec:young}

Objects with spectral features indicating low surface gravities
($log(g) < 5$ cgs) are of particular interest as low gravity
implies low mass. In addition, low-mass brown dwarfs with
late-type M and early-type L spectral types must be younger than
their equivalently-classified higher-mass counterparts. Thus, low
gravity suggests both low mass and youth. A growing number of
late-type dwarfs with low-gravity features are now being uncovered
in field samples (\citealt{Gizis02, Kirkpatrick06}; \citeauthor[in
prep.]{DavyLs}) and their study is emerging as new way to probe
the evolutionary properties of brown dwarfs and to study their age
and mass distributions.

Through comparison to both late-type giants and objects found in
young clusters, several gravity-sensitive spectral features have
been identified. In particular, we have used enhanced VO bands
(7330--7530, 7850--8000~\AA), less-broad \ion{K}{1} doublet (7665,
7699~\AA), and weaker \ion{Na}{1} doublet (8183, 8195~\AA) as
diagnostics \citep{Martin96,Gizis99,Gorlova03,McGovern04,
Kirkpatrick06}.

\begin{figure}
\epsscale{1}
\plotone{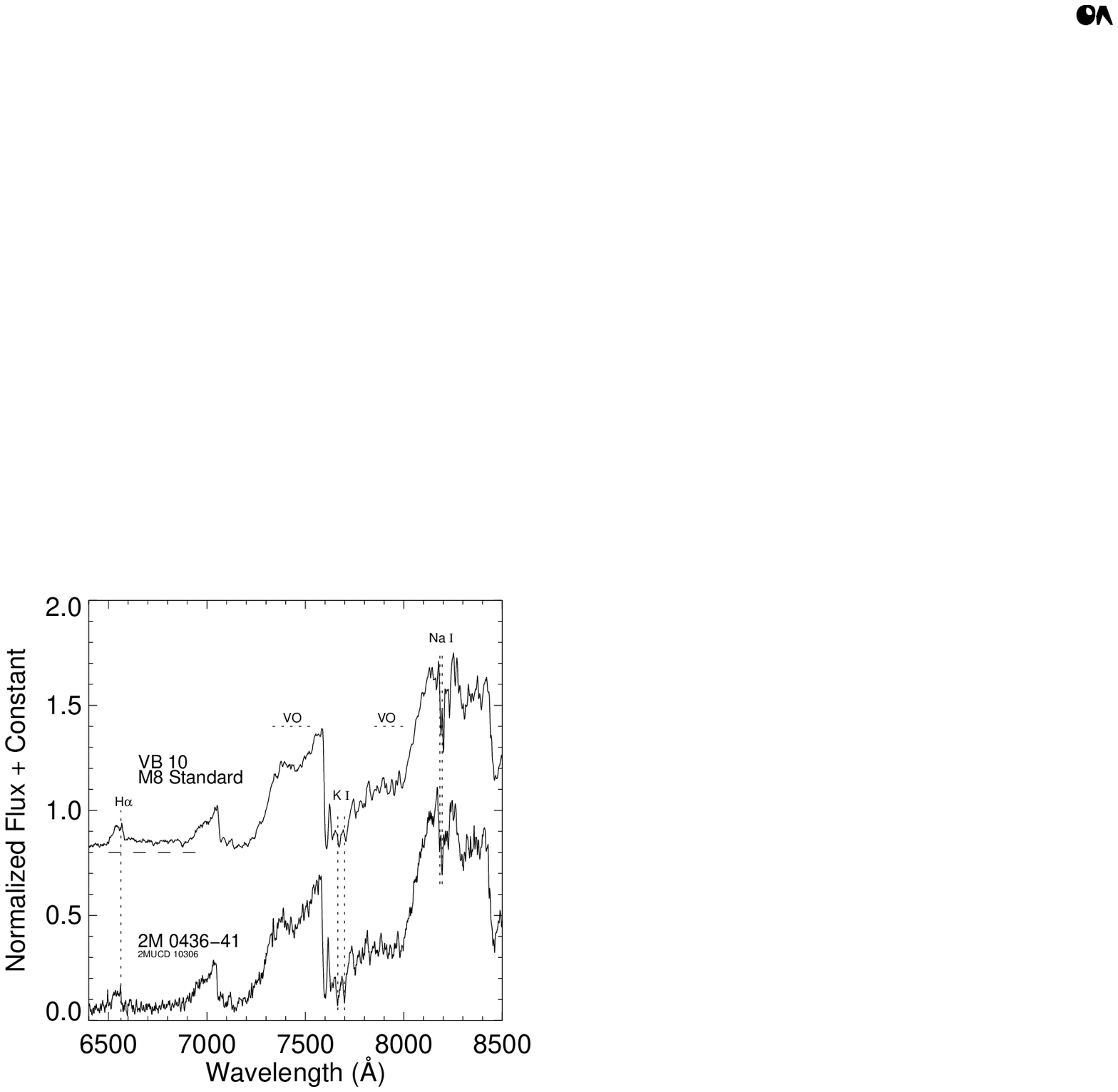}
\caption[Spectrum of Low-Gravity M Dwarf 2M~0436$-$41]{Spectrum of
the low-gravity dwarf \object[2MASSI
J0436278-411446]{2M~0436$-$41} with the M8 standard \object{VB
10}. Pressure/gravity sensitive features are marked. The bottom
spectrum is not offset and the zero point of VB~10 is shown by a
dashed line.\label{fig:youngM8}}
\end{figure}

\begin{figure}
\plotone{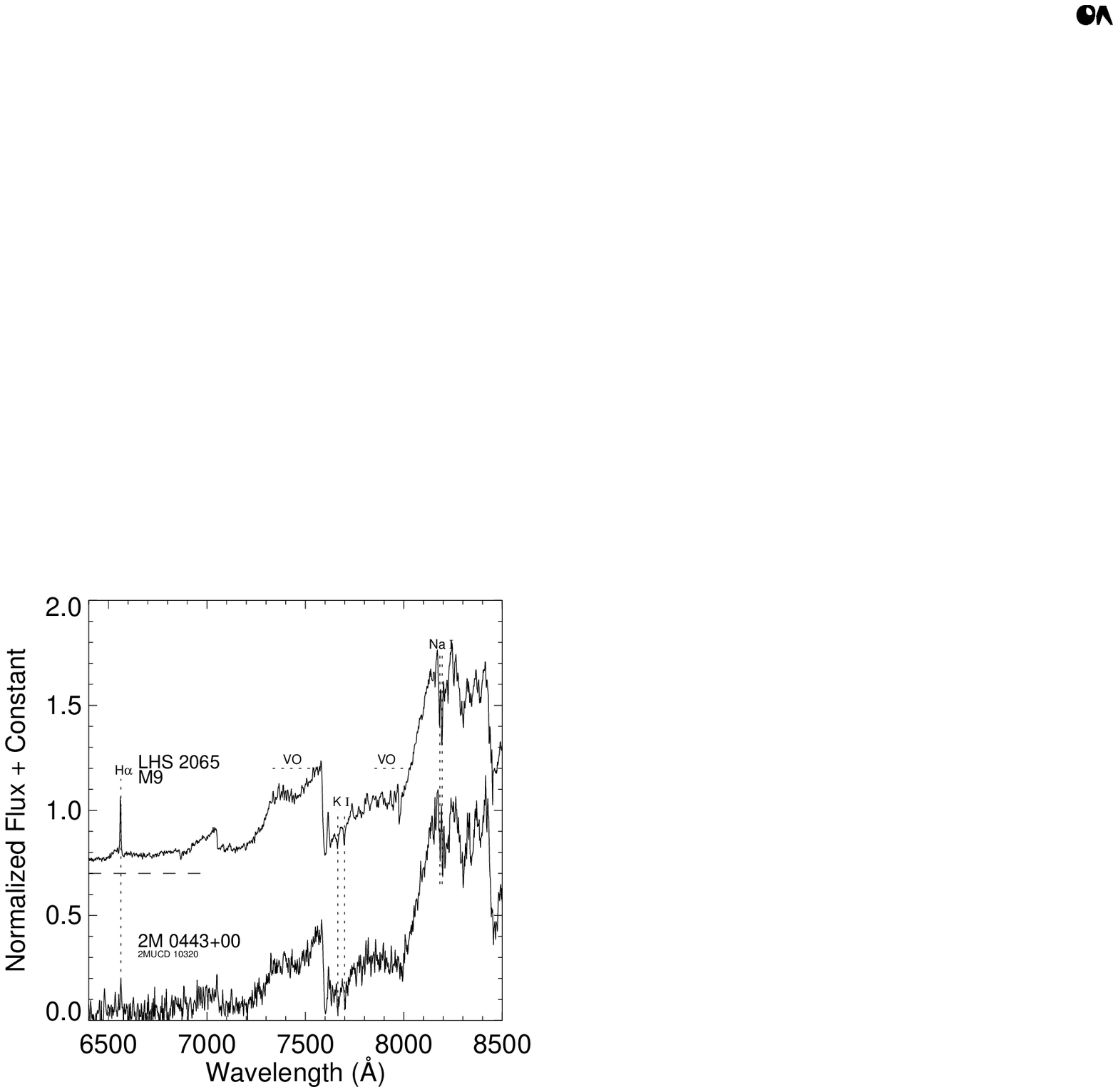}
\caption[Spectrum of Low-Gravity M Dwarf 2M~0443+00]{Spectrum of
the low gravity dwarf \object[2MASSI J0443376+000205]{2M~0443+00}
with the M9 standard \object{LHS 2065}. Pressure/gravity sensitive
features are marked. The bottom spectrum is not offset and the
zero point of LHS~2065 is shown by a dashed
line.\label{fig:youngM9}}
\end{figure}

\begin{figure}
\plotone{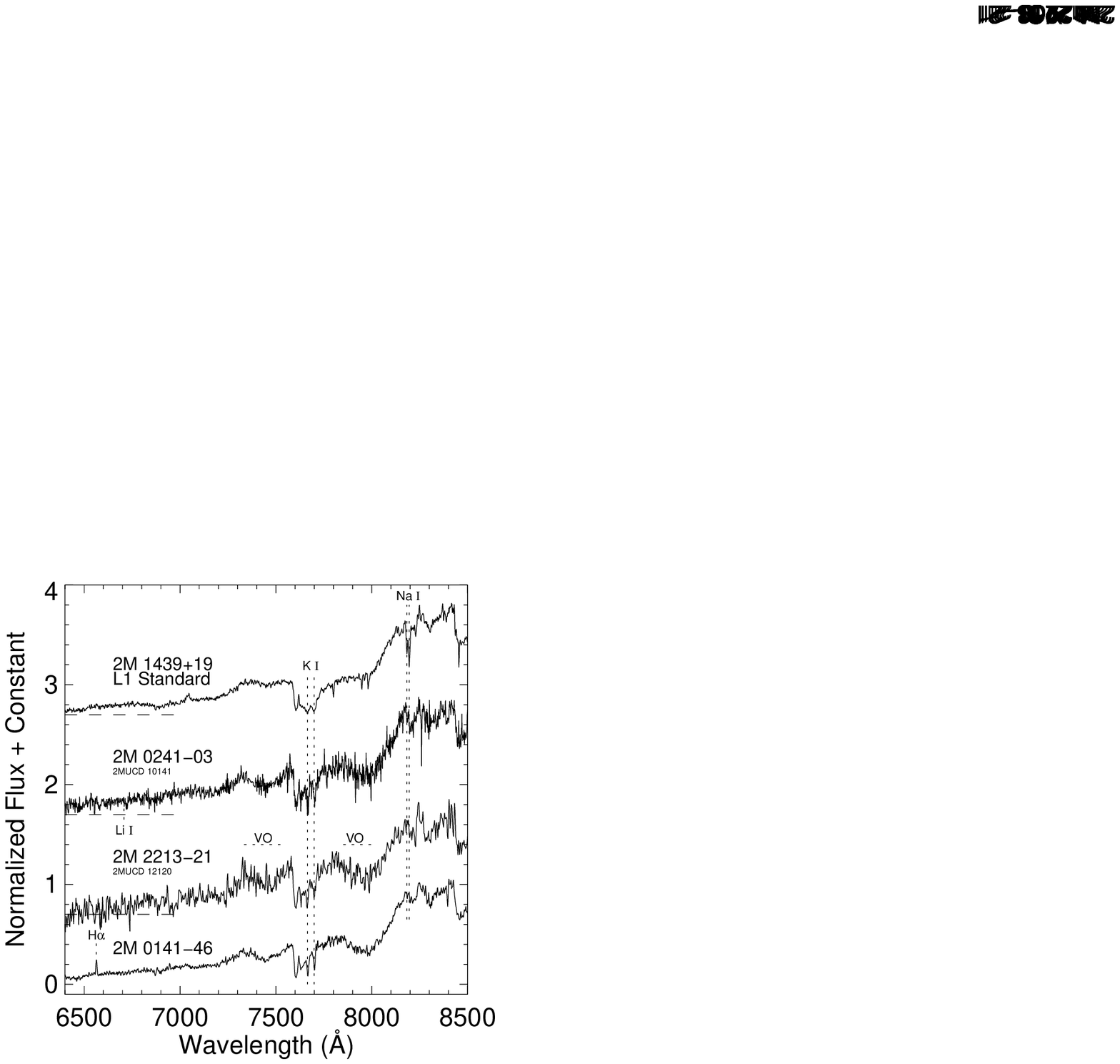}
\caption[Spectrum of Young L Dwarfs 2M~0241$-$03 and
2M~2213+19]{Spectrum of the low-gravity dwarfs, \objectname[2MASSI
J0241115-032658]{2M~0241$-$03} (\textit{second-to-top}) and
\objectname[2MASSI J2213449-213607]{2M~2213$-$21}
(\textit{second-to-bottom}), with the L1 standard
\objectname[2MASS J14392836+1929149]{2M~1439+19}
\citep[\textit{top},][]{K99} and the young field dwarf
\objectname[2MASS J01415823-4633574]{2M~0141-46}
\citep[\textit{bottom},][]{Kirkpatrick06}. Pressure/gravity
sensitive features are marked. The locations of the H$\alpha$
emission line and \ion{Li}{1} absorption line are labeled but
these features are not present in all spectra. The bottom spectrum
is not offset and the zero points of the offset spectra are shown
by dashed lines.\label{fig:young}}
\end{figure}

\begin{figure}
\epsscale{0.75}
\plotone{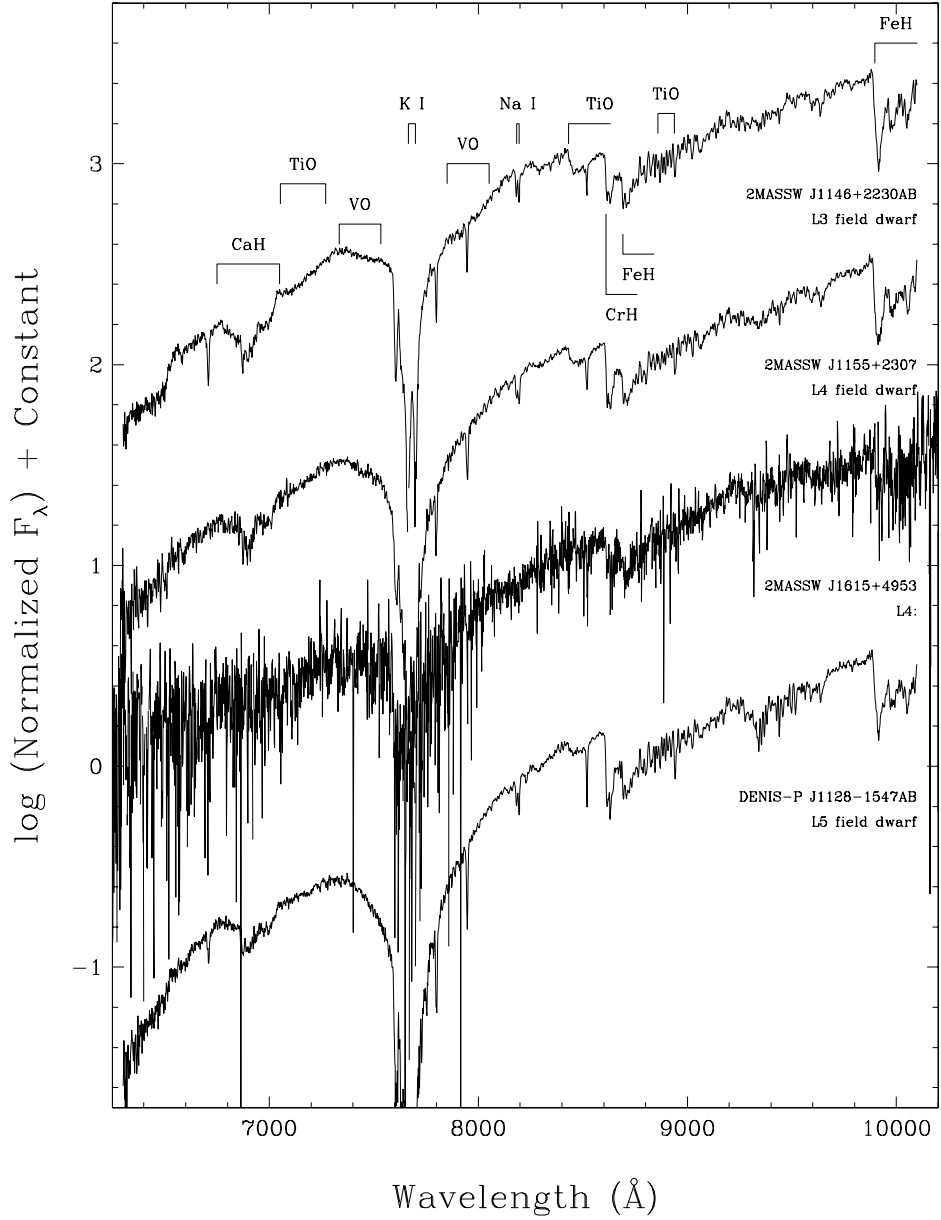}
\caption[Spectrum of the Low-gravity Dwarf 2M~1615+49]{Spectrum of
the low-gravity dwarf \objectname[]{2M~1615+49} compared to
spectral standards of type L3, L4, and L5 from \citet{K99}. The
flux of each spectrum has been normalized to unity at 8250~\AA,
and constant offsets have been added.\label{fig:youngL4}}
\end{figure}

Five objects with these low-gravity features were presented in
\citetalias{Cruz03} and we present an additional five here. They
are listed, along with their proper motions, in
Table~\ref{tab:9young} and their spectra are displayed in
Figures~\ref{fig:youngM8}--\ref{fig:youngL4}.

The spectrum of of \object{2MASSI J0443376+000205} most resembles
an M9 and using this spectral type results in a distance estimate
within 20~pc. Even though this object has low-gravity features, we
include it in the sample used to measure the luminosity function.

The spectra of the two objects in Figure~\ref{fig:young},
\objectname[2MASSI J0241115-032658]{2MASSI J0241115$-$032658} and
\objectname[2MASSI J2213449-213607]{2MASSI J2213449$-$213607}, are
later than any of the low-gravity objects from
\citetalias{Cruz03}. While noisy, these spectra are comparable to
the spectrum of \objectname[2MASS J01415823-4633574]{2MASS
J01415823$-$4633574}, recently described by \citet{Kirkpatrick06}
and estimated to have an age of 1--50~Myr and a mass of
6--25~$M_{Jup}$. All three objects are most comparable to an L1
dwarf but exhibit strong VO molecular absorption and weak
\ion{K}{1} and \ion{Na}{1} absorption lines. Unlike 2M~0141$-$46,
H$\alpha$ emission is not detected in either of our new objects.

The spectrum of \object{2MASSI J1615425+495321} most resembles an
L4 dwarf (Figure~\ref{fig:youngL4}) but, despite the poorer signal
of the spectrum, departures from the spectral signatures of normal
(old) field dwarfs are evident. In particular, the hydride bands
of CaH, CrH, and FeH are much weaker as are the core and wings of
the normally strong \ion{K}{1} doublet. The bluer slope at the
shortest wavelengths can likewise be attributed to weaker
absorption by the normally broad \ion{Na}{1} ``D'' doublet located
off the short-wavelength end of the plot. Weaker alkali lines and
hydride bands are hallmarks of lower gravity, leading us to
believe that this is a young, field L dwarf.

As with 2M~0141$-$46, we suspect these candidate young objects to
be possible members of the nearby young associations such as the
Tucana/Horologium association or the $\beta$ Pic moving group. We
are currently in the process of measuring their UVW space motions
to definitively test for membership---proper motions are already
in hand and radial velocity measurements of the brighter
candidates are underway. In addition, these objects are the focus
of high-resolution imaging programs searching for even fainter
companions.

\subsection{Two Unusually Blue L Dwarfs: Metal Poor?}
\label{sec:blueLs}

We have identified two objects in the 20-pc 2MU2 sample (included
in Table~\ref{tab:9lf}) that have unusually blue colors for their
spectral type: \object{2MASSI J1300425+191235} (L1) and
\object{2MASSI J1721039+334415} (L3). We first pointed out these
objects in \citetalias{Cruz03}. The ($J-K_S$) colors for these two
objects are 0.3 and 0.6 magnitudes bluer than the mean for their
spectral types \citep{K00}, a significant deviation suggesting
unusual atmospheric properties.

In L dwarfs, variations in ($J-K_S$) color for a given T$_{eff}$
can be linked to differences in condensate opacity and
metallicity. Condensates are largely responsible for the red
near-infrared colors of these objects, with the collective dust
particles acting as a warm pseudo-blackbody source that radiates
predominately at near- and mid-infrared wavelengths
\citep{Ackerman01}. A reduction in the condensates in the
photosphere, due perhaps to more efficient sedimentation to lower
layers in the atmosphere, can lead to bluer near-infrared colors
\citep[see][Figure 1]{Marley02}. \citet{Knapp04} and
\citet{Chiu06} suggest this scenario for seven blue L dwarfs
identified in the Sloan Digital Sky Survey.

As one of the brightest L dwarfs, 2M~1300+19 has been moderately
well studied. \citet{Gelino02} and \citet{Maiti05} both found
nonperiodic variability, with the former study suggesting that
such variability is evidence for an atmospheric event such as the
creation or dissipation of a large storm. If the condensate cloud
layer on this object is optically thinner (less condensates), then
variations in cloud coverage may be more readily detectable.
\citet{Mclean03} acquired a \textit{J}-band spectrum of 2M~1300+19
and noted that this object has ``the highest equivalent widths in
\ion{K}{1} of any L1, or indeed almost any other L type.'' As the
\textit{J} band is highly sensitive to condensate opacity in L
dwarfs \citep{Ackerman01}, a reduction in overall condensates may
explain the greater optical depth of this line.

A second possibility is reduced atmospheric metallicity, a
characteristic of low-mass subdwarfs. With fewer metal species,
most molecular opacity is diminished with the exception of H$_2$
absorption at \textit{K$_S$} band. The combination of weaker
opacity from metal molecules and enhanced H$_2$ leads to bluer
($J-K_S$) colors \citep{Linsky69, Saumon94, Borysow97}.

Two metal-poor L subdwarfs have already been identified in 2MASS
data, \object{2MASS J05325346+8246465}
\citep{Burgasser03_subdwarf} and \object{2MASS J16262034+3925190}
\citep{Burgasser04_subdwarf}, and both are quite blue,
$(J-K)=0.26$ and $-0.03$, respectively\footnote{LSR 1610-0040 was
identified as an L subdwarf by \citet{Lepine03_sdL} but recent
work has concluded that it is more likely to be a metal-poor M6
dwarf \citep{Cushing06, Reiners06} or a peculiar binary
(A.~Burgasser 2006, private communication).}. The redder colors of
2M~1300+19 and 2M~1721+33 suggest that they are not as metal-poor
as these L subdwarfs, and their optical spectra do not exhibit the
stronger metal-hydride bands (6750~\AA\ CaH, 8611~\AA\ CrH,
8692~\AA\ FeH) and weaker metal oxides bands (7053, 8432~\AA\ TiO)
expected for metal-poor ultracool dwarfs
\citep{Burgasser06_subdwarfs}. Nevertheless, the possibility
remains that these objects could be ``mild'' subdwarfs, with
sufficient metal-deficiency to enhance H$_2$ absorption.

In addition to being outliers in color space, 2M~1300+19 and
2M~1721+33 stand out kinematically. By combining our
spectrophotometric distance with proper motions, we have been
investigating the kinematics of the 20-pc 2MU2 sample \citep[in
prep.]{Sarah}. These two objects have two of the three highest
tangential velocities in the entire sample:
$V_{tan}=139\pm15$~km~s$^{-1}$ and 98$\pm8$~km~s$^{-1}$ for
2M~1721+33 and 2M~1300+19, respectively. (The second fastest is
\objectname[2MASSI J0251148-035245]{2MASSI J0251148$-$035245} with
$V_{tan}=125\pm13$~km~s$^{-1}$.) These kinematics suggest thick
disk membership, which would be consistent with slight
metal-deficiency ([Fe/H] $\approx-0.5$).

The blue colors of these peculiar L dwarfs may also result from a
\textit{combination} of slight metal deficiency and reduced
condensate formation. \citet{Burgasser03_subdwarf} and
\citet{Burgasser06_subdwarfs} found evidence that condensate
formation was inhibited in L subdwarf atmospheres due to the
persistence of TiO molecular bands and \ion{Ti}{1} and \ion{Ca}{1}
lines in their optical spectra. These species are generally absent
in L dwarf spectra due to their incorporation into condensates.
While we find no evidence for these features in the spectra of
2M~1300+19 and 2M~1721+33, the combination of slight
metal-deficiency and somewhat reduced condensate formation may tip
the near-infrared colors of these objects significantly blueward.
Because metallicity significantly impacts the composition and
chemistry of low temperature atmospheres, identifying and studying
metallicity effects in blue L dwarfs provide critical empirical
constraints for the next generation of theoretical atmosphere
models.

\subsection{Candidate Wide Ultracool Companions}
\label{sec:widebinaries}

Low-mass stars and brown dwarfs found as wide companions to
higher-mass main sequence stars provide opportunities to test
evolutionary and atmospheric models since the objects are assumed
to be coeval and thus have the same age and metallicity. We have
uncovered three ultracool dwarfs, included in
Table~\ref{tab:9latefar}, that are likely wide companions to
higher-mass stars. These discoveries, and others, will be
discussed by \citet[in prep.]{Solomon}.

\begin{description}

\item{\object{HD 225118} and \objectname[2MASSI
J0003422-282241]{2MASSI J0003422$-$282241}:} This G8 and M7.5
constitute a common proper motion pair separated by 1.1~arcminutes
(1700--2600~AU). We measured the proper motion of the M7.5 to be
($\mu_{\alpha}, \mu_{\delta})=228\pm57, -135\pm37$~mas~yr$^{-1}$.
HD~225118 was observed by the Hipparcos satellite
\citep{Hipparcos} and was found to have ($\mu_{\alpha}$,
$\mu_{\delta})=280\pm1.2, -143\pm0.73$~mas~yr$^{-1}$; in good
agreement with the motion of the ultracool dwarf. The parallax of
HD~225118 measured by Hipparcos yields a distance of
$39.5\pm1.7$~pc. Adopting this distance for the ultracool dwarf
yields $M_J=10.1\pm0.1$, nearly one magnitude brighter than what
is expected for an M7.5. Thus, we suspect the wide ultracool
companion might be an unresolved binary itself. An M7 or M8-type
companion would bring the spectrophotometric distance into
agreement with the Hipparcos distance estimate. If resolved, this
would further support the higher binary fraction among widely
separated companions found by \citet{Burgasser05_binary}. For
HD~225118, \citet{Geneva} estimate a mass of $\sim0.9~M_{\sun}$
and the age is not well constrained (upper limit of 15.8~Gyr).

\item{\object{HD 2057} and \object{2MASSI J0025036+475919}:} These
L4 and F8 dwarfs appear to be common proper motion companions
separated by 3.6~arcminutes (7000--9000~AU). The Hipparcos proper
motion for the F8, ($\mu_{\alpha}, \mu_{\delta})= 274\pm0.31,
11\pm0.87$~mas~yr$^{-1}$, is consistent with our measured value
for the ultracool dwarf, ($\mu_{\alpha}, \mu_{\delta})= 312\pm39,
-9\pm44$~mas~yr$^{-1}$. While the projected physical separation is
double that of the widest currently known multiple system with an
ultracool dwarf component, \object{Gl 584C} at 3600~AU
\citep{K01_gstar,Reid01_binary}, it is not unreasonable based on
the log-normal relation between maximum separation and the total
mass of the system found by \citet{Reid01_binary}.

Again, our spectrophotometric distance and the Hipparcos distance
are discrepant. This time, however, we have already resolved the
ultracool wide companion into an equal luminosity binary
\citep{Reid06_binary}. Taking this binarity into account yields a
spectrophotometric distance of $32\pm7$~pc, within $1.5\sigma$ of
the Hipparcos distance of $42\pm2$~pc. Additionally, as mentioned
above in \S~\ref{sec:lithium}, lithium absorption present in the
spectrum of the ultracool dwarf implies an age less than 1~Gyr and
probably closer to 500~Myr. For HD~2057, \citet{Geneva} find an
age of $\sim1.1$~Gyr. However, with an upper limit of 3.6~Gyr and
no lower limit given, this age is uncertain.

\item{\object{BD+13 1727} and \object{2MASSI J0739438+130507}:}
These M8 and K5 dwarfs are a common proper motion pair with a
separation of 10.5~arcseconds (380~AU). Their common proper
motions were recognized while cross-referencing our sample with
the LSPM catalog \citep{LSPM_North}. The Tycho-2 \citep{Tycho2}
proper motion for \object{BD+13 1727}, $(\mu_{\alpha},
\mu_{\delta})=-76.1\pm1.3, -156.5\pm1.2$~mas~yr$^{-1}$, agrees
well with the LSPM proper motion of the ultracool dwarf,
$(\mu_{\alpha}, \mu_{\delta})=-69\pm8, -145\pm8$~mas~yr$^{-1}$.
Our spectrophotometric distance for the M8 is $36.3\pm3.1$~pc. No
parallax or additional information on the primary could be found
in the literature.

\end{description}

\subsection{Suspected Unresolved Ultracool Binaries}
\label{sec:toobright}

Ultracool binaries with small separations, like wide systems,
place constraints on star formation models
\citep{Burgasser06_PPV}. We have noticed two objects, originally
presented in \citetalias{Cruz03}, that, based on trigonometric
parallax measurements, appear to be significantly overluminous,
suggesting that they are unresolved binaries. Both of these
objects are being targeted for ground-based, high-resolution
imaging.

These two objects were also misclassified by \citet[hereafter
PMSU]{PMSU} as having earlier spectral types than revealed by our
new spectra. The PMSU spectral types are based on the TiO5
index/spectral type relation which turns around at M7
\citep[Figure 3]{Cruz02}. Using the early-type branch of the
calibration resulted in the assignment of a too-early spectral
type. The spectral types estimated from our new spectra agree well
with those predicted by the late-type TiO5/spectral type relation
derived by \citeauthor{Cruz02}.

\begin{description}

\item{\objectname[LHS 1604]{LHS 1604 (2M~0351$-$00)}:} While PMSU
assigned a spectral type of M6 ($TiO5 = 0.18$), our observations
reveal a spectral type of M7.5 \citepalias{Cruz03}. Using our
spectral type/$M_J$ relation, we estimate $M_J=10.96\pm0.21$.
However, the parallax of $68.1\pm1.8$~mas \citep{vanAltena}
combined with the the 2MASS \textit{J} magnitude of
$11.262\pm0.023$, implies $M_J=10.43\pm0.06$. This
$\sim0.6$~magnitude overluminosity suggests that the object might
be an unresolved binary. We find that an M7.5/M9 pair would be
consistent with the astrometric results. While this object is
included in our total space density estimates, it is not included
in our analysis of space density per magnitude because the
parallax indicates a magnitude brighter than our brightest $M_J$
bin ($M_J=10.5$).

\item{\objectname[LHS 3406]{LHS 3406 (2M~1843+40)}:} For this
object, PMSU estimated a spectral type of M5.5 ($TiO5=0.24$) while
we find a spectral type of M8, indicating $M_J=11.16\pm0.18$
\citepalias{Cruz03}. However, the parallax of $70.7\pm0.8$
\citep{Monet92} and the 2MASS \textit{J} magnitude of
$11.299\pm0.028$ yield $M_J=10.55\pm0.04$. This overluminousity by
$\sim0.6$~magnitudes could be explained if the object is an M8/M9
unresolved binary pair. Additionally, this object was
misclassified as a dwarf nova \citep[U Gem
variable,][]{Downes97_CV} but has since been recognized as an M
dwarf by the cataclysmic variable community \citep{Liu99_CV}.
Unfortunately, we were unable to locate the original data that the
dwarf nova classification was based on to investigate the
possibility of a flare event.

\end{description}

\section{The 20-pc 2MU2 Sample and the Luminosity Function}
\label{sec:9LF}

As described below, the 2MU2 sample covers 36\% of the celestial
sphere. Within that area, we identify \LFobjects\ objects in
\LFsystems\ systems with spectral types between M7 and L8 and
estimated distances within 20~pc of the Sun---we dub these the
20-pc 2MU2 sample. Over half of these systems were added to this
sample through the observations described in this series of
papers. Table~\ref{tab:9lf} lists relevant data for all objects in
the 20-pc 2MU2 sample. Additionally, we list systems with distance
estimates within 1$\sigma$ of 20~parsecs in
Table~\ref{tab:onesigma}.

In Figure~\ref{fig:stdist}, we show the spectral type distribution
of the 20-pc sample with both the distance source and multiplicity
properties distinguished. Trigonometric parallaxes are available
for 30 systems; distances for the remaining sources are
spectrophotometric. Among the 91 systems, ten are multiple and
contribute eight additional objects. The color-magnitude and
color-color diagrams of the sample are shown in
Figure~\ref{fig:LFcolor}. In Figure~\ref{fig:jkdist}, we show the
$(J-K_S)$ color distribution for different spectral types.

We use these data to build on the space density analysis given in
\citetalias{Cruz03}, and derive an improved estimate of the
\textit{J}- and \textit{K}-band luminosity functions for ultracool
dwarfs. We describe how we estimate our sky coverage in
\S~\ref{sec:skycoverage}. Sample completeness is clearly an
important factor in this analysis, and we address that issue in
\S~\ref{sec:complete}. The Malmquist bias corrections appropriate
to the sample are described in \S~\ref{sec:malmquist}, binarity is
discussed in \S~\ref{sec:binary}, and the derived luminosity
function is presented in \S~\ref{sec:lf2}.

\subsection{Areal Sky Coverage}
\label{sec:skycoverage}

The total areal sky coverage of the 2MASS Second Release PSC is
19,641.6~deg$^2$ imaged in 27,493 $6\degr\times8\farcm5$ tiles.
From this, we excluded 3807 tiles with central galactic latitudes
within 10 degrees of the Galactic plane, reducing the coverage by
2878.1~deg$^2$. Parts of the remaining 16,763.5~deg$^2$ were
excluded to eliminate star formation regions, the Magellanic
clouds, and other highly reddened or crowded regions
\citepalias[listed in][Tables 2 and 3]{Cruz03}. However, these
regions were selected by galactic coordinates, whereas the 2MASS
imaging tiles are mapped in equatorial coordinates. Additionally,
many of these regions had only partial tile coverage in the 2MASS
Second Release PSC.

In order to quantify the areal coverage excluded in these regions,
we made use of the HIST\_ND algorithm written by J.D. Smith to
histogram the coordinates of the point sources eliminated by the
positional cuts onto a two-dimensional grid in a rectangular
region aligned with the coordinate axes and of easily calculable
area. The fraction of the non-zero elements in the resultant array
was used as the estimate of the fraction of the area filled by the
eliminated point sources and thus provided a measure of the area
of the excluded region, taking into account the discontinuous sky
coverage of the 2MASS Second Release PSC.

While the resulting measurement of the area is very sensitive to
the histogram bin size chosen, the results are accurate to a few
degrees---sufficient for our purposes. As many of the regions are
aligned with the coordinate axes in one system, but not in the
other, the areas were measured in both galactic and celestial
systems to give an estimate of the uncertainties. The resulting
measurement of the area removed due to positional cuts is
1940$\pm$10~deg$^2$.

Bright stars ($K < 4$) represent a further minor source of
confusion, since their extended halo on the 2MASS scans rules out
the possibility of detecting faint sources within as much as a few
arcminutes. However, the total area obscured in this manner is
less than 15~deg$^2$, producing negligible impact on our
statistics.

The final areal coverage of the 2MU2 sample is 14,823.5~deg$^2$,
or 36\% of the celestial sphere (9\% smaller than the
16,350~deg$^2$ previously stated in \citetalias{Cruz03}).

\subsection{Completeness}
\label{sec:complete}

The 2MU2 sample has been carefully constructed and extensive
spectroscopic follow-up observations have been completed to yield
a complete sample of M7--L8 type dwarfs to 20~parsecs over 36\% of
the sky. Below we discuss our observational, spectral type and the
resulting volume completeness.

\subsubsection{Observational Completeness}
\label{sec:obs_complete}

As discussed in \S~\ref{sec:9sample}, we lack sufficient data for
one object, \objectname[2MASS J00282091+2249050]{2MASSI
J0028208+224905}, that infrared spectroscopy identifies as a
late-type L dwarf at 20--25~pc. It is possible that optical
spectroscopy might result in a spectral type of L7 or later,
moving it within the 20-pc limit. If this is the case, it would
contribute to the 14.25 or 14.75 $M_J$ bin of \LFj. For this
reason, and other more significant ones discussed in
\S~\ref{sec:lf2}, our measurement of $\Phi(14 <M_J<15)$ and
$\Phi(12.5<M_{K_S}<13)$ are lower limits.

\subsubsection{Spectral Type Incompleteness}
\label{sec:st_complete}

The 2MU2 sample is a color- and magnitude-defined sample. These
selection criteria result in incompleteness at the extremes of the
range of spectral type covered by our survey. For example, we are
incomplete for M7 dwarfs due to the $(J-K_S) > 1.0$ selection
criterion; several M7 dwarfs are known with $(J-K_S)$ colors bluer
than this limit, including \object{VB 8}, the M7 archetype.

We have quantified this incompleteness through analysis of a
sample of ultracool dwarfs with $J<16.5$ and trigonometric
parallax measurements from \citet{Dahn02} or \citet{Vrba04}. We
obtained photometry for these objects from the 2MASS Second
Incremental Data Release PSC where available and otherwise used
the data from the 2MASS All-Sky PSC. We refer to this sample as
the ``ultracool dwarf trigonometric'' (UCDt) sample. Adjusting the
measured $M_J$ to a distance of 20~pc shows the systems as they
would appear at the far edge of our volume limit.
Figure~\ref{fig:missing} displays those data against our selection
criteria, marking objects that are excluded. Blue M7 and M8
dwarfs, some L7 and L8 dwarfs, and unusual objects are likely to
be excluded from the 2MU2 sample.

\paragraph{Late-M Dwarfs}

Considering earlier types, two M7 dwarfs in the UCDt sample,
\objectname[2MASS J22104999-1952249]{GRH~2208$-$2007} (M7.5) and
\object{VB 8} (M7), are excluded by both the $J/(J-K_S)$ and
$(J-H)/(H-K_S)$ selection criteria. While all of the M8 dwarfs in
the UCDt sample fall within our criteria, they also appear to be
cut-off by the $(J-K_S)=1.0$ limit (Figure~\ref{fig:LFcolor}). To
see this more clearly, in Figure~\ref{fig:jkdist} we show the
$(J-K_S)$ distribution for M7, M8, and M9 dwarfs in the 20-pc 2MU2
sample---while the M7 and M8 distributions appear truncated, the
M9 distribution does not.

We cannot use the UCDt sample to estimate the resultant
incompleteness, since it includes only three M7 and six M8 dwarfs.
Instead, we have compiled data for M7 and M8 dwarfs listed in the
literature that were not selected using $(J-K_S)$ color as a
criterion. In Figure~\ref{fig:m7correction} we examine the
$(J-K_S)$ distribution of these objects. Three of the fourteen M7
dwarfs have $(J-K_S)< 1$ and we use a binomial distribution to
estimate 78.6$^{+7}_{-14.3}$\% of M7 dwarfs are redder than
$(J-K)=1$. Applying this correction to the observed number of 21
M7 dwarfs in the 20-pc 2MU2 sample yields a corrected number of
26.7 dwarfs, and the luminosity functions are changed accordingly.
None of the twelve M8 dwarfs found in the literature have
$(J-K_S)< 1$. While the 20-pc 2MU2 sample is likely incomplete for
M8 dwarfs, the effect is small ($\sim10$\%) and we do not correct
the luminosity function for missing M8 dwarfs.

\paragraph{Late-L Dwarfs}

At later types, the 2MU2 sample is incomplete since
pressure-induced H$_2$ absorption and perhaps dust settling
\citep{Knapp04} leads to some L7 and L8 dwarfs having colors bluer
than our selection criteria. The UCDt sample includes four L7 and
six L8 dwarfs---one L7.5 and four L8 dwarfs are excluded in the
$J/(J-K_S)$ plane. While this incompleteness affects $14<M_J<15$
and $12.5<M_{K_S}<13$, we have not computed explicit completeness
corrections for those spectral types for two reasons. First, the
incompleteness is by magnitude, and therefore distance dependent;
there is no incompleteness for late-type sources within 11~pc, for
example. Second, and more significantly, early T dwarfs, which are
not included in our survey, contribute to the space density of
ultracool dwarfs in the same $M_J$ range as L7 and L8 dwarfs
($14<M_J<15$). In the \textit{K$_S$} band, on the other hand, T
dwarfs are fainter than L dwarfs and do not contribute to the
space density in our $M_{K_S}$ range of interest \citep{Vrba04}.
Thus, we take our measured $\Phi(14<M_J<15)$ as a lower-limit due
to both missing L and T dwarfs while $\Phi(12.5<M_{K_S}<13)$ is
incomplete only due to missing late-L dwarfs.

\paragraph{Unusual M9 and L0 Dwarfs}

One M9 dwarf and two L0 dwarfs in the UCDt sample have unusual
$(H-K_S)$ colors ($\sim0.35$): \object{PC~0025+0447} (M9.5),
\objectname[SDSS J143517.20-004612.9]{SDSS J143517.20$-$004612.9}
(L0), and \objectname[SDSS J225529.09-003433.4]{SDSS
J225529.09$-$003433.4} (L0:). PC~0025+0447 is recognized as a
highly unusual object that is young, has low-gravity features, and
has persistent, extremely strong H$\alpha$ emission
(EW$\sim$100~\AA). The two SDSS objects are faint ($J>15.6$) and
the photometric errors on SDSS~1435$-$00 are substantial
($\sim$0.1). We propose that either these three objects are
unusual in a similar, as-yet-unrecognized way, or more likely,
that the uncertainties in the 2MASS photometry, either due to
systematic errors or coincidence, have resulted in a similar blue
$(H-K_S)$ color. No correction is applied to the luminosity
functions based on these three objects.


\paragraph{Other Excluded Objects}

Two additional objects in the UCDt sample are excluded by the
$(J-H)/(H-K_S)$ selection criteria: \object{GJ 1048B} (L1) with
$(J-H)=0.69$, $(H-K_S)=0.67$ and \object{SDSS J144600.60+002452.0}
(L6) with $(J-H)=1.38$, $(H-K_S)=0.58$. The $(H-K_S)$ color of
GJ~1048B is unusually red for its spectral type and is probably
due to its proximity to GJ~1048A resulting in $\sim0.1$
photometric uncertainties. For an L6, SDSS~1446+00 appears
unusually red in $(J-H)$ and the 2MASS photometry is fairly robust
with uncertainties ranging from 0.035 in $H$ band to 0.082
magnitudes in \textit{J} band. However, the 2MASS flag `ndet'
equals 264566 indicating that there were only 2 detections at
\textit{J} band out of 6 possible, 4 detections out of 5 possible
at \textit{H} band, and 6 out of 6 at \textit{K$_S$} band. Thus,
with only 2 detections, the quoted \textit{J} magnitude is highly
suspect. Using the transformation given in the 2MASS Explanatory
Supplement\footnote{\url{http://www.ipac.caltech.edu/2mass/releases/allsky/doc/sec6\_4b.html}}
we convert the UKIRT MKO photometry listed in \citet{Geballe02} to
the 2MASS system and find $JHK_S$ colors that are not anomalous
when compared to other late-L dwarfs. A more accurate 2MASS
\textit{J} magnitude is probably 15.6 rather than the 15.9 listed
in both the 2MASS Second Release PSC and the All-Sky PSC. No
correction is applied to the luminosity functions based on these
two objects.

\subsubsection{Volume Completeness}
\label{sec:vol_complete}

We show the space densities for two sets of 4-pc thick spherical
shells for five absolute magnitude bins in
Figure~\ref{fig:complete}. (Note that these bins are coarser than
those used in \LF\ and the last two bins overlap.) The distance at
which the sample begins to be incomplete is indicated by a
downturn in the measured space densities. The 2MU2 sample appears
to be complete to 20~parsecs at all magnitudes except for the two
faintest ($14<M_J<15$). This is likely due to our incompleteness
at the latest-L types and, as discussed above, we take our
measurement of the luminosity functions to be a lower limit at the
relevant magnitudes.

\subsection{Malmquist Bias}
\label{sec:malmquist}

We have adopted a unique value of $M_J$ for each spectral type,
$M_{obs}$, where the true situation is a dispersion of absolute
magnitudes about some average for each spectral type. As a result,
the sample is biased towards more luminous objects at a given
spectral type. The intrinsically less luminous objects (where the
true intrinsic absolute magnitude, $M_0 > M_{obs}$) are
systematically excluded because using $M_{obs}$ overestimates
their distance and thus they are more likely to fall outside of
the distance limit. Similarly, a greater number of overluminous
objects are included in the sample because their distances are
underestimated. This is classical Malmquist Bias and we must
correct the estimated absolute magnitude for those 69 objects in
the sample without trigonometric parallax data \citep{Malmquist}.

Since we expect nearby M and L dwarfs to be uniformly distributed
throughout the Solar Neighborhood we can use Malmquist's formula
to correct the absolute magnitudes,
\begin{equation}
M_0=M_{obs}-1.38\sigma^2
\end{equation}
where $\sigma$ is the uncertainty in $M_{obs}$. The uncertainty in
$M_{obs}$ depends partly on the scatter in the $M_J$/spectral-type
relation, but mostly on the uncertainties in the spectral type
estimates. Combining these, we derive uncertainties of $0.13 <
\sigma < 0.42$ mags resulting in corrections of 0.02 to 0.24 mags.
These corrections are applied to our measured \LF.

\subsection{Unresolved Binary Systems}
\label{sec:binary}

Unrecognized binarity is likely to affect the statistical
properties of the 20-pc 2MU2 sample. The sample includes members
of 10 known multiple systems. Eight of those systems have
trigonometric parallax data. Four are systems where one component
falls beyond the spectral type limits of the present sample. Those
systems are: \objectname[GJ 2005]{LHS 1070 (2M~0024$-$27)}, where
the primary is an M5.5 dwarf; \objectname[G 216-7]{G 216-7
(2M~2237+29)}, where G~216-7A is an M0 dwarf; \objectname[SDSS
J042348.57-041403.5]{SDSS J042348.57$-$041403.5}, where the
secondary is an early-type T dwarf \citep{Burgasser05_0423}; and
\objectname[DENIS-P J020529.0-115925]{DENIS-P J020529.0$-$115925},
where a possible T-type component has been reported
\citep{Bouy05}.

More than half of the sources in the 20-pc 2MU2 sample have either
high spatial-resolution ground-based observations
\citep{Koerner99, Close03, Siegler05} or imaging by the Hubble
Space Telescope \citep{Reid01_binary, Gizis03, Reid06_binary}. Of
the 55 systems that have been targeted, nine are resolved as close
binaries, yielding an observed binary fraction of
17$^{+6}_{-4}$\%, consistent with prior analysis
\citep{Gizis03,Bouy03,Burgasser06_binary}. (See
\citet{Reid06_binary} for a thorough discussion of the binary
frequency of our 20-pc sample.) Applying this binary fraction to
the 36 unobserved objects suggests that there are approximately
five currently unresolved binaries included in our measurement of
\LF. Indeed, as discussed in \S~\ref{sec:toobright}, we already
suspect two objects to be unresolved binaries based on their
overluminosity.

\citet{Burgasser04} has modeled multiplicity corrections for a
wide range of scenarios for a magnitude-limited observational
sample. The vast majority (34/36) of sources that lack
high-resolution imaging also lack trigonometric parallax data. As
a result, these sources are effectively a magnitude-limited
sample, and \citeauthor{Burgasser04}'s analysis is appropriate. In
the most extreme case of a binary fraction of 50\%, the correction
is only 20\% to the late-L dwarfs. Under more likely circumstances
(binary fraction $\sim$20\%), the space density of late-M dwarfs
is overestimated by 5\% while the density of late-L dwarfs is
underestimated by about 8\%. We also note that $\sim$5\% of the
sources in our sample may prove to be unrecognized spectroscopic
($\Delta < 2$~AU) binaries \citep{Reid02_highres,Basri06}.
Nonetheless, we expect the effects on \LF\ due to unrecognized
binarity to be small ($<$10\%) and we do not apply any corrections
for them.

\subsection{The Luminosity Function}
\label{sec:lf2}

We show the Malmquist-corrected \textit{J}- and
\textit{K$_S$}-band luminosity functions derived from the 20-pc
2MU2 sample in Figure~\ref{fig:LF2} and list the measured space
densities in Table~\ref{tab:lf}. As outlined earlier in this
section (\S~\ref{sec:st_complete}), the color-magnitude criteria
used to define the sample lead to incomplete sampling at spectral
types M7/M8 and L7/L8. We have made explicit allowance for the
incompleteness at M7, increasing the space density at
$M_J=10.75\pm0.25$ by 41\%.

The correction factors are more difficult to assess at the latest
spectral types, however. Due to the incompleteness at late-L
types, we take our measured \LFk\ as a lower limit in the faintest
magnitude bin ($12.5<M_{K_S}<13$). In addition to late-L dwarfs, T
dwarfs contribute to the faintest two bins ($14<M_J<15$) of \LFj.
Trigonometric parallax results show that brown dwarfs {\sl
brighten} from $M_J\sim15$ at spectral type T0 to $M_J \sim 14.5$
at type T3/4 \citep{Dahn02,Tinney03,Vrba04}. This behavior is
probably related to the clearing of dust clouds within the
atmosphere, allowing the $\tau=1$ photospheric level to descend to
greater physical depths (and higher temperatures) at wavelengths
near 1.2~$\micron$ \citep{Burgasser02_clouds,Knapp04}. In any
case, our survey was not designed to identify field T dwarfs, and
our measured space densities must represent a lower limit to \LFj\
at $M_J>14$.

\section{Discussion}
\label{sec:9discussion}

Integrating our results for the luminosity function, we derive a
space density of $8.7\pm0.8 \times 10^{-3}$~pc$^{-3}$ for M7--L8
dwarfs. Ultracool M dwarfs, spectral types M7 to M9.5, contribute
a density of $4.9\pm0.6 \times 10^{-3}$~pc$^{-3}$. We find the
space density of L dwarfs to be $\ge3.8\pm0.6\times
10^{-3}$~pc$^{-3}$, with L0--L3 dwarfs contributing
$1.7\pm0.4\times 10^{-3}$~pc$^{-3}$ and late-type L dwarfs
responsible for at least $2.2\pm0.4 \times 10^{-3}$~pc$^{-3}$. Our
results are in good agreement with the space density of
$4.5\times10^{-3}$~pc$^{-3}$ derived for ultracool M dwarfs by
\citet{NN}, but exceed by almost a factor of two their initial
estimate of the local L dwarf density. Since we are using a
significantly larger sample than \citeauthor{NN}, our measured
densities are likely more accurate.

Putting these results in a wider context, Figure~\ref{fig:8pcLF}
superimposes \LFj\ as derived from our ultracool sample against
results for the 8-pc sample \citep{Paper4}. As discussed in
\citet{Paper8}, 95\% of the 8-parsec stars have reliable
trigonometric parallaxes; the \textit{J}-band photometry is taken
either from 2MASS or, for bright stars, drawn from the literature
and transformed, if necessary, to the CIT system. The derived
space densities are listed in Table~\ref{tab:8pclf}. While the
8-pc sample includes only a handful of ultracool dwarfs, it is
clear that the two datasets are in excellent agreement, with the
new data confirming the sharp decline in number densities at $M_J
> 10$. In order to get an idea of the space densities in the \textit{I}
band, $M_I\sim13$ for M7 and $(I-J)$ ranges from 2.4--4 for M7--L8
\citep{Dahn02}.

The overall morphology of the \textit{J}-band luminosity function
reflects the convolution of the underlying mass function,
$\Psi(M)$, and the $M_J$-mass relation. Qualitatively, \LFj\
increases for $0<M_J<7$ since the mass function increases with
decreasing mass. At fainter magnitudes, \LFj\ turns over not
because $\Psi(M)$ changes drastically, but because the slope of
the $M_J$-mass relation changes---while ${{\delta Mass}\over
{\delta M_J}} \sim 0.4 M_\sun$ mag$^{-1}$ for $M_J < 7$, ${{\delta
Mass}\over {\delta M_J}} \sim 0.07 M_\sun$ mag$^{-1}$ for
$7<M_J<10$ \citep{Delfosse00}.

Our analysis shows that \LFj\ declines sharply with decreasing
luminosity beyond $M_J=10$, reaching a minimum at $M_J \sim13$,
roughly corresponding to spectral type L4. Formally, our data
indicate that the space densities remain approximately constant at
fainter magnitudes; however, since our measurements are lower
limits, \LFj\ likely increases at $M_J > 14$.

Our field survey has uncovered both very low-mass stars and
sub-stellar mass brown dwarfs from the local population. Assuming
a typical age of 2--5 Gyrs for a field dwarf, even the most
massive brown dwarfs have cooled to L-dwarf temperatures.
Moreover, theoretical models \citep[e.g.,][]{Burrows01} indicate
that the hydrogen-burning limit (for solar abundances) corresponds
to a temperature of $\sim$1700~K, or spectral type $\sim$L4
\citep{Golimowski04}. Thus, early-type L dwarfs include a mix of
stars and brown dwarfs, with brown dwarfs acquiring increasing
dominance from L0 to L4 until they are the sole contributors at
$M_J\ga13.5$.

As discussed elsewhere \citep{Chabrier03, Burgasser04, Allen05},
the morphology of the luminosity function is due to this mix of
stars and brown dwarfs in the ultracool regime and is
qualitatively in accord with theoretical expectations. The drop in
number density from M7 to L4 reflects a further contraction in
${{\delta Mass}\over {\delta M_J}}$. The population of the very
lowest-mass stars that appear as L dwarfs span an extremely small
range in mass $(0.075<M_\sun<0.085$) and, as a result, are rare.
The brown dwarfs in this effective temperature regime are
relatively young and are at the high-mass extreme, near the
hydrogen-burning limit.

Brown dwarfs dominate the counts beyond the plateau in \LFj\ at
$M_J \sim 13.5$ ($\sim$L4), and the upturn in number densities
reflects the slow down in cooling rates at lower temperatures. For
example, a 0.07~$M_\sun$ brown dwarf takes 2.7~Gyrs to evolve down
the L dwarf sequence, but remains a (cooling) T dwarf
(T$_{eff}\approx$1400--600~K) for $\sim$30~Gyrs, or more than 2
Hubble times; a low-mass, 0.025~$M_\sun$ brown dwarf spends only
120~Myrs as an L dwarf, but 1.5~Gyrs as a T dwarf
\citep{Burrows01}.

As noted in \S~\ref{sec:lf2}, early-type T dwarfs also contribute
to the \LFj\ at magnitudes fainter than $M_J=14$. Evolutionary
models suggest that there may be as many as 80 T0--T5 dwarfs
within 20 parsecs \citep{Burgasser04}, as compared with the 12
late-type L dwarfs contributing to our luminosity function at
$M_J>14$. Incorporating those cooler dwarfs in the analysis is
likely to lead to a steeply increasing \LFj.

Clearly, one of the aims in deriving the luminosity function for
ultracool dwarfs is setting constraints on the mass function,
usually parameterized as a power-law, ${dN \over dM} \propto
M^{-\alpha}$, where the Salpeter value is $\alpha = 2.5$.
\citet{Schultheis06} have recently proposed that $\alpha
>2$ at low masses (implying that brown dwarfs and low-mass stars
essentially account for dark matter). This is in contrast to
values of $\alpha \sim 1$ derived in most other analysis, such as
\citet{PMSU4}\footnote{Note that \citeauthor{Schultheis06} are
incorrect in suggesting that the nearby star sample employed in
this analysis has significant incompleteness.}. A full discussion
of this issue is beyond the scope of the present paper. However,
we note that the \citeauthor{Schultheis06} analysis rests on
matching models against the \textit{V}-band luminosity function
for nearby stars, and is therefore very weakly constrained for
$0.1 M_\sun < M < 0.15 M_\sun$ and essentially unconstrained at
lower masses. \citet{Burgasser04} and \citet{Allen05} have shown
that our initial results presented in \citetalias{Cruz03} (which
are in general agreement with the luminosity function derived
here) are consistent with a range of values of $\alpha < 1.5$.
While this upper limit is not fully satisfying, it is in
disagreement with the steeply re-rising mass function as proposed
by \citeauthor{Schultheis06}.

\section{Summary}
\label{sec:9summary}

We have mined the 2MASS Second Incremental Data Release PSC for
ultracool dwarfs within 20~parsecs of the Sun. Extensive
spectroscopic follow-up has led to the discovery of $\sim$100 L
dwarfs and $\sim$200~late-M dwarfs---over 50 of these are within
20~pc \citepalias[and this paper]{Cruz03}, doubling the local
census of ultracool dwarfs. In these data we have also uncovered
several wide binaries, a handful of young objects, and two
slightly metal-poor L dwarfs. Combining our data with previously
known nearby late-type dwarfs, we have used \LFobjects~objects in
\LFsystems~systems to estimate the \textit{J} and
\textit{K$_S$}-band luminosity functions of ultracool dwarfs in
the Solar Neighborhood. This work has provided the first robust
estimate of the luminosity function of late-type stars and brown
dwarfs. We have measured the density of late-M dwarfs (M7--M9.5)
to be $4.9\pm0.6\times10^{-3}$~pc$^{-3}$ and the density of L
dwarfs to be at least $3.8\pm0.6\times10^{-3}$~pc$^{-3}$. Reliable
discrimination between different models for the underlying mass
function must await observational surveys that probe brown dwarfs
at temperatures below $\sim$600~K, likely near the boundary
between T and Y dwarfs.

\acknowledgments

We would like to acknowledge the numerous NOAO telescope operators
and support staff at Kitt Peak, Cerro Tololo, Cerro Panchon, and
Mauna Kea that made this work possible and endured our busy
observing program: S.~Adams, A.~Alvarez, T.~Beck, M.~Bergmann,
R.~Carrasco, G.~Doppmann, E.~Eastburn, A.~Fhima, B.~Gillespie,
P.~Gomez, A.~Guerra, M.~Hainaut-Rouelle, H.~Halbedel, D.~Harmer,
K.~Labrie, L.~Macri, H.~Mathis, A.~Matulonis, D.~Maturana,
S.~Pizarro, P.~Prado, K.~Roth, K.~Volk, and D.~Willmarth. We also
thank the NOAO Telescope Allocation committees for their enduring
support of this project. We acknowledge Finlay Mungall for
observing assistance. K.~L.~C is supported by a NSF Astronomy and
Astrophysics Postdoctoral Fellowship under AST-0401418. This
research was partially supported by a grant from the NASA/NSF
NStars initiative, administered by JPL, Pasadena, CA. This
publication makes use of data products from the Two Micron All Sky
Survey, which is a joint project of the University of
Massachusetts and Infrared Processing and Analysis
Center/California Institute of Technology, funded by the National
Aeronautics and Space Administration and the National Science
Foundation; the NASA/IPAC Infrared Science Archive, which is
operated by the Jet Propulsion Laboratory/California Institute of
Technology, under contract with the National Aeronautics and Space
Administration. Based on observations obtained with the Apache
Point Observatory 3.5-meter telescope, which is owned and operated
by the Astrophysical Research Consortium. Based on observations
obtained at the Gemini Observatory, which is operated by the
Association of Universities for Research in Astronomy, Inc., under
a cooperative agreement with the NSF on behalf of the Gemini
partnership: the National Science Foundation (United States), the
Particle Physics and Astronomy Research Council (United Kingdom),
the National Research Council (Canada), CONICYT (Chile), the
Australian Research Council (Australia), CNPq (Brazil) and CONICET
(Argentina) This research has made use of the SIMBAD database,
operated at CDS, Strasbourg, France.

\textit{Facilities}: \facility{FLWO:2MASS}, \facility{CTIO:2MASS},
\facility{Mayall (MARS)}, \facility{Blanco (RC Spec)},
\facility{Gemini:South (GMOS)}, \facility{Gemini:Gillett (GMOS),
\facility{KPNO:2.1m (GoldCam)}, \facility{CTIO:1.5m (RC Spec)},
\facility{ARC (DIS II)}}



\clearpage

\begin{figure}
\epsscale{1.0}
\plotone{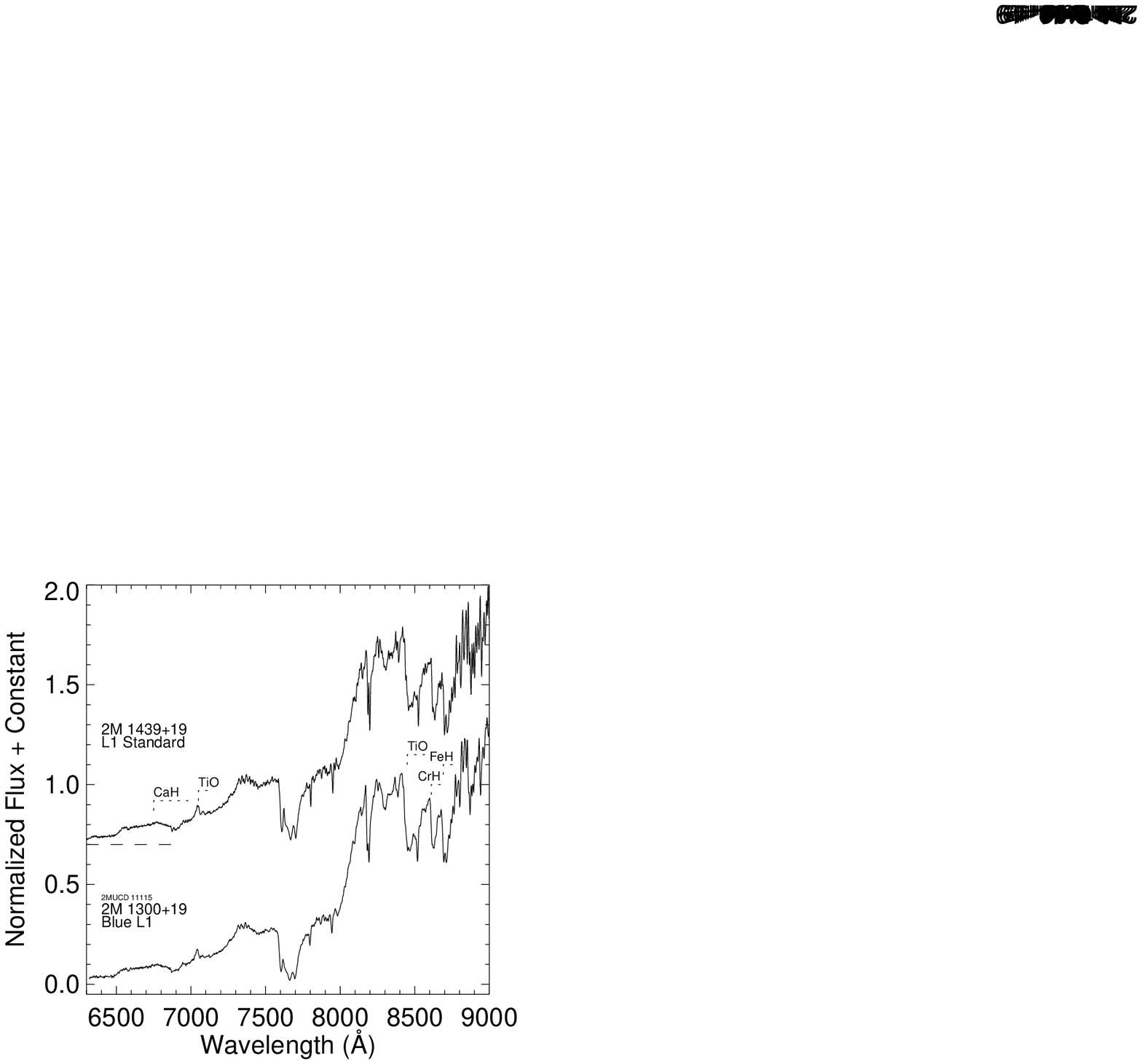}
\caption[Spectrum of blue L1 dwarf.]{Spectrum of blue L1 dwarf,
\objectname[2MASSI J1300425+191235]{2M~1300+19} with the L1
standard \objectname[2MASS J14392836+1929149]{2M~1439+19}
\citep{K99}. Metallicity-sensitive spectral features are marked.
Despite blue near-infrared colors and fast kinematics, the
spectrum of 2M~1300+19 does not exhibit any unusual features. The
bottom spectrum is not offset and the zero point of the spectral
standard is shown by a dashed line. \label{fig:blueL1}}
\end{figure}

\clearpage

\begin{figure}
\epsscale{1.0}
\plotone{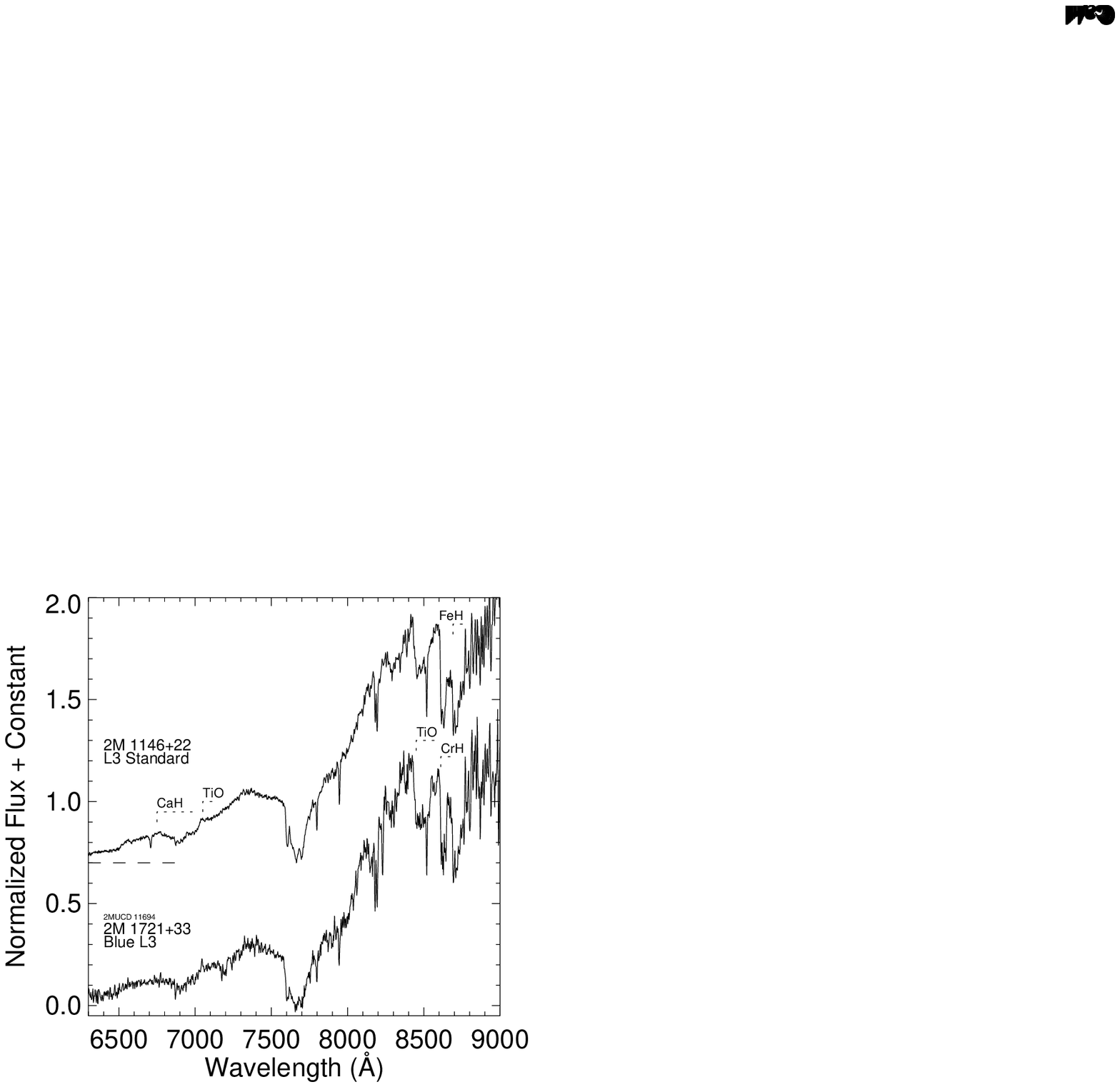}
\caption[Spectrum of blue L3 dwarf.]{Spectrum of blue L3 dwarf,
\objectname[2MASSI J1721039+334415]{2M~1721+33} with the L3
standard \objectname[2MASSI J1146344+223052]{2M~1146+22}
\citep{K99}. Metallicity-sensitive spectral features are marked.
Despite blue near-infrared colors and fast kinematics, the
spectrum of 2M~1721+33 does not exhibit any unusual features. The
bottom spectrum is not offset and the zero point of the spectral
standard is shown by a dashed line. \label{fig:blueL3}}
\end{figure}

\clearpage

\begin{figure}
\plotone{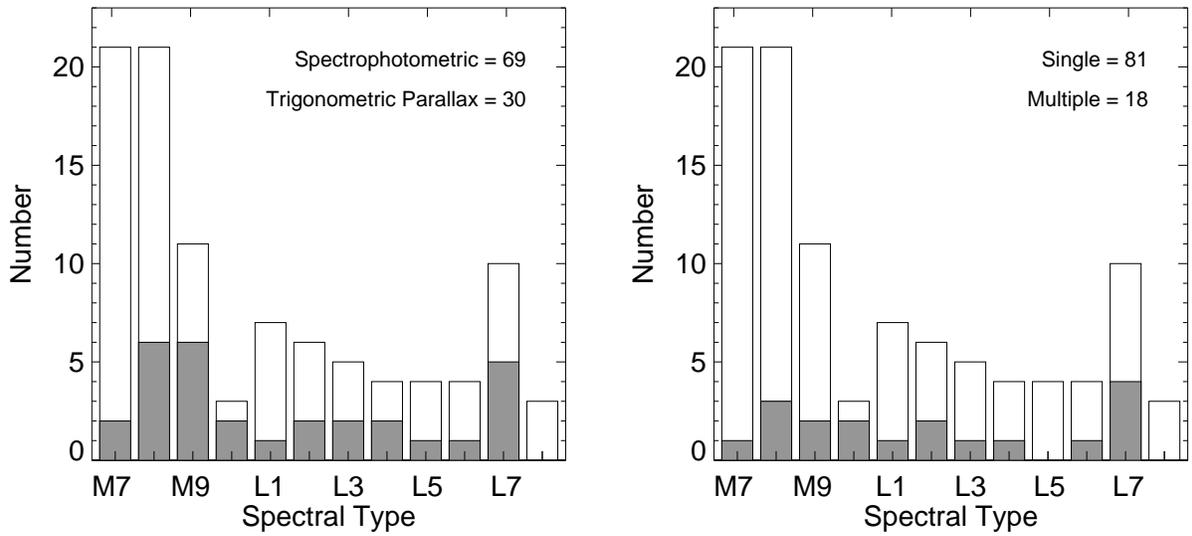}
\caption[]{Spectral type distribution of the 20-pc 2MU2 sample. In
the left panel, objects with distances based on trigonometric
parallax measurements (\textit{shaded}) are distinguished from
those with spectrophotometric distance estimates (\textit{not
shaded}). In the right panel, objects in multiple systems
(\textit{shaded}) are distinguished from single objects
(\textit{not shaded}). \label{fig:stdist}}
\end{figure}

\clearpage

\begin{figure}
\plotone{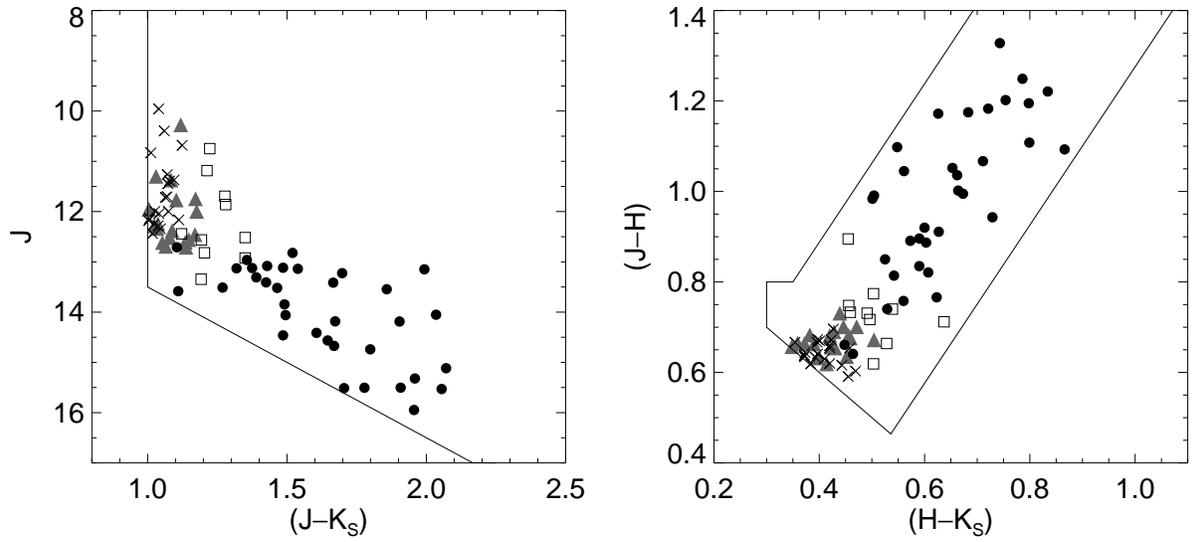}\caption[Color-magnitude and color-color
diagrams of the 2MU2 sample]{Color-magnitude and color-color
selection criteria (\textit{solid line}) and distribution of the
20-pc 2MU2 sample grouped by spectral type: M7 and M7.5
(\textit{crosses}), M8 and M8.5 (\textit{triangles}), M9 and M9.5
(\textit{squares}), and L dwarfs (\textit{circles}). The two blue
L dwarfs near $(J-K_S) = 1.1$ in the left panel and $(H-K_S) =
0.45$ in the right panel are discussed in \S~\ref{sec:blueLs}.
\label{fig:LFcolor}}
\end{figure}

\clearpage

\begin{figure}
\plotone{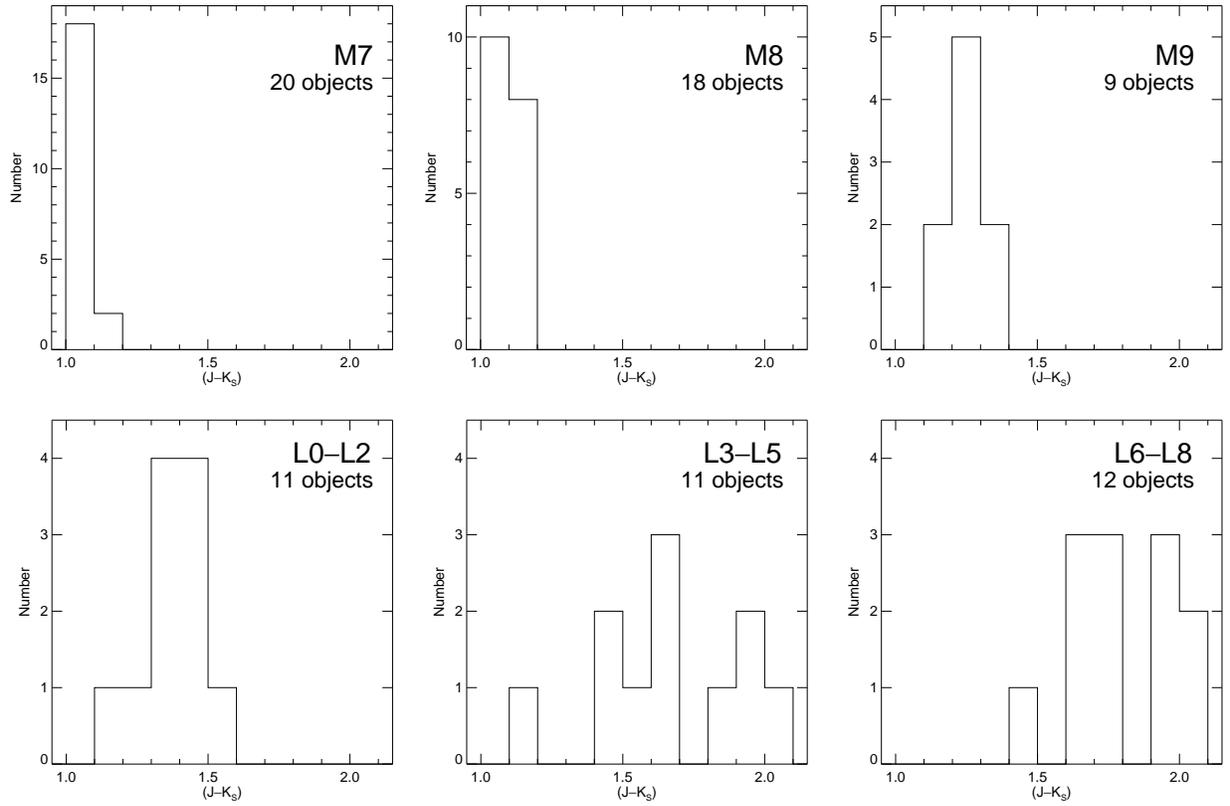}
\caption[$(J-K_S)$ distribution for single dwarfs in the 20-pc
2MU2 sample.]{$(J-K_S)$ distribution for single dwarfs in the
20-pc 2MU2 sample.\label{fig:jkdist}}
\end{figure}

\clearpage

\begin{figure}
\epsscale{1.0}
\plotone{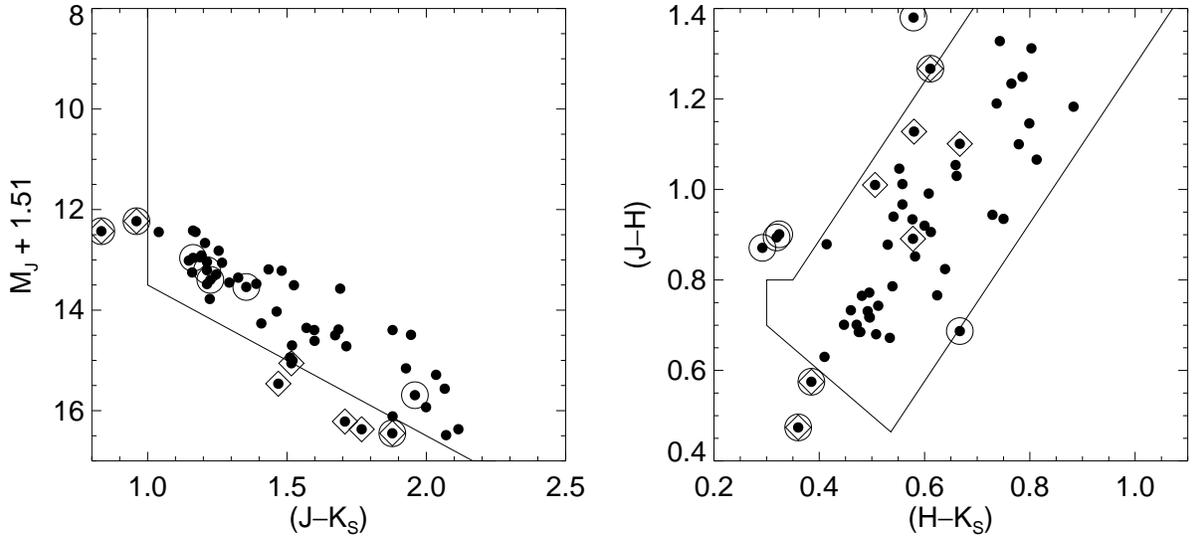}
\caption[Color-magnitude and color-color diagram of ultracool
dwarfs with parallaxes]{Color-magnitude and color-color diagram of
M7--L8 dwarfs with parallax measurements and $J<16.5$ (UCDt
sample, \textit{filled circles}), and our selection criteria
(\textit{solid lines}). $M_J$ has been shifted by 1.51 to reflect
our distance limit of 20~pc. Objects that are excluded by our
selection criteria in $J/(J-K_S)$ (\textit{diamonds}) and
$(J-H)/(H-K_S)$ (\textit{open circles}) are marked in both panels
and discussed in \S~\ref{sec:st_complete}.\label{fig:missing}}
\end{figure}

\clearpage

\begin{figure}
\plotone{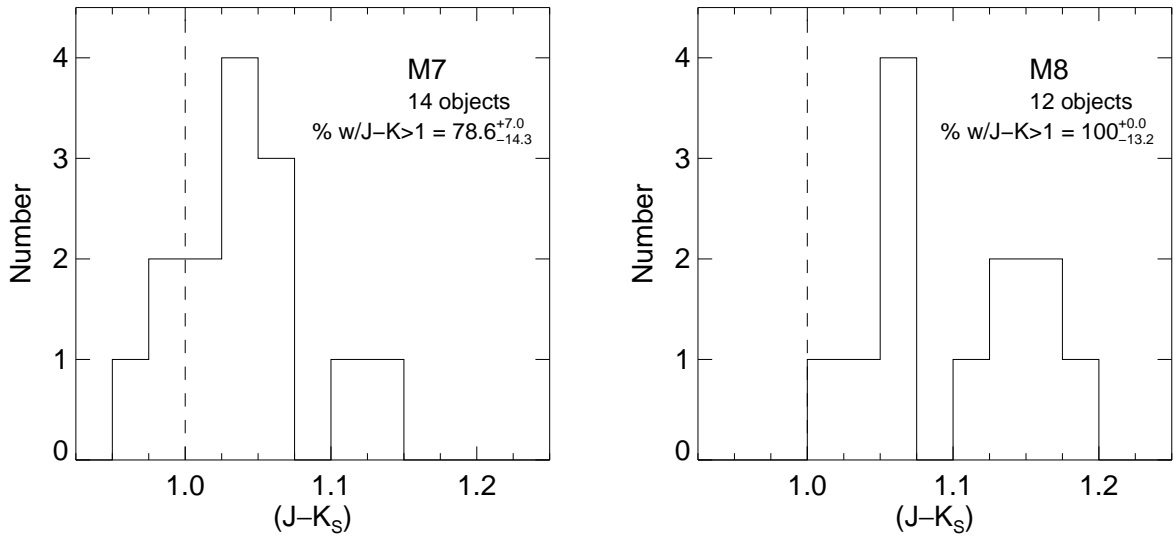}
\caption[]{$(J-K_S)$ distribution for M7 and M8 dwarfs not
discovered using $(J-K_S)$ color as a search criterion. The
resulting fraction of M7 dwarfs with $(J-K_S)>1$ is applied as a
correction to our derived \LF.\label{fig:m7correction}}
\end{figure}

\clearpage

\begin{figure}
\epsscale{0.5}\plotone{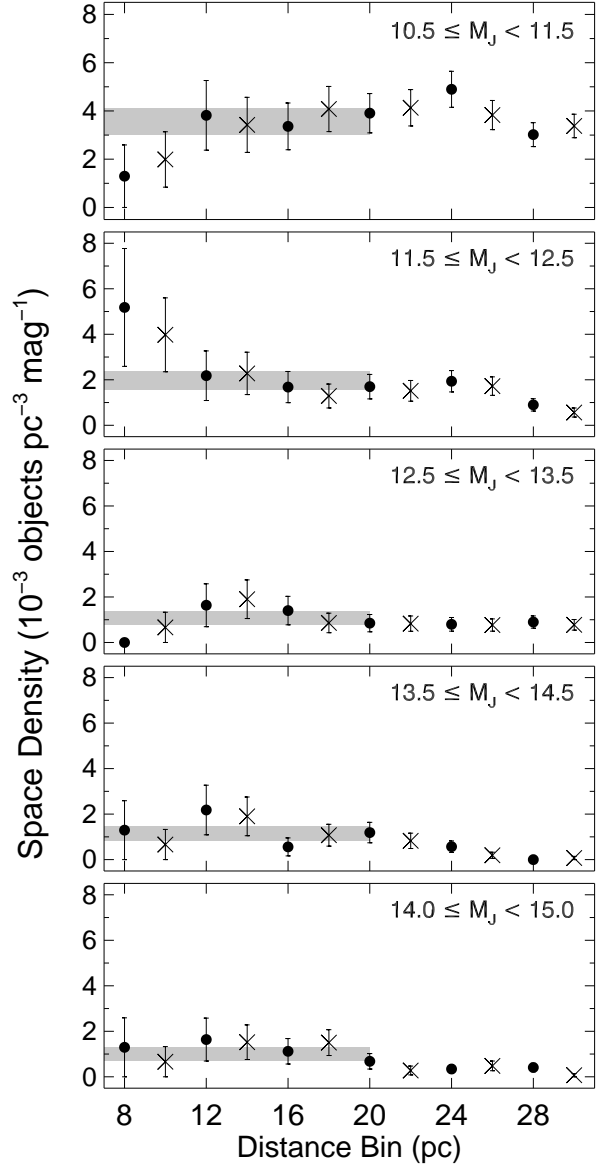}\caption[Volume completeness
of the 2MU2 sample]{Malmquist corrected space densities for two
sets of spherical shells for the 2MU2 sample as a function of
$M_J$. The densities are shown for spherical shells with inner and
outer radii from 0--8, 8--12, 12--16, 16--20, 20--24, 24--28~pc
(\textit{circles}) and 0--10, 10--14, 14--18, 18--22, 22--26,
26--30~pc (\textit{crosses}). The density for each shell is
plotted at the distance of the outer radius (e.g., the density for
the 8--12~pc shell is plotted at 12~pc). The shaded bar indicates
the overall density for each $M_J$ bin and the associated Poisson
uncertainty. These $M_J$ bins are coarser than those used in \LF,
one~magnitude wide instead of 0.5~magnitudes, and the two faintest
overlap due to the odd number of bins.\label{fig:complete}}
\end{figure}

\clearpage

\begin{figure}
\epsscale{1}
\plotone{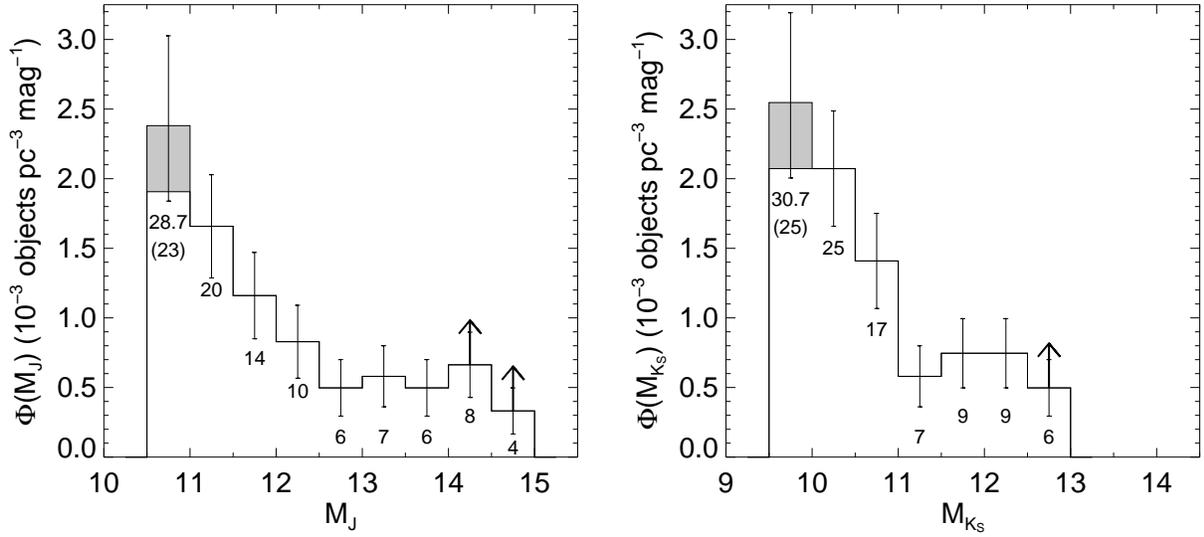}\caption[$J$- and $K_S$-band luminosity function of
ultracool dwarfs]{Malmquist corrected near-infrared luminosity
function with Poisson uncertainties (\textit{unshaded}) and the
correction for missing M7 dwarfs (\textit{shaded},
\S~\ref{sec:st_complete}). Each bin is labeled with the number of
objects used to calculate the space density. The space densities
are listed in Table~\ref{tab:lf}. Our measurement is a lower limit
for $14<M_J<15$ and $12.5<M_{K_S}<13$ because the 2MU2 sample is
incomplete for L7 and L8 dwarfs and does not include early and
mid-T dwarfs. \label{fig:LF2}}
\end{figure}

\clearpage

\begin{figure}
\epsscale{1}
\plotone{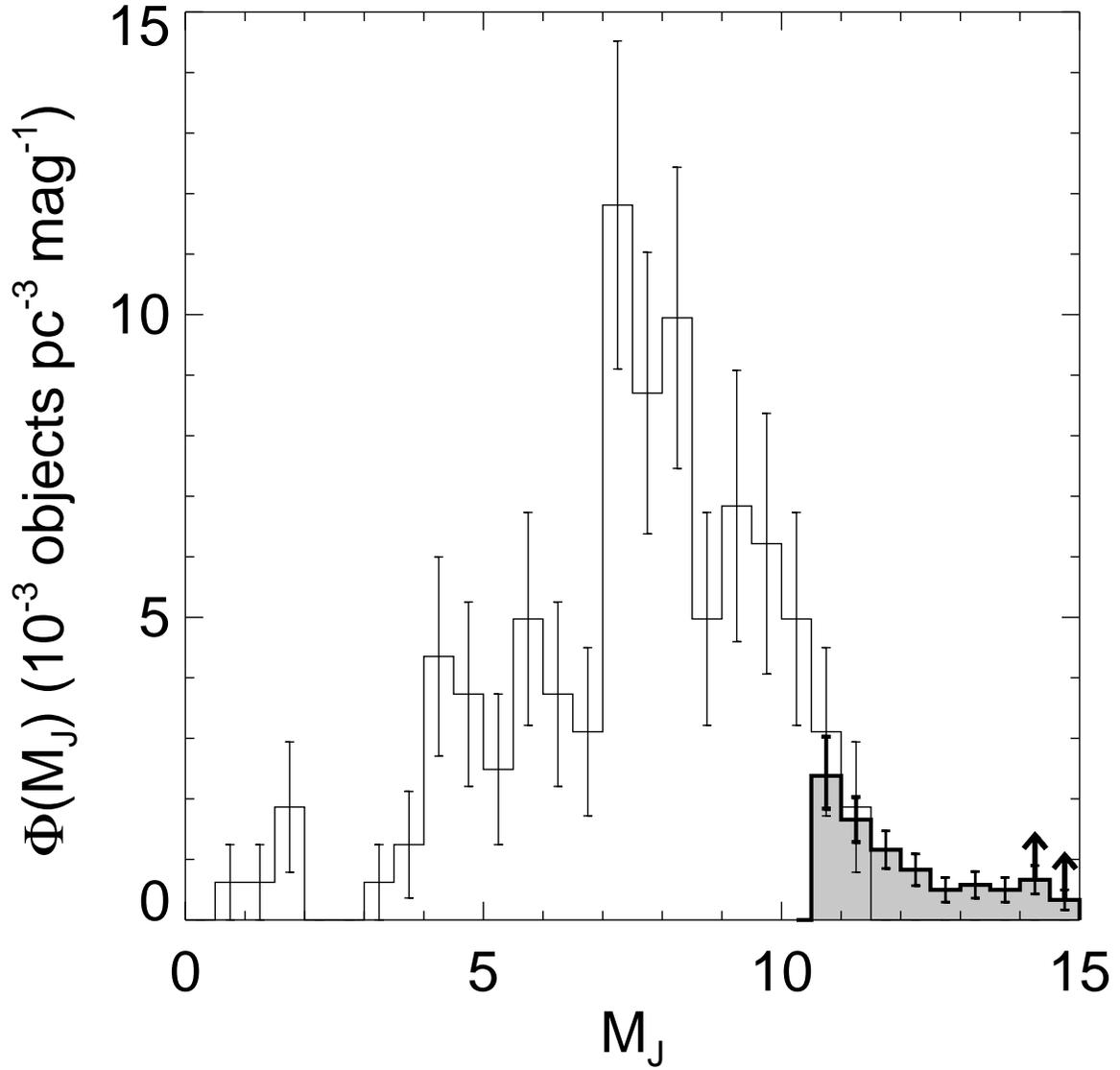}\caption[$J$-band luminosity function of the
8-pc and 20-pc 2MU2 samples]{$J$-band luminosity function of the
8-pc \citep[\textit{unshaded}]{Paper4} and 20-pc 2MU2
(\textit{shaded}) samples. The plotted densities are listed in
Tables~\ref{tab:lf} and \ref{tab:8pclf}.\label{fig:8pcLF}}
\end{figure}

\clearpage

\input{tab1.tex}
\clearpage
\input{tab2.tex}
\clearpage
\input{tab3.tex}
\clearpage
\input{tab4.tex}
\clearpage
\input{tab5.tex}
\clearpage
\input{tab6.tex}
\clearpage
\input{tab7.tex}
\clearpage
\input{tab8.tex}
\clearpage
\input{tab9.tex}
\clearpage
\input{tab10.tex}
\clearpage
\input{tab11.tex}
\clearpage
\input{tab12.tex}

\end{document}

%% file: tab1.tex
\begin{deluxetable}{llll}
\tablecaption{Spectral Standards Used\label{tab:standards}}
\tablehead{ \colhead{Optical} & \colhead{} & \colhead{} \\
             \colhead{Sp. Type} & \colhead{Object} & \colhead{Telescope} & \colhead{Ref.}}
\startdata

M1.5 & \object{Gl 205}   & KP 2.1 m & 1 \\
M2.5 & \object{Gl 250B}  & CT 1.5 m & 1 \\
M4.5 & \object{Gl 83.1}  & CT 4 m & 1 \\
M5.5 & \object{Gl 65}    & CT 4 m & 1  \\
M6   & \object{LHS 1326}\tablenotemark{a} & KP 2.1 m &  2  \\
M7   & \object{VB 8}     &  KP 4 m & 1  \\
M8   & \object{VB 10}    &  KP 4 m &  1 \\
M9   & \object{LHS 2065} &  CT 4 m & 1  \\
L0   & \object{2MASS J03454316+2540233} &  Keck I  &  3   \\
L1   & \object{2MASS J14392837+1929150} &  Keck I  &  3   \\
L2   & \objectname[2MASS J13054019-2541059]{Kelu 1}      &  Keck I &  3  \\
L3   & \objectname[DENIS-P J1058.7-1548]{DENIS-P J1058.7$-$1548}  &  Keck I     &  3 \\
L4   & \objectname[2MASSW J1155008+230705]{2MASS J11550087+2307058}  &  Keck I  &  3  \\
L4.5 & \object{2MASS J06523073+4710348}   & GN     & 4 \\
L5   & \objectname[DENIS-P J122813.8-154711]{DENIS-P J1228.2$-$1547}  & Keck I & 3 \\
     &                                    & GS     &  \\
L6   & \object{2MASS J08503593+1057156}   & Keck I & 3 \\
L7   & \objectname[DENIS-P J020529.0-115925]{DENIS-P J0205.4$-$1159}  & Keck I & 3 \\
     &                                    & GS     &  \\
L8   & \object{2MASS J16322911+1904407}   & Keck I & 3  \\
     &                                    & GN     &   \\
\enddata

\tablecomments{While M and L dwarf spectral standards were taken
during the course of our observations, we supplemented our data
with the publicly available high signal-to-noise Keck I + LRIS
spectra (\url{DwarfArchives.org}).}

\tablenotetext{a}{Not actually a spectral standard.}

\tablerefs{(1)~\citet{KHM91}; (2)~\cite{PMSU}; (3)~\cite{K99}; (4)~\cite{Paper10}
}

\end{deluxetable}

%% file: tab2.tex
\begin{deluxetable}{lllllllllllc}
\tablewidth{0pt}
\tabletypesize{\scriptsize}
\rotate
\tablecaption{M6--L8 Dwarfs Discovered Within 20 pc\label{tab:9latenear}}
\tablehead{ \colhead{2MASSI} & \colhead{2MUCD} & \colhead{Other} & \colhead{2MASS} &
\colhead{2MASS} & \colhead{2MASS} & \colhead{Obs.} & \colhead{} &
\colhead{Optical}& \colhead{}& \colhead{$d$}& \colhead{Other}\\
\colhead{Designation\tablenotemark{a}} &  \colhead{\#} & \colhead{Name} &
\colhead{$J$} & \colhead{$J-H$} & \colhead{$J-K_S$} & \colhead{Date (UT)} & \colhead{Telescope}
& \colhead{Sp. Type}& \colhead{$M_J$}& \colhead{(pc)}& \colhead{Refs.}
}
\startdata
\object[2MASS J01231125-6921379]{0123112$-$692138} & 13056 &                 & 12.335 & 0.619 & 1.034 & 2003 Nov 8  & CT 1.5~m & M8                     & 11.16$\pm$0.18                  & 17.2$\pm$1.4                  &  \\ 
                   &       &              &        &       &       & 2006 Jan 15 & CT 4~m   &                        &                                 &                               &  \\ 
\object[2MASS J01382152-7320584]{0138215$-$732058} & 13063 &                 & 11.508 & 0.720 & 1.056 & 2003 Nov 9  & CT 1.5~m & M6                     & 10.12$\pm$0.37                  & 19.0$\pm$3.3                  &  \\
                    &       &              &        &       &       & 2006 Jan 15 & CT 4~m   &                        &                                 &                               &  \\
\object[2MASS J03185403-3421292]{0318540$-$342129} & 10176 &                 & 15.530 & 1.221 & 2.055 &             & Keck~I   & L7                     & 14.45$\pm$0.30                  & 16.5$\pm$2.3                  &  1 \\
\object[2MASS J03400942-6724051]{0340094$-$672405} & 10202 &                 & 14.740 & 1.172 & 1.799 & 2006 Jan 14 & CT 4~m   & L7::                   & 14.45$\pm$0.90                  & 11.4$\pm$4.7                  & \\
\object[2MASS J05441150-2433018]{0544115$-$243301} & 10444 &                 & 12.518 & 0.656 & 1.079 & 2002 Sep 25 & KP 4~m   & M8                     & 11.16$\pm$0.18                  & 18.7$\pm$1.6                  &  \\ 
\object[2MASS J07003664+3157266]{0700366+315726}   & 10617 &                 & 12.922 & 0.967 & 1.614 & 2003 Oct 12 & KP 2.1~m & L3.5comb\tablenotemark{b} & 12.49$\pm$0.06\tablenotemark{b} & 12.2$\pm$0.3\tablenotemark{b} & 2, 3 \\
\enddata

\tablenotetext{a}{The 2MASS designation is 2MASSI
Jhhmmss[.]s$\pm$ddmmss.}

\tablenotetext{b}{This object has been resolved into two
components \citep{Reid06_binary} and the spectral type listed here
is for the unresolved system. The distance and $M_J$ for the
system is based on a trigonometric parallax \citep{TK03}.}

\tablerefs{(1)~\citeauthor[in prep.]{DavyLs} (2)~\citet{TK03};
(3)~\citet{Reid06_binary}; (4)~\citet{LSPM_North};
(5)~\citet{Lepine05}; (6)~\citet{Gizis02}.}

\end{deluxetable}

%% file: tab3.tex
\begin{deluxetable}{lllllllllllc}
\tablewidth{0pt}
\tabletypesize{\scriptsize}
\setlength{\tabcolsep}{0.04in}
\rotate
\tablecaption{M7--L8 Dwarfs Discovered Outside 20 pc\label{tab:9latefar}}
\tablehead{ \colhead{2MASSI} & \colhead{2MUCD} &\colhead{Other} & \colhead{2MASS} &
\colhead{2MASS} & \colhead{2MASS} & \colhead{Obs.} & \colhead{} &
\colhead{Optical}& \colhead{}& \colhead{$d$}& \colhead{Other}\\
\colhead{Designation\tablenotemark{a}} & \colhead{\#} &\colhead{Names} &
\colhead{$J$} & \colhead{$J-H$} & \colhead{$J-K_S$} & \colhead{Date (UT)} & \colhead{Telescope}
& \colhead{Sp. Type}& \colhead{$M_J$}& \colhead{(pc)}& \colhead{Refs.}
}
\startdata
\object[2MASS J00002867-1245153]{0000286$-$124515}  & 10001 &                              & 13.166 & 0.746 & 1.223 & 2002 Jul 07 & KP 2.1 m   & M8.5                 & 11.32$\pm$0.15 & 23.4$\pm$1.7   & \\ 
\object[2MASS J00034227-2822410]{0003422$-$282241}  & 10002 & HD 225118B?\tablenotemark{b} & 13.063 & 0.699 & 1.108 & 2002 Sep 25 & KP 4 m     & M7.5                 & 10.96$\pm$0.21 & 26.3$\pm$2.6   & \\ 
\object[2MASS J00070787-2458042]{0007078$-$245804}  & 10003 & LEHPM 193                    & 13.147 & 0.684 & 1.095 & 2002 Sep 26 & KP 4 m     & M7                   & 10.73$\pm$0.25 & 30.4$\pm$3.5   & 1, 2 \\ 
\object[2MASS J00100009-2031122]{0010001$-$203112}  & 10006 &                              & 14.153 & 0.761 & 1.285 & 2002 Sep 25 & KP 4 m     & L0                   & 11.73$\pm$0.13 & 30.5$\pm$1.9   & \\ 
\object[2MASS J00211624-6644052]{0021162$-$664405}  & 13013 &                              & 13.663 & 0.715 & 1.269 & 2004 Aug 09 & CT 4 m     & M7::                 & 10.73$\pm$1.01 & 38.5$\pm$17.9  & \\ 
\object[2MASS J00234754-3239254]{0023475$-$323925}  & 10017 &                              & 12.646 & 0.641 & 1.002 & 2002 Sep 28 & KP 4 m     & M7                   & 10.73$\pm$0.25 & 24.1$\pm$2.8   & \\ 
\object[2MASS J00250365+4759191]{0025036+475919  }  & 13016 & HD 2057B?\tablenotemark{b}   & 14.859 & 1.210 & 1.953 & 2004 Sep 22 & GN         & L4:\tablenotemark{c} & 13.09$\pm$0.22 & 22.5$\pm$2.4   & \\
\enddata

\tablenotetext{a}{The 2MASS designation is 2MASSI
Jhhmmss[.]s$\pm$ddmmss.}

\tablenotetext{b}{Ultracool common proper motion companion to a
star as discussed in \S~\ref{sec:widebinaries}.}

\tablenotetext{c}{Spectrum shows lithium absorption line as shown
in Figure~\ref{fig:spec} and discussed in \S~\ref{sec:lithium}.}

\tablenotetext{d}{Probable L/T binary
\citep{Cruz04,Burgasser06_binary}. Combined-light optical spectral
type from \citeauthor[in prep.]{Davy0518}.}

\tablenotetext{e}{Variable and/or flaring object that will be
discussed in detail in \citet[in prep.]{Sarah}.}

\tablerefs{(1)~\citet{Reyle04}; (2)~\citet{LEHPM}; (3)~\citet{Deacon05};
(4)~\citet{Cruz04}; (5)~\citeauthor[in prep.]{Davy0518};
(6)~\citet{Burgasser06_binary}; (7)~\citet{LSPM_North};
(8)~\citet{Sarah}; (9)~\citet{Lepine05}; (10)~\citet{NN4};
(11)~\citet{CESO}; (12)~\citeauthor[in prep.]{DavyLs};
(13)~\citet{Lodieu02}.}

\end{deluxetable}

%% file: tab4.tex
\begin{deluxetable}{llllllllllc}
\tablewidth{0pt}
\rotate
\tabletypesize{\footnotesize}
\setlength{\tabcolsep}{0.06in}
\tablecaption{Early-to-Mid Type M Dwarfs Discovered Outside 20 pc\label{tab:9earlyfar}}
\tablehead{ \colhead{2MASSI} & \colhead{Other} & \colhead{2MASS} &
\colhead{2MASS} & \colhead{2MASS} & \colhead{Obs.} & \colhead{} &
\colhead{Optical}& \colhead{}& \colhead{$d$}& \colhead{Other}\\
\colhead{Designation\tablenotemark{a}} & \colhead{Names} &
\colhead{$J$} & \colhead{$J-H$} & \colhead{$J-K_S$} & \colhead{Date (UT)} & \colhead{Telescope}
& \colhead{Sp. Type}& \colhead{$M_J$} & \colhead{(pc)}&
\colhead{Refs.} }
\startdata
\object[2MASS J00171857-0406062]{0017185$-$040606} &                           & 12.485 & 0.642 &  1.025 & 2002 Jul 07 &  KP 2.1 m & M6.5 &   10.5$\pm$0.3 &\phn25$\pm$4   &   \\
\object[2MASS J01075903-2004237]{0107590$-$200423} &                           & 12.817 & 0.614 &  1.024 & 2002 Sep 27 &  KP 4 m   & M6   &   10.1$\pm$0.4 &\phn35$\pm$6   &   \\
\object[2MASS J03554031-1123105]{0355403$-$112310} &                           & 12.947 & 0.645 &  1.010 & 2002 Sep 27 &  KP 4 m   & M6   &   10.1$\pm$0.4 &\phn37$\pm$6   &   \\
\object[2MASS J04110642+1247481]{0411063+124748  } &                           & 12.777 & 0.652 &  1.015 & 2002 Sep 28 &  KP 4 m   & M6   &   10.1$\pm$0.4 &\phn34$\pm$6   &   \\
\object[2MASS J05102396-2800539]{0510239$-$280053} &                           & 12.964 & 0.646 &  1.010 & 2002 Sep 28 &  KP 4 m   & M4   &\phn8.5$\pm$0.7 &\phn60--100    &   \\
\object[2MASS J05441677-2049098]{0544167$-$204909} &                           & 14.394 & 1.002 &  1.515 & 2002 Sep 25 &  KP 4 m   & M5   &\phn8.5$\pm$0.7 &   110--180    &   \\
\object[2MASS J08183325-0036282]{0818332$-$003628} &                           & 16.639 & 1.302 &  2.110 & 2004 Dec 12 &  GS       & M6   &   10.1$\pm$0.4 &   202$\pm$38  &   \\
\object[2MASS J09411157+4254031]{0941115+425403  } &                           & 16.753 & 1.288 &  2.101 & 2004 Nov 07 &  GN       & M5   &\phn9.3$\pm$0.4 &   312$\pm$59  &   \\
\enddata

\tablenotetext{a}{The 2MASS designation is 2MASSI
Jhhmmss[.]s$\pm$ddmmss.}

\tablenotetext{b}{A spectrum of this object is shown by \cite{B91}
but not mentioned in the text nor assigned a spectral type. It
appears to have been overlooked in the subsequent literature on
late-type M dwarfs. To our knowledge, it was first formally
assigned a spectral type by \citet[Paper VII]{Paper7} using the
same data presented here.}

\tablerefs{(1)~\citet[LHS]{LHS}; (2)~\citet{B91}; (3)~\citet[Paper~I]{Paper1};
(4)~\citet[Paper VII]{Paper7}; (5)~\citet{RG05};
(6)~\citet{LSPM_North}; (7)~\citet{Lepine05}; (8)~\citet{NN4};
(9)~\citet{NN6}; (10)~\citet[NLTT]{NLTT}.}

\end{deluxetable}

%% file: tab5.tex
\begin{deluxetable}{llllllr}
\tablewidth{0pt}
\tabletypesize{\small}
\tablecaption{Spectroscopically Confirmed Giants\label{tab:9giants}}
\tablehead{\colhead{2MASSI} & \colhead{2MASS} & \colhead{2MASS} &
\colhead{2MASS} & \colhead{Obs.} & \colhead{} & \colhead{Optical} \\
\colhead{Designation\tablenotemark{a}} & \colhead{$J$} & \colhead{$J-H$} &
\colhead{$J-K_S$} & \colhead{Date (UT)} & \colhead{Telescope} & \colhead{Sp. Type} }
\startdata
\object[2MASS J00570171+4509495]{0057017+450949  } &  10.002     & 0.776 &  1.202 & 2002 Jul 08 &  KP 2.1 m & M8 III    \\
\object[2MASS J04262583+1545029]{0426258+154502  } &  11.860     & 0.695 &  1.021 & 2002 Sep 26 &  KP 4 m   & M5 III    \\
\object[2MASS J05575096-1359503]{0557509$-$135950} &  12.873     & 0.741 &  1.135 & 2002 Sep 27 &  KP 4 m   & M6 pec III\tablenotemark{b} \\
\object[2MASS J07124352+3958313]{0712435+395831  } &  \phn9.156  & 0.743 &  1.119 & 2002 Sep 25 &  KP 4 m   & K5 III    \\
\object[2MASS J11190511+7006099]{1119051+700609  } &  10.276     & 0.741 &  1.096 & 2002 Jul 08 &  KP 2.1 m & M4 III    \\
\object[2MASS J13403717-0116045]{1340371$-$011604} &  13.492     & 0.619 &  1.106 & 2002 Jul 07 &  KP 2.1 m & $<$K5 III  \\
\object[2MASS J15502490+8210093]{1550248+821009  } &  \phn9.667  & 0.778 &  1.095 & 2002 Jul 10 &  APO      & M7 III    \\
\object[2MASS J17414944+1523172]{1741494+152317  } &  \phn9.173  & 0.758 &  1.071 & 2002 Sep 25 &  KP 4 m   & M3 III    \\
\enddata

\tablenotetext{a}{The 2MASS designation is 2MASSI
Jhhmmss[.]s$\pm$ddmmss.}

\tablenotetext{b}{Spectrum displays H$\alpha$ in emission.}

\end{deluxetable}

%% file: tab6.tex
\begin{deluxetable}{llllll}
\tablewidth{0pt}
\tablecaption{Spectroscopically Confirmed Carbon Stars\label{tab:9carbon}}
\tablehead{ \colhead{2MASSI} & \colhead{2MASS} & \colhead{2MASS} &
\colhead{2MASS} & \colhead{Obs.} & \colhead{} \\
\colhead{Designation\tablenotemark{a}} & \colhead{$J$} & \colhead{$J-H$} &
\colhead{$J-K_S$} & \colhead{Date (UT)} & \colhead{Telescope}}
\startdata
\object[2MASS J00081707-7537276]{0008170$-$753727} & 13.826     & 1.259 &  2.224 & 2006 Jan 15 &  CT 4~m    \\
\object[2MASS J00275240-7052312]{0027523$-$705231} & 13.368     & 1.382 &  2.320 & 2003 Nov 09 &  CT 1.5 m  \\
\object[2MASS J00441550-7115401]{0044155$-$711540} & 13.398     & 1.189 &  2.006 & 2003 Nov 09 &  CT 1.5 m  \\
\object[2MASS J00460080-7521128]{0046008$-$752112} & 13.182     & 1.071 &  1.922 & 2003 Nov 09 &  CT 1.5 m  \\
\object[2MASS J01075228-6921365]{0107522$-$692136} & 13.153     & 1.288 &  2.177 & 2003 Nov 09 &  CT 1.5 m  \\
\object[2MASS J01214338-7403460]{0121433$-$740346} & 13.688     & 1.488 &  2.604 & 2006 Jan 15 &  CT 4 m    \\
\object[2MASS J01263491-7039476]{0126348$-$703947} & 12.817     & 1.478 &  2.494 & 2003 Nov 09 &  CT 1.5 m  \\
\object[2MASS J01480783-7155220]{0148078$-$715521} & 12.471     & 1.271 &  2.199 & 2003 Nov 10 &  CT 1.5 m  \\
\enddata

\tablenotetext{a}{The 2MASS designation is 2MASSI
Jhhmmss[.]s$\pm$ddmmss.}

\end{deluxetable}

%% file: tab7.tex
\begin{deluxetable}{lllllllllllc}
\tablewidth{0pt}
\rotate
\tabletypesize{\scriptsize}
\tablecaption{Objects with New Spectra\label{tab:newdata}}
\tablehead{ \colhead{2MASSI} & \colhead{2MUCD} & \colhead{Other} & \colhead{2MASS} &
\colhead{2MASS} & \colhead{2MASS} & \colhead{Obs.} & \colhead{} &
\colhead{Optical}& \colhead{}& \colhead{$d$}& \colhead{Other}\\
\colhead{Designation\tablenotemark{a}} & \colhead{\#} &\colhead{Names} &
\colhead{$J$} & \colhead{$J-H$} & \colhead{$J-K_S$} & \colhead{Date (UT)} & \colhead{Telescope}
& \colhead{Sp. Type}& \colhead{$M_J$} & \colhead{(pc)}& \colhead{Refs.}
}
\startdata
\object[2MASS J01410321+1804502]{0141032+180450  }   & 10085 &                              & 13.822 & 0.772 & 1.315 & 2004 Feb 11 & KP 4 m   & L1                   & 12.00$\pm$0.14 & 23.1$\pm$1.6   & 1 \\ 
\object[2MASS J03264225-2102057]{0326422$-$210205}   & 10184 &                              & 16.111 & 1.337 & 2.226 & 2005 Oct 10 & GS       & L4\tablenotemark{b}  & 13.09$\pm$0.22 & 40.1$\pm$4.5   & 2 \\
\object[2MASS J04285096-2253227]{0428509$-$225322}   & 10286 &                              & 13.579 & 0.882 & 1.459 & 2004 Feb 11 & KP 4 m   & L1                   & 12.00$\pm$0.14 & 20.7$\pm$1.4   & 3 \\
\object[2MASS J10031918-0105079]{1003191$-$010507}   & 10873 & LHS 5165/DENIS-P J1003$-$0105& 12.352 & 0.667 & 1.085 & 2003 Mar 15 & KP 2.1 m & M7            & 10.73$\pm$0.25 & 21.1$\pm$2.5   & 4, 5 \\
\object[2MASS J10110024+4245035]{1011002+424503  }   & 10885 & TVLM 263-71765               & 13.357 & 0.657 & 1.056 & 2004 Feb 10 & KP 4 m   & M7.5          & 10.96$\pm$0.21 & 30.1$\pm$2.9   & 6 \\
\object[2MASS J15150083+4847416]{1515009+484739  }   & 11314 &                              & 14.060 & 0.991 & 1.495 & 2002 Jul 10 & APO      & L6            & 14.02$\pm$0.30 & 10.2$\pm$1.4   & 1 \\
                    &       &                              &        &       &       & 2004 Feb 10 & KP 4~m   &                     &                &                &  \\
\enddata

\tablenotetext{a}{The 2MASS designation is 2MASSI
Jhhmmss[.]s$\pm$ddmmss.}

\tablenotetext{b}{Spectrum shows lithium absorption line as shown
in Figure~\ref{fig:spec} and discussed in \S~\ref{sec:lithium}.}

\tablerefs{(1)~\citet{Wilson}; (2)~\citetalias{Cruz03}; (3)~\citet{Kendall03};
(4)~\citet{Gizis02}; (5)~\citet{NN4}; (6)~\citet{Tinney95}}

\end{deluxetable}

%% file: tab8.tex
\begin{deluxetable}{lllllllllllcc}
\tablewidth{0pt}
\tabletypesize{\scriptsize}
\rotate
\tablecaption{Low-Gravity Objects\label{tab:9young}}

\tablehead{\colhead{2MASSI} & \colhead{2MUCD} &\colhead{} & \colhead{2MASS} & \colhead{2MASS} &
\colhead{2MASS} & \colhead{Obs.} & \colhead{} & \colhead{Optical}&
\colhead{$\mu$} & \colhead{PA} & \colhead{$\mu$} & \colhead{Other} \\

\colhead{Designation\tablenotemark{a}} &  \colhead{\#} & \colhead{Other Name} &
\colhead{$J$} & \colhead{$J-H$}& \colhead{$J-K_S$} & \colhead{Date (UT)} & \colhead{Telescope} &
\colhead{Sp. Type\tablenotemark{b}} & \colhead{(mas yr$^{-1}$)} & \colhead{($\degr$)} & \colhead{Ref.} & \colhead{Refs.}}

\startdata
\object[2MASS J02411151-0326587]{ 0241115$-$032658} & 10141 &                      & 15.831 & 1.019 & 1.807 & 2002 Sep 27 & KP 4~m   & (L1)                  & \phn57$\pm$18 & \phn88  & 1  & \\
  &       &                     &        &       &       & 2005 Oct 10 & GS       &                       &               &         &   &   \\
\object[2MASS J04362788-4114465]{ 0436278$-$411446} & 10306 & DENIS-P J0436$-$4114 & 13.105 & 0.689 & 1.043 & 2003 Nov 08 & CT 1.5~m & (M8)\tablenotemark{c} & \phn22$\pm$18 & \phn80  & 2 & 1, 2 \\
  &       &                      &        &       &       & 2006 Jan 15 & CT 4~m   &                       &               &         &   & \\
\object[2MASS J04433761+0002051]{ 0443376+000205  } & 10320 & SDSS 0443+0002       & 12.517 & 0.713 & 1.350 & 2002 Jan 28 & CT 1.5~m & (M9)\tablenotemark{d} & 159$\pm$78    & 146     & 1 & 3, 4\\
\object[2MASS J16154255+4953211]{ 1615425+495321  } & 11538 &                      & 16.732 & 1.441 & 2.413 & 2004 Sep 12 & GN       & (L4)                  & 115$\pm$83    & 287     & 1 & \\
\object[2MASS J22134491-2136079]{ 2213449$-$213607} & 12120 &                      & 15.392 & 0.975 & 1.647 & 2002 Sep 27 & KP 4~m   & (L1)                  & \nodata       & \nodata &   &   \\
                  &       &                      &        &       &       & 2005 Sep 08 & GS       &                       &               &         &   &   \\
\enddata

\tablecomments{Spectra are plotted in
Figures~\ref{fig:youngM8}--\ref{fig:youngL4}.}

\tablenotetext{a}{The 2MASS designation is 2MASSI
Jhhmmss[.]s$\pm$ddmmss.}

\tablenotetext{b}{The spectral types should be regarded as
preliminary since all of these objects have low-gravity spectral
features and do not fit into the dwarf spectral sequence as it is
currently defined.}

\tablenotetext{c}{These new spectral data supplements the
photometric data of \citet{NN5} listed in \citetalias{Cruz03}.}

\tablenotetext{d}{This is a reclassification based on a reanalysis
of the spectrum presented in \citetalias{Cruz03}.}

\tablerefs{Proper Motion: (1)~This Paper; (2)~\citet{NN5}.}

\tablerefs{Other: (1)~\citet{NN5}; (2)~\citet{NN6}; (3)~\citet{Wilson}; (4)~\citet{Hawley02}.}

\end{deluxetable}

%% file: tab9.tex
\begin{deluxetable}{lllllllllllcc}
\tablewidth{0pt}
\rotate
\tabletypesize{\scriptsize}
\setlength{\tabcolsep}{0.05in}
\tablecaption{The 20~pc 2MU2 Sample Used to Estimate the Luminosity Function\label{tab:9lf}}

\tablehead{ \colhead{2MASSI} & \colhead{2MUCD} & \colhead{Other} & \colhead{2MASS} &
\colhead{2MASS} & \colhead{2MASS} & \colhead{Optical}& \colhead{} &  \colhead{} &
\colhead{$d$} & \colhead{d} & \colhead{Discovery} & \colhead{Parallax}\\
\colhead{Designation\tablenotemark{a}} & \colhead{\#} & \colhead{Names} &
\colhead{$J$} & \colhead{$J-H$} & \colhead{$J-K_S$} & \colhead{Sp. Type\tablenotemark{b}}& \colhead{$M_J$} &
\colhead{$M_{K_S}$} & \colhead{(pc)\tablenotemark{c}} & \colhead{source\tablenotemark{c}} & \colhead{Refs.}
& \colhead{Refs.} }

\startdata
\object[2MASS J00192626+4614078]{0019262+461407   }                  & 10013 &                                         &   12.609 & 0.676 &  1.135 &  M8    &               11.16$\pm$0.18   &   10.02$\pm$0.18 & 19.5\phn$\pm$1.6  & ST       & 1        & \\
\object[2MASS J00194579+5213179]{0019457+521317   }                  & 10014 &                                         &   12.820 & 0.748 &  1.204 &  M9    &               11.47$\pm$0.14   &   10.26$\pm$0.15 & 18.7\phn$\pm$1.2  & ST       & 1        & \\
\object[2MASS J00242463-0158201]{0024246$-$015819 }                  & 10018 & BRI 0021-0214                           &   11.860 & 0.740 &  1.280 &  M9.5  &               11.55$\pm$0.10   &   10.27$\pm$0.10 &    11.55$\pm$0.53 & Parallax & 2        & 1 \\
\object[2MASS J00244419-2708242]{0024442$-$270825B}                  & 10019 & LHS 1070B\tablenotemark{d}              &\phn9.262 & 0.730 &  1.030 &  M8.5  &               11.64$\pm$0.06   &   10.49$\pm$0.05 & \phn7.71$\pm$0.14 & Parallax & 3, 4, 5  & 2, 3 \\
\object[2MASS J00244419-2708242]{0024442$-$270825C}                  &       & LHS 1070C                               &          &       &        &  L0    &               11.97$\pm$0.08   &   10.81$\pm$0.07 & \phn7.71$\pm$0.14 & Parallax & 3, 4, 5  & 2, 3 \\
\object[2MASS J00275592+2219328]{0027559+221932A  }                  & 10022 & LP 349-25A                              &   10.608 & 0.638 &  1.047 &  M8:   &               11.16$\pm$0.35   &   10.13$\pm$0.24 & 10.3\phn$\pm$1.7  & ST       & 6, 7     &  \\
\object[2MASS J00275592+2219328]{0027559+221932B  }                  &       & LP 349-25B                              &          &       &        &  M9:   &               11.47$\pm$0.28   &   10.39$\pm$0.25 & 10.3\phn$\pm$1.7  & ST       & 8        &  \\
\enddata
\end{deluxetable}

%% file: tab10.tex
\begin{deluxetable}{llllllllllc}
\tablewidth{0pt}
\rotate
\tabletypesize{\scriptsize}
\setlength{\tabcolsep}{0.05in}
\tablecaption{Objects with distance estimates within 1$\sigma$ of
20 parsecs\label{tab:onesigma}}
\tablehead{ \colhead{2MASSI} & \colhead{2MUCD} & \colhead{Other} & \colhead{2MASS} &
\colhead{2MASS} & \colhead{2MASS} & \colhead{Optical}&  \colhead{} &
\colhead{$d$} & \colhead{d} & \colhead{} \\
\colhead{Designation\tablenotemark{a}} &  \colhead{\#} & \colhead{Names} & \colhead{$J$} & \colhead{$J-H$} & \colhead{$J-K_S$} &
\colhead{Sp. Type\tablenotemark{b}}& \colhead{$M_J$} &  \colhead{(pc)\tablenotemark{c}} &
\colhead{source\tablenotemark{c}} & \colhead{Refs.}  }

\startdata
\object[2MASS J00502444-1538184]{0050244$-$153818 }   & 10045 &   SIPS 0050$-$1538            & 13.765 & 0.673 & 1.126 & L1   &  12.00$\pm$0.29 & 22.5$\pm$3.0 & ST       &  1, 2                          \\
\object[2MASS J01204916-0741036]{0120491$-$074103 }   & 10076 &                               & 12.976 & 0.684 & 1.122 & M8   &  11.16$\pm$0.35 & 23.1$\pm$3.8 & ST       &  3                          \\
\object[2MASS J01311838+3801554]{0131183+380155   }   & 13061 &                               & 14.694 & 1.022 & 1.666 & L4   &  13.09$\pm$0.44 & 20.9$\pm$4.3 & ST       &  2                          \\
\object[2MASS J02402950+2832576]{0240295+283257   }   & 10138 &                               & 12.679 & 0.701 & 1.103 & M7.5 &  10.96$\pm$0.21 & 22.0$\pm$2.2 & ST       &  4                               \\
\object[2MASS J03144011-0450316]{0314401$-$045031 }   & 10172 &                               & 12.656 & 0.653 & 1.035 & M7.5 &  10.96$\pm$0.21 & 21.8$\pm$2.1 & ST       &  3                         \\
\object[2MASS J04285096-2253227]{0428509$-$225322 }   & 10286 &                               & 13.579 & 0.882 & 1.459 & L1   &  12.00$\pm$0.14 & 20.7$\pm$1.4 & ST       &  5                        \\
\object[2MASS J06145280+4536556]{0614528+453655   }   & 10529 &                               & 13.003 & 0.755 & 1.234 & M9   &  11.47$\pm$0.14 & 20.3$\pm$1.3 & ST       &  3                       \\
\object[2MASS J08105865+1420390]{0810586+142039   }   & 10699 &                               & 12.714 & 0.669 & 1.104 & M8   &  11.16$\pm$0.18 & 20.5$\pm$1.7 & ST       &  3                       \\
\enddata


\tablenotetext{a}{The 2MASS designation is 2MASSI
Jhhmmss[.]s$\pm$ddmmss.}

\tablenotetext{b}{A single or double colon indicates an
uncertainty on the spectral type of $\pm$1 or $\pm$2 types
respectively due to a low signal-to-noise spectrum. A single colon
is also used to indicate uncertainty on the spectral type of
individual objects in multiple systems where resolved spectroscopy
has not been obtained.}

\tablenotetext{c}{Distance are spectrophotometric (d source = ST)
except where a trigonometric parallax is available (d source =
Parallax).}

\tablerefs{(1)~\citet{Deacon05}; (2)~This Paper; (3)~\citetalias{Cruz03};
(4)~\citet{NN}; (5)~\citet{Kendall03}; (6)~\citet{Reid02_highres};
(7)~\citet{LHS};(8)~\citet{NN4}; (9)~\citet{Gizis02};
(10)~\citet{Delfosse97}; (11)~\citet{Dahn02};
(12)~\citet{Koerner99};(13)~\citet{Martin99_D1228};
(14)~\citet{Fan00}; (15)~\citet{Geballe02}; (16)~\citet{Vrba04};
(17)~\citet{K92_thesis}; (18)~\citet{Siegler05};
(19)~\citeauthor[in prep.]{DavyLs}; (20)~\citet{NLTT};
(21)~\citet{Close02}; (22)~\citet{Close03}.}

\end{deluxetable}

%% file: tab11.tex
\begin{deluxetable}{llc|llc}
\tablewidth{0pt}
\tablecaption{The 20-pc Ultracool Dwarf \textit{J}- and \textit{K$_S$}-band Luminosity Functions\label{tab:lf}}
\tablehead{ \colhead{} & \colhead{} & \colhead{$\rho_{tot}$} &
            \colhead{} & \colhead{} & \colhead{$\rho_{tot}$} \\
             \colhead{$M_J$} & \colhead{N$_{tot}$} & \colhead{(10$^{-3}$ dwarfs pc$^{-3}$)} &
             \colhead{$M_{K_S}$} & \colhead{N$_{tot}$} & \colhead{(10$^{-3}$ dwarfs pc$^{-3}$)}}
\startdata
10.75  &    28.7\tablenotemark{a} & \phm{$>$}2.38$\pm$0.64\tablenotemark{a}  & \phn9.75 & 30.7\tablenotemark{a}  & \phm{$>$}2.55$\pm$0.41\tablenotemark{a} \\
11.25  &    20 &                    \phm{$>$}1.66$\pm$0.37                   &    10.25 & 25                     & \phm{$>$}2.07$\pm$0.41 \\
11.75  &    14 &                    \phm{$>$}1.16$\pm$0.31                   &    10.75 & 17                     & \phm{$>$}1.41$\pm$0.34 \\
12.25  &    10 &                    \phm{$>$}0.83$\pm$0.26                   &    11.25 & \phn7                  & \phm{$>$}0.58$\pm$0.22 \\
12.75  & \phn6 &                    \phm{$>$}0.50$\pm$0.20                   &    11.75 & \phn9                  & \phm{$>$}0.75$\pm$0.25 \\
13.25  & \phn7 &                    \phm{$>$}0.58$\pm$0.22                   &    12.25 & \phn9                  & \phm{$>$}0.75$\pm$0.25 \\
13.75  & \phn6 &                    \phm{$>$}0.50$\pm$0.20                   &    12.75 & \phn6                  &       $>$0.50$\pm$0.20 \\
14.25  & \phn8 &                    $>$0.66$\pm$0.23                         &          &                        &  \\
\enddata

\tablecomments{Values for the space densities plotted in
Figure~\ref{fig:LF2}. The absolute magnitudes listed are the center of 0.5~magnitude wide bin.}

\tablenotetext{a}{Corrected value as discussed in
\S~\ref{sec:st_complete}.}


\end{deluxetable}

%% file: tab12.tex
\begin{deluxetable}{lllcc}
\tablewidth{0pt}
\tablecaption{The 8-pc \textit{J}-band Luminosity Function\label{tab:8pclf}}
\tablehead{
\colhead{} & \colhead{} & \colhead{} &
\colhead{$\rho_{sys}$} & \colhead{$\rho_{tot}$} \\

\colhead{$M_J$} & \colhead{N$_{sys}$} & \colhead{N$_{sec}$} & \colhead{($10^{-3}$ dwarfs pc$^{-3}$)} &
\colhead{($10^{-3}$ dwarfs pc$^{-3}$)} \\
\colhead{(1)} & \colhead {(2)} & \colhead {(3)} & \colhead{(4)} &
\colhead{(5)}
}
\startdata
    0.75&   1 &   0 &   0.62 &   0.62 \\
    1.25&   1 &   0 &   0.62 &   0.62 \\
    1.75&   3 &   0 &   1.87 &   1.87 \\
    2.25&   0 &   0 &  \nodata&  \nodata\\
    2.75&   0 &   0 &  \nodata&  \nodata\\
    3.25&   1 &   0 &   0.62 &   0.62 \\
    3.75&   2 &   0 &   1.24 &   1.24 \\
    4.25&   6 &   1 &   3.73 &   4.35 \\
    4.75&   6 &   0 &   3.73 &   3.73 \\
    5.25&   3 &   1 &   1.87 &   2.49 \\
    5.75&   3 &   5 &   1.87 &   4.97 \\
    6.25&   6 &   0 &   3.73 &   3.73 \\
    6.75&   5 &   0 &   3.11 &   3.11 \\
    7.25&  15 &   4 &   9.33 &  11.81 \\
    7.75&  10 &   4 &   6.22 &   8.70 \\
    8.25&  12 &   4 &   7.46 &   9.95 \\
    8.75&   6 &   2 &   3.73 &   4.97 \\
    9.25&  11 &   2 &   5.60 &   6.84 \\
    9.75&   6 &   5 &   3.11 &   6.22 \\
   10.25&   3 &   5 &   1.87 &   4.97 \\
   10.75&   3 &   2 &   1.87 &   3.11 \\
   11.25&   1 &   2 &   0.62 &   1.87 \\
\enddata
\tablecomments{Column 2 lists the number of systems (single stars
and primaries) and column 3 gives the number of secondary
components; column 4 lists the corresponding system space density,
and column 5 lists the total space density, including the
contribution from secondaries. The resulting \LFj\ (total space
density) is shown in Figure~\ref{fig:8pcLF}.}

\end{deluxetable}

%% file: ms.bbl
\begin{thebibliography}{140}
\expandafter\ifx\csname natexlab\endcsname\relax\def\natexlab#1{#1}\fi

\bibitem[{{Ackerman} \& {Marley}(2001)}]{Ackerman01}
{Ackerman}, A.~S. \& {Marley}, M.~S. 2001, \apj, 556, 872

\bibitem[{{Allen} {et~al.}(2005){Allen}, {Koerner}, {Reid}, \&
  {Trilling}}]{Allen05}
{Allen}, P.~R., {Koerner}, D.~W., {Reid}, I.~N., \& {Trilling}, D.~E. 2005,
  \apj, 625, 385

\bibitem[{{Basri} \& {Reiners}(2006)}]{Basri06}
{Basri}, G. \& {Reiners}, A. 2006, \aj, 132, 663

\bibitem[{{Bessell}(1991)}]{B91}
{Bessell}, M.~S. 1991, \aj, 101, 662

\bibitem[{{Borysow} {et~al.}(1997){Borysow}, {Jorgensen}, \&
  {Zheng}}]{Borysow97}
{Borysow}, A., {Jorgensen}, U.~G., \& {Zheng}, C. 1997, \aap, 324, 185

\bibitem[{{Bouy} {et~al.}(2003){Bouy}, {Brandner}, {Mart{\'{\i}}n}, {Delfosse},
  {Allard}, \& {Basri}}]{Bouy03}
{Bouy}, H., {Brandner}, W., {Mart{\'{\i}}n}, E.~L., {Delfosse}, X., {Allard},
  F., \& {Basri}, G. 2003, \aj, 126, 1526

\bibitem[{{Bouy} {et~al.}(2005){Bouy}, {Mart{\'{\i}}n}, {Brandner}, \&
  {Bouvier}}]{Bouy05}
{Bouy}, H., {Mart{\'{\i}}n}, E.~L., {Brandner}, W., \& {Bouvier}, J. 2005, \aj,
  129, 511

\bibitem[{{Burgasser}(2002)}]{Burgasser02_thesis}
{Burgasser}, A.~J. 2002, Ph.D.~Thesis

\bibitem[{{Burgasser}(2004{\natexlab{a}})}]{Burgasser04_subdwarf}
---. 2004{\natexlab{a}}, \apjl, 614, L73

\bibitem[{{Burgasser}(2004{\natexlab{b}})}]{Burgasser04}
---. 2004{\natexlab{b}}, \apjs, 155, 191

\bibitem[{{Burgasser} {et~al.}(2006{\natexlab{a}}){Burgasser}, {Cruz}, \&
  {Kirkpatrick}}]{Burgasser06_subdwarfs}
{Burgasser}, A.~J., {Cruz}, K.~L., \& {Kirkpatrick}, J.~D. 2006{\natexlab{a}},
  \apj, submitted

\bibitem[{{Burgasser} {et~al.}(2003{\natexlab{a}}){Burgasser}, {Kirkpatrick},
  {Burrows}, {Liebert}, {Reid}, {Gizis}, {McGovern}, {Prato}, \&
  {McLean}}]{Burgasser03_subdwarf}
{Burgasser}, A.~J., {Kirkpatrick}, J.~D., {Burrows}, A., {Liebert}, J., {Reid},
  I.~N., {Gizis}, J.~E., {McGovern}, M.~R., {Prato}, L., \& {McLean}, I.~S.
  2003{\natexlab{a}}, \apj, 592, 1186

\bibitem[{{Burgasser} {et~al.}(2006{\natexlab{b}}){Burgasser}, {Kirkpatrick},
  {Cruz}, {Reid}, {Leggett}, {Liebert}, {Burrows}, \&
  {Brown}}]{Burgasser06_binary}
{Burgasser}, A.~J., {Kirkpatrick}, J.~D., {Cruz}, K.~L., {Reid}, I.~N.,
  {Leggett}, S.~K., {Liebert}, J., {Burrows}, A., \& {Brown}, M.~E.
  2006{\natexlab{b}}, \apj, in press (astro-ph/0605577)

\bibitem[{{Burgasser} {et~al.}(2003{\natexlab{b}}){Burgasser}, {Kirkpatrick},
  {Liebert}, \& {Burrows}}]{Burgasser03_optical}
{Burgasser}, A.~J., {Kirkpatrick}, J.~D., {Liebert}, J., \& {Burrows}, A.
  2003{\natexlab{b}}, \apj, 594, 510

\bibitem[{{Burgasser} {et~al.}(2005{\natexlab{a}}){Burgasser}, {Kirkpatrick},
  \& {Lowrance}}]{Burgasser05_binary}
{Burgasser}, A.~J., {Kirkpatrick}, J.~D., \& {Lowrance}, P.~J.
  2005{\natexlab{a}}, \aj, 129, 2849


\bibitem[{{Burgasser} {et~al.}(2002){Burgasser}, {Marley}, {Ackerman},
  {Saumon}, {Lodders}, {Dahn}, {Harris}, \& {Kirkpatrick}}]{Burgasser02_clouds}
{Burgasser}, A.~J., {Marley}, M.~S., {Ackerman}, A.~S., {Saumon}, D.,
  {Lodders}, K., {Dahn}, C.~C., {Harris}, H.~C., \& {Kirkpatrick}, J.~D. 2002,
  \apjl, 571, L151

\bibitem[{{Burgasser} {et~al.}(2005{\natexlab{b}}){Burgasser}, {Reid},
  {Leggett}, {Kirkpatrick}, {Liebert}, \& {Burrows}}]{Burgasser05_0423}
{Burgasser}, A.~J., {Reid}, I.~N., {Leggett}, S.~K., {Kirkpatrick}, J.~D.,
  {Liebert}, J., \& {Burrows}, A. 2005{\natexlab{b}}, \apjl, 634, L177

\bibitem[{{Burgasser} {et~al.}(2006{\natexlab{c}}){Burgasser}, {Reid},
  {Siegler}, {Close}, {Allen}, {Lowrance}, \& {Gizis}}]{Burgasser06_PPV}
{Burgasser}, A.~J., {Reid}, I.~N., {Siegler}, N., {Close}, L., {Allen}, P.,
  {Lowrance}, P., \& {Gizis}, J. 2006{\natexlab{c}}, in Planets and Protostars
  V, ed. B. Reipurth, D. Jewitt, \& K. Keil (Tucson: University of Arizona
  Press)  (astro-ph/0602122)

\bibitem[{{Burrows} {et~al.}(2001){Burrows}, {Hubbard}, {Lunine}, \&
  {Liebert}}]{Burrows01}
{Burrows}, A., {Hubbard}, W.~B., {Lunine}, J.~I., \& {Liebert}, J. 2001,
  Reviews of Modern Physics, 73, 719

\bibitem[{{Chabrier}(2003)}]{Chabrier03}
{Chabrier}, G. 2003, \pasp, 115, 763

\bibitem[{{Chabrier} \& {Baraffe}(2000)}]{Chabrier00_Review}
{Chabrier}, G. \& {Baraffe}, I. 2000, \araa, 38, 337

\bibitem[{{Chabrier} {et~al.}(2000){Chabrier}, {Baraffe}, {Allard}, \&
  {Hauschildt}}]{Chabrier00}
{Chabrier}, G., {Baraffe}, I., {Allard}, F., \& {Hauschildt}, P. 2000, \apj,
  542, 464

\bibitem[{{Chiu} {et~al.}(2006){Chiu}, {Fan}, {Leggett}, {Golimowski}, {Zheng},
  {Geballe}, {Schneider}, \& {Brinkmann}}]{Chiu06}
{Chiu}, K., {Fan}, X., {Leggett}, S.~K., {Golimowski}, D.~A., {Zheng}, W.,
  {Geballe}, T.~R., {Schneider}, D.~P., \& {Brinkmann}, J. 2006, \aj, 131, 2722

\bibitem[{{Close} {et~al.}(2003){Close}, {Siegler}, {Freed}, \&
  {Biller}}]{Close03}
{Close}, L.~M., {Siegler}, N., {Freed}, M., \& {Biller}, B. 2003, \apj, 587,
  407

\bibitem[{{Close} {et~al.}(2002){Close}, {Siegler}, {Potter}, {Brandner}, \&
  {Liebert}}]{Close02}
{Close}, L.~M., {Siegler}, N., {Potter}, D., {Brandner}, W., \& {Liebert}, J.
  2002, \apjl, 567, L53

\bibitem[{{Costa} {et~al.}(2005){Costa}, {M{\'e}ndez}, {Jao}, {Henry},
  {Subasavage}, {Brown}, {Ianna}, \& {Bartlett}}]{Costa05}
{Costa}, E., {M{\'e}ndez}, R.~A., {Jao}, W.-C., {Henry}, T.~J., {Subasavage},
  J.~P., {Brown}, M.~A., {Ianna}, P.~A., \& {Bartlett}, J. 2005, \aj, 130, 337

\bibitem[{{Crifo} {et~al.}(2005){Crifo}, {Phan-Bao}, {Delfosse}, {Forveille},
  {Guibert}, {Mart{\'{\i}}n}, \& {Reyl{\'e}}}]{NN6}
{Crifo}, F., {Phan-Bao}, N., {Delfosse}, X., {Forveille}, T., {Guibert}, J.,
  {Mart{\'{\i}}n}, E.~L., \& {Reyl{\'e}}, C. 2005, \aap, 441, 653

\bibitem[{{Cruz} {et~al.}(2004){Cruz}, {Burgasser}, {Reid}, \&
  {Liebert}}]{Cruz04}
{Cruz}, K.~L., {Burgasser}, A.~J., {Reid}, I.~N., \& {Liebert}, J. 2004, \apjl,
  604, L61

\bibitem[{{Cruz} \& {Reid}(2002)}]{Cruz02}
{Cruz}, K.~L. \& {Reid}, I.~N. 2002, \aj, 123, 2828

\bibitem[{{Cruz} {et~al.}(2006){Cruz}, {Reid}, {Liebert}, {Allen},
  {Kirkpatrick}, {Burgasser}, \& {Solomon}}]{CruzNIR}
{Cruz}, K.~L., {Reid}, I.~N., {Liebert}, J., {Allen}, P.~R., {Kirkpatrick},
  J.~D., {Burgasser}, A.~J., \& {Solomon}, A.~R. 2006, in prep.

\bibitem[{{Cruz} {et~al.}(2003){Cruz}, {Reid}, {Liebert}, {Kirkpatrick}, \&
  {Lowrance}}]{Cruz03}
{Cruz}, K.~L., {Reid}, I.~N., {Liebert}, J., {Kirkpatrick}, J.~D., \&
  {Lowrance}, P.~J. 2003, \aj, 126, 2421

\bibitem[{{Cushing} \& {Vacca}(2006)}]{Cushing06}
{Cushing}, M.~C. \& {Vacca}, W.~D. 2006, \aj, 131, 1797

\bibitem[{{Dahn} {et~al.}(2002){Dahn}, {Harris}, {Vrba}, {Guetter}, {Canzian},
  {Henden}, {Levine}, {Luginbuhl}, {Monet}, {Monet}, {Pier}, {Stone}, {Walker},
  {Burgasser}, {Gizis}, {Kirkpatrick}, {Liebert}, \& {Reid}}]{Dahn02}
{Dahn}, C.~C., {Harris}, H.~C., {Vrba}, F.~J., {Guetter}, H.~H., {Canzian}, B.,
  {Henden}, A.~A., {Levine}, S.~E., {Luginbuhl}, C.~B., {Monet}, A.~K.~B.,
  {Monet}, D.~G., {Pier}, J.~R., {Stone}, R.~C., {Walker}, R.~L., {Burgasser},
  A.~J., {Gizis}, J.~E., {Kirkpatrick}, J.~D., {Liebert}, J., \& {Reid}, I.~N.
  2002, \aj, 124, 1170

\bibitem[{{Deacon} {et~al.}(2005){Deacon}, {Hambly}, \& {Cooke}}]{Deacon05}
{Deacon}, N.~R., {Hambly}, N.~C., \& {Cooke}, J.~A. 2005, \aap, 435, 363

\bibitem[{{Delfosse} {et~al.}(2000){Delfosse}, {Forveille}, {S{\'e}gransan},
  {Beuzit}, {Udry}, {Perrier}, \& {Mayor}}]{Delfosse00}
{Delfosse}, X., {Forveille}, T., {S{\'e}gransan}, D., {Beuzit}, J.-L., {Udry},
  S., {Perrier}, C., \& {Mayor}, M. 2000, \aap, 364, 217

\bibitem[{{Delfosse} {et~al.}(1997){Delfosse}, {Tinney}, {Forveille},
  {Epchtein}, {Bertin}, {Borsenberger}, {Copet}, {de Batz}, {Fouque},
  {Kimeswenger}, {Le Bertre}, {Lacombe}, {Rouan}, \& {Tiphene}}]{Delfosse97}
{Delfosse}, X., {Tinney}, C.~G., {Forveille}, T., {Epchtein}, N., {Bertin}, E.,
  {Borsenberger}, J., {Copet}, E., {de Batz}, B., {Fouque}, P., {Kimeswenger},
  S., {Le Bertre}, T., {Lacombe}, F., {Rouan}, D., \& {Tiphene}, D. 1997, \aap,
  327, L25

\bibitem[{{Downes} {et~al.}(1997){Downes}, {Webbink}, \& {Shara}}]{Downes97_CV}
{Downes}, R., {Webbink}, R.~F., \& {Shara}, M.~M. 1997, \pasp, 109, 345

\bibitem[{{Fan} {et~al.}(2000){Fan}, {Knapp}, {Strauss}, {Gunn}, {Lupton},
  {Ivezi{\' c}}, {Rockosi}, {Yanny}, {Kent}, {Schneider}, {Kirkpatrick},
  {Annis}, {Bastian}, {Berman}, {Brinkmann}, {Csabai}, {Federwitz}, {Fukugita},
  {Gurbani}, {Hennessy}, {Hindsley}, {Ichikawa}, {Lamb}, {Lindenmeyer},
  {Mantsch}, {McKay}, {Munn}, {Nash}, {Okamura}, {Pauls}, {Pier},
  {Rechenmacher}, {Rivetta}, {Sergey}, {Stoughton}, {Szalay}, {Szokoly},
  {Tucker}, {York}, \& {The SDSS Collaboration}}]{Fan00}
{Fan}, X., {Knapp}, G.~R., {Strauss}, M.~A., {Gunn}, J.~E., {Lupton}, R.~H.,
  {Ivezi{\' c}}, {\v Z}., {Rockosi}, C.~M., {Yanny}, B., {Kent}, S.,
  {Schneider}, D.~P., {Kirkpatrick}, J.~D., {Annis}, J., {Bastian}, S.,
  {Berman}, E., {Brinkmann}, J., {Csabai}, I., {Federwitz}, G.~R., {Fukugita},
  M., {Gurbani}, V.~K., {Hennessy}, G.~S., {Hindsley}, R.~B., {Ichikawa}, T.,
  {Lamb}, D.~Q., {Lindenmeyer}, C., {Mantsch}, P.~M., {McKay}, T.~A., {Munn},
  J.~A., {Nash}, T., {Okamura}, S., {Pauls}, A.~G., {Pier}, J.~R.,
  {Rechenmacher}, R., {Rivetta}, C.~H., {Sergey}, G., {Stoughton}, C.,
  {Szalay}, A.~S., {Szokoly}, G.~P., {Tucker}, D.~L., {York}, D.~G., \& {The
  SDSS Collaboration}. 2000, \aj, 119, 928

\bibitem[{{Forveille} {et~al.}(2005){Forveille}, {Beuzit}, {Delorme},
  {S{\'e}gransan}, {Delfosse}, {Chauvin}, {Fusco}, {Lagrange}, {Mayor},
  {Montagnier}, {Mouillet}, {Perrier}, {Udry}, {Charton}, {Gigan}, {Conan},
  {Kern}, \& {Michet}}]{Forveille05}
{Forveille}, T., {Beuzit}, J.-L., {Delorme}, P., {S{\'e}gransan}, D.,
  {Delfosse}, X., {Chauvin}, G., {Fusco}, T., {Lagrange}, A.-M., {Mayor}, M.,
  {Montagnier}, G., {Mouillet}, D., {Perrier}, C., {Udry}, S., {Charton}, J.,
  {Gigan}, P., {Conan}, J.-M., {Kern}, P., \& {Michet}, G. 2005, \aap, 435, L5

\bibitem[{{Freed} {et~al.}(2003){Freed}, {Close}, \& {Siegler}}]{LHS2397aB}
{Freed}, M., {Close}, L.~M., \& {Siegler}, N. 2003, \apj, 584, 453

\bibitem[{{Garcia}(1989)}]{Garcia89} {Garcia}, B. 1989, Bull. Inf.
Centre Donnees Stellaires, 36, 27

\bibitem[{{Geballe} {et~al.}(2002){Geballe}, {Knapp}, {Leggett}, {Fan},
  {Golimowski}, {Anderson}, {Brinkmann}, {Csabai}, {Gunn}, {Hawley},
  {Hennessy}, {Henry}, {Hill}, {Hindsley}, {Ivezi{\' c}}, {Lupton}, {McDaniel},
  {Munn}, {Narayanan}, {Peng}, {Pier}, {Rockosi}, {Schneider}, {Smith},
  {Strauss}, {Tsvetanov}, {Uomoto}, {York}, \& {Zheng}}]{Geballe02}
{Geballe}, T.~R., {Knapp}, G.~R., {Leggett}, S.~K., {Fan}, X., {Golimowski},
  D.~A., {Anderson}, S., {Brinkmann}, J., {Csabai}, I., {Gunn}, J.~E.,
  {Hawley}, S.~L., {Hennessy}, G., {Henry}, T.~J., {Hill}, G.~J., {Hindsley},
  R.~B., {Ivezi{\' c}}, {\v Z}., {Lupton}, R.~H., {McDaniel}, A., {Munn},
  J.~A., {Narayanan}, V.~K., {Peng}, E., {Pier}, J.~R., {Rockosi}, C.~M.,
  {Schneider}, D.~P., {Smith}, J.~A., {Strauss}, M.~A., {Tsvetanov}, Z.~I.,
  {Uomoto}, A., {York}, D.~G., \& {Zheng}, W. 2002, \apj, 564, 466

\bibitem[{{Gelino} {et~al.}(2006){Gelino}, {Kulkarni}, \&
  {Stephens}}]{Gelino06}
{Gelino}, C.~R., {Kulkarni}, S.~R., \& {Stephens}, D.~C. 2006, \pasp, 118, 611

\bibitem[{{Gelino} {et~al.}(2002){Gelino}, {Marley}, {Holtzman}, {Ackerman}, \&
  {Lodders}}]{Gelino02}
{Gelino}, C.~R., {Marley}, M.~S., {Holtzman}, J.~A., {Ackerman}, A.~S., \&
  {Lodders}, K. 2002, \apj, 577, 433

\bibitem[{{Gizis}(2002)}]{Gizis02}
{Gizis}, J.~E. 2002, \apj, 575, 484

\bibitem[{{Gizis} {et~al.}(2000){Gizis}, {Monet}, {Reid}, {Kirkpatrick},
  {Liebert}, \& {Williams}}]{NN}
{Gizis}, J.~E., {Monet}, D.~G., {Reid}, I.~N., {Kirkpatrick}, J.~D., {Liebert},
  J., \& {Williams}, R.~J. 2000, \aj, 120, 1085

\bibitem[{{Gizis} \& {Reid}(1997)}]{GR97}
{Gizis}, J.~E. \& {Reid}, I.~N. 1997, \pasp, 109, 849

\bibitem[{{Gizis} {et~al.}(2003){Gizis}, {Reid}, {Knapp}, {Liebert},
  {Kirkpatrick}, {Koerner}, \& {Burgasser}}]{Gizis03}
{Gizis}, J.~E., {Reid}, I.~N., {Knapp}, G.~R., {Liebert}, J., {Kirkpatrick},
  J.~D., {Koerner}, D.~W., \& {Burgasser}, A.~J. 2003, \aj, 125, 3302

\bibitem[{{Gizis} {et~al.}(1999){Gizis}, {Reid}, \& {Monet}}]{Gizis99}
{Gizis}, J.~E., {Reid}, I.~N., \& {Monet}, D.~G. 1999, \aj, 118, 997

\bibitem[{{Golimowski} {et~al.}(2004){Golimowski}, {Leggett}, {Marley}, {Fan},
  {Geballe}, {Knapp}, {Vrba}, {Henden}, {Luginbuhl}, {Guetter}, {Munn},
  {Canzian}, {Zheng}, {Tsvetanov}, {Chiu}, {Glazebrook}, {Hoversten},
  {Schneider}, \& {Brinkmann}}]{Golimowski04}
{Golimowski}, D.~A., {Leggett}, S.~K., {Marley}, M.~S., {Fan}, X., {Geballe},
  T.~R., {Knapp}, G.~R., {Vrba}, F.~J., {Henden}, A.~A., {Luginbuhl}, C.~B.,
  {Guetter}, H.~H., {Munn}, J.~A., {Canzian}, B., {Zheng}, W., {Tsvetanov},
  Z.~I., {Chiu}, K., {Glazebrook}, K., {Hoversten}, E.~A., {Schneider}, D.~P.,
  \& {Brinkmann}, J. 2004, \aj, 127, 3516

\bibitem[{{Gorlova} {et~al.}(2003){Gorlova}, {Meyer}, {Rieke}, \&
  {Liebert}}]{Gorlova03}
{Gorlova}, N.~I., {Meyer}, M.~R., {Rieke}, G.~H., \& {Liebert}, J. 2003, \apj,
  593, 1074

\bibitem[{{Hamuy} {et~al.}(1994){Hamuy}, {Suntzeff}, {Heathcote}, {Walker},
  {Gigoux}, \& {Phillips}}]{Hamuy}
{Hamuy}, M., {Suntzeff}, N.~B., {Heathcote}, S.~R., {Walker}, A.~R., {Gigoux},
  P., \& {Phillips}, M.~M. 1994, \pasp, 106, 566

\bibitem[{{Hartwick} {et~al.}(1984){Hartwick}, {Cowley}, \&
  {Mould}}]{Hartwick84}
{Hartwick}, F.~D.~A., {Cowley}, A.~P., \& {Mould}, J.~R. 1984, \apj, 286, 269

\bibitem[{{Hawkins} \& {Bessell}(1988)}]{LHS2065}
{Hawkins}, M.~R.~S. \& {Bessell}, M.~S. 1988, \mnras, 234, 177

\bibitem[{{Hawley} {et~al.}(2002){Hawley}, {Covey}, {Knapp}, {Golimowski},
  {Fan}, {Anderson}, {Gunn}, {Harris}, {Ivezi{\' c}}, {Long}, {Lupton},
  {McGehee}, {Narayanan}, {Peng}, {Schlegel}, {Schneider}, {Spahn}, {Strauss},
  {Szkody}, {Tsvetanov}, {Walkowicz}, {Brinkmann}, {Harvanek}, {Hennessy},
  {Kleinman}, {Krzesinski}, {Long}, {Neilsen}, {Newman}, {Nitta}, {Snedden}, \&
  {York}}]{Hawley02}
{Hawley}, S.~L., {Covey}, K.~R., {Knapp}, G.~R., {Golimowski}, D.~A., {Fan},
  X., {Anderson}, S.~F., {Gunn}, J.~E., {Harris}, H.~C., {Ivezi{\' c}}, {\v
  Z}., {Long}, G.~M., {Lupton}, R.~H., {McGehee}, P.~M., {Narayanan}, V.,
  {Peng}, E., {Schlegel}, D., {Schneider}, D.~P., {Spahn}, E.~Y., {Strauss},
  M.~A., {Szkody}, P., {Tsvetanov}, Z., {Walkowicz}, L.~M., {Brinkmann}, J.,
  {Harvanek}, M., {Hennessy}, G.~S., {Kleinman}, S.~J., {Krzesinski}, J.,
  {Long}, D., {Neilsen}, E.~H., {Newman}, P.~R., {Nitta}, A., {Snedden}, S.~A.,
  \& {York}, D.~G. 2002, \aj, 123, 3409

\bibitem[{{Henry} {et~al.}(1999){Henry}, {Franz}, {Wasserman}, {Benedict},
  {Shelus}, {Ianna}, {Kirkpatrick}, \& {McCarthy}}]{Henry99}
{Henry}, T.~J., {Franz}, O.~G., {Wasserman}, L.~H., {Benedict}, G.~F.,
  {Shelus}, P.~J., {Ianna}, P.~A., {Kirkpatrick}, J.~D., \& {McCarthy}, D.~W.
  1999, \apj, 512, 864

\bibitem[{{H{\o}g} {et~al.}(2000){H{\o}g}, {Fabricius}, {Makarov}, {Urban},
  {Corbin}, {Wycoff}, {Bastian}, {Schwekendiek}, \& {Wicenec}}]{Tycho2}
{H{\o}g}, E., {Fabricius}, C., {Makarov}, V.~V., {Urban}, S., {Corbin}, T.,
  {Wycoff}, G., {Bastian}, U., {Schwekendiek}, P., \& {Wicenec}, A. 2000, \aap,
  355, L27

\bibitem[{{Hook} {et~al.}(2004){Hook}, {J{\o}rgensen}, {Allington-Smith},
  {Davies}, {Metcalfe}, {Murowinski}, \& {Crampton}}]{GMOS}
{Hook}, I.~M., {J{\o}rgensen}, I., {Allington-Smith}, J.~R., {Davies}, R.~L.,
  {Metcalfe}, N., {Murowinski}, R.~G., \& {Crampton}, D. 2004, \pasp, 116, 425

\bibitem[{{Irwin} {et~al.}(1991){Irwin}, {McMahon}, \& {Reid}}]{BRI0021}
{Irwin}, M., {McMahon}, R.~G., \& {Reid}, N. 1991, \mnras, 252, 61P

\bibitem[{{Kendall} {et~al.}(2003){Kendall}, {Mauron}, {Azzopardi}, \&
  {Gigoyan}}]{Kendall03}
{Kendall}, T.~R., {Mauron}, N., {Azzopardi}, M., \& {Gigoyan}, K. 2003, \aap,
  403, 929

\bibitem[{{Kirkpatrick}(1992)}]{K92_thesis}
{Kirkpatrick}, J.~D. 1992, Ph.D.~Thesis

\bibitem[{{Kirkpatrick} {et~al.}(2006{\natexlab{a}}){Kirkpatrick}, {Barman},
  {Burgasser}, {McGovern}, {McLean}, {Tinney}, \& {Lowrance}}]{Kirkpatrick06}
{Kirkpatrick}, J.~D., {Barman}, T.~S., {Burgasser}, A.~J., {McGovern}, M.~R.,
  {McLean}, I.~S., {Tinney}, C.~G., \& {Lowrance}, P.~J. 2006{\natexlab{a}},
  \apj, 639, 1120

\bibitem[{{Kirkpatrick} {et~al.}(2001{\natexlab{a}}){Kirkpatrick}, {Dahn},
  {Monet}, {Reid}, {Gizis}, {Liebert}, \& {Burgasser}}]{K01_gstar}
{Kirkpatrick}, J.~D., {Dahn}, C.~C., {Monet}, D.~G., {Reid}, I.~N., {Gizis},
  J.~E., {Liebert}, J., \& {Burgasser}, A.~J. 2001{\natexlab{a}}, \aj, 121,
  3235

\bibitem[{{Kirkpatrick} {et~al.}(1997){Kirkpatrick}, {Henry}, \&
  {Irwin}}]{KHI97}
{Kirkpatrick}, J.~D., {Henry}, T.~J., \& {Irwin}, M.~J. 1997, \aj, 113, 1421

\bibitem[{{Kirkpatrick} {et~al.}(1991){Kirkpatrick}, {Henry}, \&
  {McCarthy}}]{KHM91}
{Kirkpatrick}, J.~D., {Henry}, T.~J., \& {McCarthy}, D.~W. 1991, \apjs, 77, 417

\bibitem[{{Kirkpatrick} {et~al.}(1995){Kirkpatrick}, {Henry}, \&
  {Simons}}]{K95}
{Kirkpatrick}, J.~D., {Henry}, T.~J., \& {Simons}, D.~A. 1995, \aj, 109, 797

\bibitem[{{Kirkpatrick} {et~al.}(2006{\natexlab{b}}){Kirkpatrick},
  {Kirkpatrick}, {Kirkpatrick}, \& {Kirkpatrick}}]{Davy0518}
{Kirkpatrick}, J.~D., {Kirkpatrick}, J.~D., {Kirkpatrick}, J.~D., \&
  {Kirkpatrick}, J.~D. 2006{\natexlab{b}}, in prep.

\bibitem[{{Kirkpatrick} {et~al.}(2006{\natexlab{c}}){Kirkpatrick},
  {Kirkpatrick}, {Kirkpatrick}, \& {Kirkpatrick}}]{DavyLs}
---. 2006{\natexlab{c}}, in prep.

\bibitem[{{Kirkpatrick} {et~al.}(2006{\natexlab{d}}){Kirkpatrick},
  {Kirkpatrick}, {Kirkpatrick}, \& {Kirkpatrick}}]{DavyTs}
---. 2006{\natexlab{d}}, in prep.

\bibitem[{{Kirkpatrick} {et~al.}(2001{\natexlab{b}}){Kirkpatrick}, {Liebert},
  {Cruz}, {Gizis}, \& {Reid}}]{K01}
{Kirkpatrick}, J.~D., {Liebert}, J., {Cruz}, K.~L., {Gizis}, J.~E., \& {Reid},
  I.~N. 2001{\natexlab{b}}, \pasp, 113, 814

\bibitem[{{Kirkpatrick} {et~al.}(1999){Kirkpatrick}, {Reid}, {Liebert},
  {Cutri}, {Nelson}, {Beichman}, {Dahn}, {Monet}, {Gizis}, \&
  {Skrutskie}}]{K99}
{Kirkpatrick}, J.~D., {Reid}, I.~N., {Liebert}, J., {Cutri}, R.~M., {Nelson},
  B., {Beichman}, C.~A., {Dahn}, C.~C., {Monet}, D.~G., {Gizis}, J.~E., \&
  {Skrutskie}, M.~F. 1999, \apj, 519, 802

\bibitem[{{Kirkpatrick} {et~al.}(2000){Kirkpatrick}, {Reid}, {Liebert},
  {Gizis}, {Burgasser}, {Monet}, {Dahn}, {Nelson}, \& {Williams}}]{K00}
{Kirkpatrick}, J.~D., {Reid}, I.~N., {Liebert}, J., {Gizis}, J.~E.,
  {Burgasser}, A.~J., {Monet}, D.~G., {Dahn}, C.~C., {Nelson}, B., \&
  {Williams}, R.~J. 2000, \aj, 120, 447

\bibitem[{{Knapp} {et~al.}(2004){Knapp}, {Leggett}, {Fan}, {Marley}, {Geballe},
  {Golimowski}, {Finkbeiner}, {Gunn}, {Hennawi}, {Ivezi{\'c}}, {Lupton},
  {Schlegel}, {Strauss}, {Tsvetanov}, {Chiu}, {Hoversten}, {Glazebrook},
  {Zheng}, {Hendrickson}, {Williams}, {Uomoto}, {Vrba}, {Henden}, {Luginbuhl},
  {Guetter}, {Munn}, {Canzian}, {Schneider}, \& {Brinkmann}}]{Knapp04}
{Knapp}, G.~R., {Leggett}, S.~K., {Fan}, X., {Marley}, M.~S., {Geballe}, T.~R.,
  {Golimowski}, D.~A., {Finkbeiner}, D., {Gunn}, J.~E., {Hennawi}, J.,
  {Ivezi{\'c}}, Z., {Lupton}, R.~H., {Schlegel}, D.~J., {Strauss}, M.~A.,
  {Tsvetanov}, Z.~I., {Chiu}, K., {Hoversten}, E.~A., {Glazebrook}, K.,
  {Zheng}, W., {Hendrickson}, M., {Williams}, C.~C., {Uomoto}, A., {Vrba},
  F.~J., {Henden}, A.~A., {Luginbuhl}, C.~B., {Guetter}, H.~H., {Munn}, J.~A.,
  {Canzian}, B., {Schneider}, D.~P., \& {Brinkmann}, J. 2004, \aj, 127, 3553

\bibitem[{{Koerner} {et~al.}(1999){Koerner}, {Kirkpatrick}, {McElwain}, \&
  {Bonaventura}}]{Koerner99}
{Koerner}, D.~W., {Kirkpatrick}, J.~D., {McElwain}, M.~W., \& {Bonaventura},
  N.~R. 1999, \apjl, 526, L25

\bibitem[{{L{\' e}pine} {et~al.}(2003{\natexlab{a}}){L{\' e}pine}, {Rich}, \&
  {Shara}}]{Lepine03_sdL}
{L{\' e}pine}, S., {Rich}, R.~M., \& {Shara}, M.~M. 2003{\natexlab{a}}, \apjl,
  591, L49

\bibitem[{{L{\' e}pine} {et~al.}(2003{\natexlab{b}}){L{\' e}pine}, {Rich}, \&
  {Shara}}]{Lepine03}
---. 2003{\natexlab{b}}, \aj, 125, 1598

\bibitem[{{L{\' e}pine} \& {Shara}(2005)}]{LSPM_North}
{L{\' e}pine}, S. \& {Shara}, M.~M. 2005, \aj, 129, 1483

\bibitem[{{Leggett} {et~al.}(2000){Leggett}, {Geballe}, {Fan}, {Schneider},
  {Gunn}, {Lupton}, {Knapp}, {Strauss}, {McDaniel}, {Golimowski}, {Henry},
  {Peng}, {Tsvetanov}, {Uomoto}, {Zheng}, {Hill}, {Ramsey}, {Anderson},
  {Annis}, {Bahcall}, {Brinkmann}, {Chen}, {Csabai}, {Fukugita}, {Hennessy},
  {Hindsley}, {Ivezi{\'c}}, {Lamb}, {Munn}, {Pier}, {Schlegel}, {Smith},
  {Stoughton}, {Thakar}, \& {York}}]{Leggett00}
{Leggett}, S.~K., {Geballe}, T.~R., {Fan}, X., {Schneider}, D.~P., {Gunn},
  J.~E., {Lupton}, R.~H., {Knapp}, G.~R., {Strauss}, M.~A., {McDaniel}, A.,
  {Golimowski}, D.~A., {Henry}, T.~J., {Peng}, E., {Tsvetanov}, Z.~I.,
  {Uomoto}, A., {Zheng}, W., {Hill}, G.~J., {Ramsey}, L.~W., {Anderson}, S.~F.,
  {Annis}, J.~A., {Bahcall}, N.~A., {Brinkmann}, J., {Chen}, B., {Csabai}, I.,
  {Fukugita}, M., {Hennessy}, G.~S., {Hindsley}, R.~B., {Ivezi{\'c}}, {\v Z}.,
  {Lamb}, D.~Q., {Munn}, J.~A., {Pier}, J.~R., {Schlegel}, D.~J., {Smith},
  J.~A., {Stoughton}, C., {Thakar}, A.~R., \& {York}, D.~G. 2000, \apjl, 536,
  L35

\bibitem[{{Leinert} {et~al.}(2000){Leinert}, {Allard}, {Richichi}, \&
  {Hauschildt}}]{Leinert00}
{Leinert}, C., {Allard}, F., {Richichi}, A., \& {Hauschildt}, P.~H. 2000, \aap,
  353, 691

\bibitem[{{Leinert} {et~al.}(1994){Leinert}, {Weitzel}, {Richichi}, {Eckart},
  \& {Tacconi-Garman}}]{Leinert94}
{Leinert}, C., {Weitzel}, N., {Richichi}, A., {Eckart}, A., \&
  {Tacconi-Garman}, L.~E. 1994, \aap, 291, L47

\bibitem[{{L{\'e}pine}(2005)}]{Lepine05}
{L{\'e}pine}, S. 2005, \aj, 130, 1680

\bibitem[{{Liebert} {et~al.}(1984){Liebert}, {Boroson}, \&
  {Giampapa}}]{LHS2924spec}
{Liebert}, J., {Boroson}, T.~A., \& {Giampapa}, M.~S. 1984, \apj, 282, 758

\bibitem[{{Liebert} {et~al.}(2003){Liebert}, {Kirkpatrick}, {Cruz}, {Reid},
  {Burgasser}, {Tinney}, \& {Gizis}}]{Liebert03}
{Liebert}, J., {Kirkpatrick}, J.~D., {Cruz}, K.~L., {Reid}, I.~N., {Burgasser},
  A., {Tinney}, C.~G., \& {Gizis}, J.~E. 2003, \aj, 125, 343

\bibitem[{{Linsky}(1969)}]{Linsky69}
{Linsky}, J.~L. 1969, \apj, 156, 989

\bibitem[{{Liu} {et~al.}(2006){Liu}, {Leggett}, {Golimowski},
    {Chiu}, {Fan},  {Geballe}, {Schneider}, \&
    {Brinkmann}}]{Liu06}
{Liu}, M.~C. and {Leggett}, S.~K. and {Golimowski}, D.~A. and
    {Chiu}, K. and {Fan}, X. and {Geballe}, T.~R. and {Schneider}, D.~P. and
    {Brinkmann}, J. 2006, \apj, 647, 1393

\bibitem[{{Liu} \& {Leggett}(2005)}]{Liu05}
{Liu}, M.~C. \& {Leggett}, S.~K. 2005, \apj, 634, 616

\bibitem[{{Liu} {et~al.}(1999){Liu}, {Hu}, {Li}, \& {Cao}}]{Liu99_CV}
{Liu}, W., {Hu}, J.~Y., {Li}, Z.~Y., \& {Cao}, L. 1999, \apjs, 122, 257

\bibitem[{{Lodieu} {et~al.}(2002){Lodieu}, {Scholz}, \&
  {McCaughrean}}]{Lodieu02}
{Lodieu}, N., {Scholz}, R.-D., \& {McCaughrean}, M.~J. 2002, \aap, 389, L20

\bibitem[{{Luyten}(1979{\natexlab{a}})}]{LHS}
{Luyten}, W.~J. 1979{\natexlab{a}}, {LHS catalogue. A catalogue of stars with
  proper motions exceeding 0"5 annually} (Minneapolis: University of Minnesota,
  1979, 2nd ed.)

\bibitem[{{Luyten}(1979{\natexlab{b}})}]{NLTT}
---. 1979{\natexlab{b}}, {New Luyten Catalog of stars with proper motions
  larger than two tenths of an arcsecond} (Minneapolis: University of
  Minnesota, 1979, 2nd ed.)

\bibitem[{{Maiti} {et~al.}(2005){Maiti}, {Sengupta}, {Parihar}, \&
  {Anupama}}]{Maiti05}
{Maiti}, M., {Sengupta}, S., {Parihar}, P.~S., \& {Anupama}, G.~C. 2005, \apjl,
  619, L183

\bibitem[{{Malmquist}(1920)}]{Malmquist}
{Malmquist}, K.~G. 1920, Lund Medd. Astron. Obs. Ser. II, 22

\bibitem[{{Marley} {et~al.}(2002){Marley}, {Seager}, {Saumon}, {Lodders},
  {Ackerman}, {Freedman}, \& {Fan}}]{Marley02}
{Marley}, M.~S., {Seager}, S., {Saumon}, D., {Lodders}, K., {Ackerman}, A.~S.,
  {Freedman}, R.~S., \& {Fan}, X. 2002, \apj, 568, 335

\bibitem[{{Martin} {et~al.}(1999){Martin}, {Brandner}, \&
  {Basri}}]{Martin99_D1228}
{Martin}, E.~L., {Brandner}, W., \& {Basri}, G. 1999, Science, 283, 1718

\bibitem[{{Mart{\'{\i}}n} {et~al.}(1999){Mart{\'{\i}}n}, {Delfosse}, {Basri},
  {Goldman}, {Forveille}, \& {Zapatero Osorio}}]{M99}
{Mart{\'{\i}}n}, E.~L., {Delfosse}, X., {Basri}, G., {Goldman}, B.,
  {Forveille}, T., \& {Zapatero Osorio}, M.~R. 1999, \aj, 118, 2466

\bibitem[{{Mart{\'{\i}}n} {et~al.}(1996){Mart{\'{\i}}n}, {Rebolo}, \&
  {Zapatero-Osorio}}]{Martin96}
{Mart{\'{\i}}n}, E.~L., {Rebolo}, R., \& {Zapatero-Osorio}, M.~R. 1996, \apj,
  469, 706

\bibitem[{{Massey} \& {Gronwall}(1990)}]{Massey90}
{Massey}, P. \& {Gronwall}, C. 1990, \apj, 358, 344

\bibitem[{{Massey} {et~al.}(1988){Massey}, {Strobel}, {Barnes}, \&
  {Anderson}}]{Massey88}
{Massey}, P., {Strobel}, K., {Barnes}, J.~V., \& {Anderson}, E. 1988, \apj,
  328, 315

\bibitem[{{McGovern} {et~al.}(2004){McGovern}, {Kirkpatrick}, {McLean},
  {Burgasser}, {Prato}, \& {Lowrance}}]{McGovern04}
{McGovern}, M.~R., {Kirkpatrick}, J.~D., {McLean}, I.~S., {Burgasser}, A.~J.,
  {Prato}, L., \& {Lowrance}, P.~J. 2004, \apj, 600, 1020

\bibitem[{{McLean} {et~al.}(2003){McLean}, {McGovern}, {Burgasser},
  {Kirkpatrick}, {Prato}, \& {Kim}}]{Mclean03}
{McLean}, I.~S., {McGovern}, M.~R., {Burgasser}, A.~J., {Kirkpatrick}, J.~D.,
  {Prato}, L., \& {Kim}, S.~S. 2003, \apj, 596, 561

\bibitem[{{Monet} {et~al.}(1992){Monet}, {Dahn}, {Vrba}, {Harris}, {Pier},
  {Luginbuhl}, \& {Ables}}]{Monet92}
{Monet}, D.~G., {Dahn}, C.~C., {Vrba}, F.~J., {Harris}, H.~C., {Pier}, J.~R.,
  {Luginbuhl}, C.~B., \& {Ables}, H.~D. 1992, \aj, 103, 638

\bibitem[{{Morrison} {et~al.}(2001){Morrison}, {R{\"o}ser}, {McLean},
  {Bucciarelli}, \& {Lasker}}]{GSC}
{Morrison}, J.~E., {R{\"o}ser}, S., {McLean}, B., {Bucciarelli}, B., \&
  {Lasker}, B. 2001, \aj, 121, 1752

\bibitem[{{Nordstr{\"o}m} {et~al.}(2004){Nordstr{\"o}m}, {Mayor}, {Andersen},
  {Holmberg}, {Pont}, {J{\o}rgensen}, {Olsen}, {Udry}, \& {Mowlavi}}]{Geneva}
{Nordstr{\"o}m}, B., {Mayor}, M., {Andersen}, J., {Holmberg}, J., {Pont}, F.,
  {J{\o}rgensen}, B.~R., {Olsen}, E.~H., {Udry}, S., \& {Mowlavi}, N. 2004,
  \aap, 418, 989

\bibitem[{{Oke} \& {Gunn}(1983)}]{OG83}
{Oke}, J.~B. \& {Gunn}, J.~E. 1983, \apj, 266, 713

\bibitem[{{Perryman} \& {ESA}(1997)}]{Hipparcos}
{Perryman}, M.~A.~C. \& {ESA}. 1997, {The HIPPARCOS and TYCHO catalogues.
  Astrometric and photometric star catalogues derived from the ESA HIPPARCOS
  Space Astrometry Mission} (The Hipparcos and Tycho catalogues.~Astrometric
  and photometric star catalogues derived from the ESA Hipparcos Space
  Astrometry Mission, Publisher: Noordwijk, Netherlands: ESA Publications
  Division, 1997, Series: ESA SP Series vol no: 1200, ISBN: 9290923997 (set))

\bibitem[{{Phan-Bao} {et~al.}(2003){Phan-Bao}, {Crifo}, {Delfosse},
  {Forveille}, {Guibert}, {Borsenberger}, {Epchtein}, {Fouqu{\' e}}, {Simon},
  \& {Vetois}}]{NN5}
{Phan-Bao}, N., {Crifo}, F., {Delfosse}, X., {Forveille}, T., {Guibert}, J.,
  {Borsenberger}, J., {Epchtein}, N., {Fouqu{\' e}}, P., {Simon}, G., \&
  {Vetois}, J. 2003, \aap, 401, 959

\bibitem[{{Phan-Bao} {et~al.}(2001){Phan-Bao}, {Guibert}, {Crifo}, {Delfosse},
  {Forveille}, {Borsenberger}, {Epchtein}, {Fouqu{\'e}}, \& {Simon}}]{NN4}
{Phan-Bao}, N., {Guibert}, J., {Crifo}, F., {Delfosse}, X., {Forveille}, T.,
  {Borsenberger}, J., {Epchtein}, N., {Fouqu{\'e}}, P., \& {Simon}, G. 2001,
  \aap, 380, 590

\bibitem[{{Pokorny} {et~al.}(2004){Pokorny}, {Jones}, {Hambly}, \&
  {Pinfield}}]{LEHPM}
{Pokorny}, R.~S., {Jones}, H.~R.~A., {Hambly}, N.~C., \& {Pinfield}, D.~J.
  2004, \aap, 421, 763

\bibitem[{{Ratzka} {et~al.}(2006){Ratzka},
  {Leinert}, \& {Allard}}]{Ratzka06}{Ratzka}, T. and {Leinert}, C. and {Allard},
  F., in prep.

\bibitem[{{Reid}(2003)}]{Paper6}
{Reid}, I.~N. 2003, \aj, 126, 2449

\bibitem[{{Reid} \& {Cruz}(2002)}]{Paper1}
{Reid}, I.~N. \& {Cruz}, K.~L. 2002, \aj, 123, 2806

\bibitem[{{Reid} {et~al.}(2003{\natexlab{a}}){Reid}, {Cruz}, {Allen},
  {Mungall}, {Kilkenny}, {Liebert}, {Hawley}, {Fraser}, {Covey}, \&
  {Lowrance}}]{Paper7}
{Reid}, I.~N., {Cruz}, K.~L., {Allen}, P., {Mungall}, F., {Kilkenny}, D.,
  {Liebert}, J., {Hawley}, S.~L., {Fraser}, O.~J., {Covey}, K.~R., \&
  {Lowrance}, P. 2003{\natexlab{a}}, \aj, 126, 3007

\bibitem[{{Reid} {et~al.}(2004){Reid}, {Cruz}, {Allen}, {Mungall}, {Kilkenny},
  {Liebert}, {Hawley}, {Fraser}, {Covey}, {Lowrance}, {Kirkpatrick}, \&
  {Burgasser}}]{Paper8}
{Reid}, I.~N., {Cruz}, K.~L., {Allen}, P., {Mungall}, F., {Kilkenny}, D.,
  {Liebert}, J., {Hawley}, S.~L., {Fraser}, O.~J., {Covey}, K.~R., {Lowrance},
  P., {Kirkpatrick}, J.~D., \& {Burgasser}, A.~J. 2004, \aj, 128, 463

\bibitem[{{Reid} {et~al.}(2003{\natexlab{b}}){Reid}, {Cruz}, {Laurie},
  {Liebert}, {Dahn}, {Harris}, {Guetter}, {Stone}, {Canzian}, {Luginbuhl},
  {Levine}, {Monet}, \& {Monet}}]{Paper4}
{Reid}, I.~N., {Cruz}, K.~L., {Laurie}, S.~P., {Liebert}, J., {Dahn}, C.~C.,
  {Harris}, H.~C., {Guetter}, H.~H., {Stone}, R.~C., {Canzian}, B.,
  {Luginbuhl}, C.~B., {Levine}, S.~E., {Monet}, A.~K.~B., \& {Monet}, D.~G.
  2003{\natexlab{b}}, \aj, 125, 354

\bibitem[{{Reid} {et~al.}(2006{\natexlab{a}}){Reid},
  {Cruz}, {Liebert}, \& {Lowrance}}]{Paper10}
---. 2006{\natexlab{a}}, in prep.

\bibitem[{{Reid} \& {Gizis}(2005)}]{RG05}
{Reid}, I.~N. \& {Gizis}, J.~E. 2005, \pasp, 117, 676

\bibitem[{{Reid} {et~al.}(2002{\natexlab{a}}){Reid}, {Gizis}, \&
  {Hawley}}]{PMSU4}
{Reid}, I.~N., {Gizis}, J.~E., \& {Hawley}, S.~L. 2002{\natexlab{a}}, \aj, 124,
  2721

\bibitem[{{Reid} {et~al.}(2001){Reid}, {Gizis}, {Kirkpatrick}, \&
  {Koerner}}]{Reid01_binary}
{Reid}, I.~N., {Gizis}, J.~E., {Kirkpatrick}, J.~D., \& {Koerner}, D.~W. 2001,
  \aj, 121, 489

\bibitem[{{Reid} {et~al.}(1995){Reid}, {Hawley}, \& {Gizis}}]{PMSU}
{Reid}, I.~N., {Hawley}, S.~L., \& {Gizis}, J.~E. 1995, \aj, 110, 1838

\bibitem[{{Reid} {et~al.}(2002{\natexlab{b}}){Reid}, {Kilkenny}, \&
  {Cruz}}]{Paper2}
{Reid}, I.~N., {Kilkenny}, D., \& {Cruz}, K.~L. 2002{\natexlab{b}}, \aj, 123,
  2822

\bibitem[{{Reid} {et~al.}(1999){Reid}, {Kirkpatrick}, {Liebert}, {Burrows},
  {Gizis}, {Burgasser}, {Dahn}, {Monet}, {Cutri}, {Beichman}, \&
  {Skrutskie}}]{Reid99}
{Reid}, I.~N., {Kirkpatrick}, J.~D., {Liebert}, J., {Burrows}, A., {Gizis},
  J.~E., {Burgasser}, A., {Dahn}, C.~C., {Monet}, D., {Cutri}, R., {Beichman},
  C.~A., \& {Skrutskie}, M. 1999, \apj, 521, 613

\bibitem[{{Reid} {et~al.}(2002{\natexlab{c}}){Reid}, {Kirkpatrick}, {Liebert},
  {Gizis}, {Dahn}, \& {Monet}}]{Reid02_highres}
{Reid}, I.~N., {Kirkpatrick}, J.~D., {Liebert}, J., {Gizis}, J.~E., {Dahn},
  C.~C., \& {Monet}, D.~G. 2002{\natexlab{c}}, \aj, 124, 519

\bibitem[{{Reid} {et~al.}(2006{\natexlab{b}}){Reid}, {Lewitus},
{Allen},
  {Cruz}, \& {Burgasser}}]{Reid06_binary}
{Reid}, I.~N., {Lewitus}, E., {Allen}, P.~R., {Cruz}, K.~L., \& {Burgasser},
  A.~J. 2006{\natexlab{b}}, \aj, 132, 891

\bibitem[{{Reid} {et~al.}(2006{\natexlab{c}}){Reid}, {Lewitus},
{Burgasser}, \&
  {Cruz}}]{Reid06_2252}
{Reid}, I.~N., {Lewitus}, E., {Burgasser}, A.~J., \& {Cruz}, K.~L.
  2006{\natexlab{c}}, \apj, 639, 1114

\bibitem[{{Reiners} \& {Basri}(2006)}]{Reiners06}
{Reiners}, A. \& {Basri}, G. 2006, \aj, 131, 1806

\bibitem[{{Reyl{\'e}} \& {Robin}(2004)}]{Reyle04}
{Reyl{\'e}}, C. \& {Robin}, A.~C. 2004, \aap, 421, 643

\bibitem[{{Ruiz} {et~al.}(1990){Ruiz}, {Anguita}, {Maza}, \& {Roth}}]{Ruiz90}
{Ruiz}, M.~T., {Anguita}, C., {Maza}, J., \& {Roth}, M. 1990, \aj, 100, 1270

\bibitem[{{Ruiz} {et~al.}(1997){Ruiz}, {Leggett}, \& {Allard}}]{Kelu1}
{Ruiz}, M.~T., {Leggett}, S.~K., \& {Allard}, F. 1997, \apjl, 491, L107+

\bibitem[{{Ruiz} {et~al.}(2001){Ruiz}, {Wischnjewsky}, {Rojo}, \&
  {Gonzalez}}]{CESO}
{Ruiz}, M.~T., {Wischnjewsky}, M., {Rojo}, P.~M., \& {Gonzalez}, L.~E. 2001,
  \apjs, 133, 119

\bibitem[{{Saumon} {et~al.}(1994){Saumon}, {Bergeron}, {Lunine}, {Hubbard}, \&
  {Burrows}}]{Saumon94}
{Saumon}, D., {Bergeron}, P., {Lunine}, J.~I., {Hubbard}, W.~B., \& {Burrows},
  A. 1994, \apj, 424, 333

\bibitem[{{Schmidt} {et~al.}(2006){Schmidt}, {Cruz}, {Bongiorno},
{Reid}, \&
  {Liebert}}]{Sarah}
{Schmidt}, S.~J., {Cruz}, K.~L., {Bongiorno}, B., {Reid}, I.~N.,
\& {Liebert}, J. 2006, \aj, submitted.

\bibitem[{{Schneider} {et~al.}(2002){Schneider}, {Knapp}, {Hawley}, {Covey},
  {Fan}, {Ramsey}, {Richards}, {Strauss}, {Gunn}, {Hill}, {MacQueen}, {Adams},
  {Hill}, {Ivezi{\'c}}, {Lupton}, {Pier}, {Saxe}, {Shetrone}, {Tufts}, {Wolf},
  {Brinkmann}, {Csabai}, {Hennessy}, \& {York}}]{Schneider02}
{Schneider}, D.~P., {Knapp}, G.~R., {Hawley}, S.~L., {Covey}, K.~R., {Fan}, X.,
  {Ramsey}, L.~W., {Richards}, G.~T., {Strauss}, M.~A., {Gunn}, J.~E., {Hill},
  G.~J., {MacQueen}, P.~J., {Adams}, M.~T., {Hill}, G.~M., {Ivezi{\'c}}, {\v
  Z}., {Lupton}, R.~H., {Pier}, J.~R., {Saxe}, D.~H., {Shetrone}, M., {Tufts},
  J.~R., {Wolf}, M.~J., {Brinkmann}, J., {Csabai}, I., {Hennessy}, G.~S., \&
  {York}, D.~G. 2002, \aj, 123, 458

\bibitem[{Schultheis} {et~al.}(2006) {Schultheis}, {Robin},
{Reyl{\'e}}, {McCracken}, {Bertin}, {Mellier}, \& {Le
F{\`e}vre}]{Schultheis06}{Schultheis}, M. and {Robin}, A.~C. and
{Reyl{\'e}}, C. and
    {McCracken}, H.~J. and {Bertin}, E. and {Mellier}, Y., \& {Le F{\`e}vre},
    O. 2006, \aap, 447, 185

\bibitem[{{Siegler} {et~al.}(2005){Siegler}, {Close}, {Cruz}, {Mart{\'{\i}}n},
  \& {Reid}}]{Siegler05}
{Siegler}, N., {Close}, L.~M., {Cruz}, K.~L., {Mart{\'{\i}}n}, E.~L., \&
  {Reid}, I.~N. 2005, \apj, 621, 1023

\bibitem[{{Skrutskie} {et~al.}(2006){Skrutskie}, {Cutri}, {Stiening},
  {Weinberg}, {Schneider}, {Carpenter}, {Beichman}, {Capps}, {Chester},
  {Elias}, {Huchra}, {Liebert}, {Lonsdale}, {Monet}, {Price}, {Seitzer},
  {Jarrett}, {Kirkpatrick}, {Gizis}, {Howard}, {Evans}, {Fowler}, {Fullmer},
  {Hurt}, {Light}, {Kopan}, {Marsh}, {McCallon}, {Tam}, {Van Dyk}, \&
  {Wheelock}}]{2MASS}
{Skrutskie}, M.~F., {Cutri}, R.~M., {Stiening}, R., {Weinberg}, M.~D.,
  {Schneider}, S., {Carpenter}, J.~M., {Beichman}, C., {Capps}, R., {Chester},
  T., {Elias}, J., {Huchra}, J., {Liebert}, J., {Lonsdale}, C., {Monet}, D.~G.,
  {Price}, S., {Seitzer}, P., {Jarrett}, T., {Kirkpatrick}, J.~D., {Gizis},
  J.~E., {Howard}, E., {Evans}, T., {Fowler}, J., {Fullmer}, L., {Hurt}, R.,
  {Light}, R., {Kopan}, E.~L., {Marsh}, K.~A., {McCallon}, H.~L., {Tam}, R.,
  {Van Dyk}, S., \& {Wheelock}, S. 2006, \aj, 131, 1163

\bibitem[{{Solomon} {et~al.}(2006){Solomon}, {Cruz}, {Schmidt}, {Reid}, \&
  {Kirkpatrick}}]{Solomon}
{Solomon}, A.~R., {Cruz}, K.~L., {Schmidt}, S.~J., {Reid}, I.~N., \&
  {Kirkpatrick}, J.~D. 2006, in prep.

\bibitem[{{Thorstensen} \& {Kirkpatrick}(2003)}]{TK03}
{Thorstensen}, J.~R. \& {Kirkpatrick}, J.~D. 2003, \pasp, 115, 1207

\bibitem[{{Tinney}(1996)}]{Tinney96}
{Tinney}, C.~G. 1996, \mnras, 281, 644

\bibitem[{{Tinney} {et~al.}(2003){Tinney}, {Burgasser}, \&
  {Kirkpatrick}}]{Tinney03}
{Tinney}, C.~G., {Burgasser}, A.~J., \& {Kirkpatrick}, J.~D. 2003, \aj, 126,
  975

\bibitem[{{Tinney} {et~al.}(1993){Tinney}, {Mould}, \& {Reid}}]{TVLM}
{Tinney}, C.~G., {Mould}, J.~R., \& {Reid}, I.~N. 1993, \aj, 105, 1045

\bibitem[{{Tinney} {et~al.}(1995){Tinney}, {Reid}, {Gizis}, \&
  {Mould}}]{Tinney95}
{Tinney}, C.~G., {Reid}, I.~N., {Gizis}, J., \& {Mould}, J.~R. 1995, \aj, 110,
  3014

\bibitem[{{van Altena} {et~al.}(1995){van Altena}, {Lee}, \&
  {Hoffleit}}]{vanAltena}
{van Altena}, W.~F., {Lee}, J.~T., \& {Hoffleit}, E.~D. 1995, {The general
  catalogue of trigonometric [stellar] paralaxes} (New Haven, CT: Yale
  University Observatory, |c1995, 4th ed., completely revised and enlarged)

\bibitem[{{Vrba} {et~al.}(2004){Vrba}, {Henden}, {Luginbuhl}, {Guetter},
  {Munn}, {Canzian}, {Burgasser}, {Kirkpatrick}, {Fan}, {Geballe},
  {Golimowski}, {Knapp}, {Leggett}, {Schneider}, \& {Brinkmann}}]{Vrba04}
{Vrba}, F.~J., {Henden}, A.~A., {Luginbuhl}, C.~B., {Guetter}, H.~H., {Munn},
  J.~A., {Canzian}, B., {Burgasser}, A.~J., {Kirkpatrick}, J.~D., {Fan}, X.,
  {Geballe}, T.~R., {Golimowski}, D.~A., {Knapp}, G.~R., {Leggett}, S.~K.,
  {Schneider}, D.~P., \& {Brinkmann}, J. 2004, \aj, 127, 2948

\bibitem[{{Wilson}(2002)}]{Wilson}
{Wilson}, J.~C. 2002, Ph.D.~Thesis, Cornell Univ.

\end{thebibliography}
